\documentclass[12pt]{article}

\usepackage{amssymb,amsmath,amsbsy,amsthm}
\usepackage{latexsym,graphicx}
\usepackage{authblk}

\title{\Large{\bf{Exact solution of a 2D interacting fermion model}}}

\date{\vspace{-1.0cm}\small November 7, 2011\vspace{0.2cm}}

\author[1,*]{Jonas de Woul}
\author[1,\dag]{Edwin Langmann}
\affil[1]{Department of Theoretical Physics, Royal Institute of Technology (KTH)\newline SE-106 91 Stockholm, Sweden \vspace{2mm}}

\newcommand{\newsection}{\setcounter{equation}{0}\section}

\renewcommand{\appendix}{\setcounter{equation}{0}\setcounter{section}{0}\renewcommand{\thesection}{\Alph{section}}}

\theoremstyle{plain}
\newtheorem{theorem}{Theorem}[section]
\newtheorem{proposition}[theorem]{Proposition}
\newtheorem{lemma}[theorem]{Lemma}
\newtheorem{corollary}[theorem]{Corollary}
\newtheorem{result}[theorem]{Result}
\newtheorem{conjecture}[theorem]{Conjecture}
\theoremstyle{definition}
\newtheorem{definition}[theorem]{Definition}
\theoremstyle{remark}
\newtheorem{remark}[theorem]{Remark}

\newcommand{\BZ}{\mathrm{BZ}}

\newcommand{\QED}{\qed}
\renewcommand{\cong}{\leftrightarrow} 
\newcommand{\pdag}{^{\phantom\dag}}
\newcommand{\define}{\stackrel{\mbox{{\tiny def}}}{=}}
\newcommand{\Ref}[1]{(\ref{#1})}
\newcommand{\tint}[1]{\;\,\tilde{\!\!\!\int_{{#1} }}}
\newcommand{\xx}{\stackrel {\scriptscriptstyle \times}{\scriptscriptstyle \times}}
\newcommand{\xxa}{\stackrel {\scriptscriptstyle \times}{\scriptscriptstyle \times} \!}
\newcommand{\xxe}{\! \stackrel {\scriptscriptstyle \times}{\scriptscriptstyle \times}}
\newcommand{\sgn}{\mathrm{sgn}}
\newcommand{\ud}{\mathrm{d}}
\newcommand{\cF}{\mathcal{F}}
\newcommand{\cA}{\mathcal{A}}
\newcommand{\cD}{\mathcal{D}}
\newcommand{\cO}{\mathcal{O}}
\newcommand{\cZ}{\mathcal{Z}}
\newcommand{\cE}{\mathcal{E}}
\newcommand{\cQ}{\mathcal{Q}}
\newcommand{\cU}{\mathcal{U}}
\newcommand{\cL}{\mathcal{L}}
\newcommand{\cV}{\mathcal{V}}

\newcommand{\tPiL}{\frac{2\pi}{L}}
\newcommand{\zeromode}{Q} 
\newcommand{\ttpV}{$t$-$t'$-$V\,$}
\newcommand{\ta}{\tilde{a}}
\newcommand{\ee}{\,{\rm e}}
\newcommand{\ii}{{\rm i}}
\newcommand{\vx}{{\bf x}}
\newcommand{\vy}{{\bf y}}
\newcommand{\vn}{{\bf n}}
\newcommand{\vm}{{\bf m}}
\newcommand{\vnu}{\boldsymbol{\nu}}
\newcommand{\valpha}{\boldsymbol{\alpha}}
\newcommand{\vlambda}{\boldsymbol{\lambda}}
\newcommand{\vQ}{{\bf Q}}
\newcommand{\vk}{{\bf k}}
\newcommand{\ve}{{\bf e}}

\newcommand{\vp}{{\bf p}}
\newcommand{\vzero}{{\bf 0}}
\newcommand{\Z}{{\mathbb Z}}
\newcommand{\R}{{\mathbb R}}
\newcommand{\N}{{\mathbb N}}
\newcommand{\C}{{\mathbb C}}

\begin{document}

\maketitle

\let\oldthefootnote\thefootnote
\renewcommand{\thefootnote}{\fnsymbol{footnote}}
\footnotetext[1]{Electronic address: {\tt jodw02@kth.se}}
\footnotetext[2]{Electronic address: {\tt langmann@kth.se}}
\let\thefootnote\oldthefootnote

\vspace{-1.5cm}

\begin{abstract}
We study an exactly solvable quantum field theory (QFT) model describing interacting fermions in 2+1 dimensions. This model is motivated by physical arguments suggesting that it provides an effective description of spinless fermions on a square lattice with local hopping and density-density interactions if, close to half filling, the system develops a partial energy gap. The necessary regularization of the QFT model is based on this proposed relation to lattice fermions.  We use bosonization methods to diagonalize the Hamiltonian and to compute all correlation functions. We also discuss how, after appropriate multiplicative renormalizations, all short- and long distance cutoffs can be removed. In particular, we prove that the renormalized two-point functions have algebraic decay with non-trivial exponents depending on the interaction strengths, which is a hallmark of Luttinger-liquid behavior. 
\end{abstract}

\newsection{Introduction} \label{sec1} 
In this paper, we study a QFT model\footnote{By this we mean a quantum model with infinitely many degrees of freedom.} describing interacting fermions in two spatial dimensions (2D). We will refer to this as {\em the Mattis model} since it is similar to a model first proposed by Mattis \cite{Mattis} as an exactly solvable model potentially relevant for 2D lattice fermion systems.\footnote{The precise relation of our model to Mattis' is explained in Section~\ref{Final remarks}, Remark~8.}  In recent work, we presented specific physical arguments that the Mattis model provides an effective description of spinless fermions on a square lattice with local density-density interactions if, close to half filling, parts of the underlying Fermi surface have no low-energy excitations \cite{EL0,EL1,dWL1}. This proposed relationship gives natural short- and long distance cutoffs for the Mattis model, and it fixes the parameters of the Mattis model in terms of the lattice model parameters \cite{EL1,dWL1}. The aim of this paper is to present a detailed and mathematically precise construction and solution of the Mattis model.

\subsection{Fermi arcs} 
\label{Physics motivation}
The derivation of the Mattis model from a lattice fermion system in \cite{EL1,dWL1} is not rigorous in the mathematical sense but relies partly on physical arguments. The latter were inspired by arguments that have been applied successfully to other low-dimensional systems; for example, the Luttinger model \cite{Tomonaga,Thirring,Luttinger,LiebMattis} can be derived in a similar manner from a system of spinless fermions on a one-dimensional lattice. A mathematically inclined reader is free to ignore the details of this physics background and take the Mattis model at face value:\ as an, in our opinion, interesting exactly solvable model for interacting fermions in 2+1 dimensions. We therefore give in the following only a short, intuitive description of the Mattis model and its relation to lattice fermions. Readers interested in further details are referred to Appendix~\ref{Lattice relation}. 

\begin{figure}[!ht]
\begin{center}
\hspace{-1cm}
\includegraphics[width=1.0\textwidth]{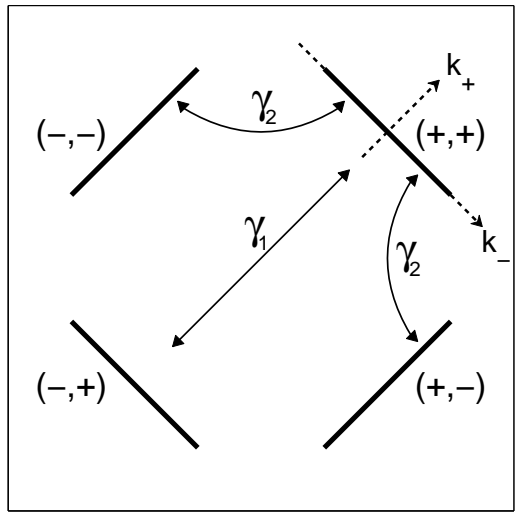} 
\end{center}
\vspace{-0.5cm}
\caption{A truncated square Fermi surface in Fourier space consisting of four disjoint Fermi arcs labeled by $(r,s)$ with $r,s=\pm$. Indicated in the figure are also the definition of momenta $k_\pm$ near the Fermi arc $(+,+)$ and the interactions of the $(+,+)$ arc fermions: they interact with opposing $(-,+)$ and neighboring $(\pm,-)$ arc fermions with couplings proportional to $\gamma_1$ and $\gamma_2$, respectively. The other interactions are fixed by invariance of the model under rotations by $\pi/2$.}
\label{Truncated Fermi surface}
\end{figure}

The Mattis model describes a coupled system of four different flavors $(r,s)$, $r,s=\pm$, of spinless fermions. It consists of two sets of parallel 1D Luttinger chains (corresponding to $s=\pm$), with each chain consisting of left- and right movers (corresponding to $r=\pm$). Its relation to the original lattice fermion model is best explained using Figure~\ref{Truncated Fermi surface}:\ shown in the figure are four line segments in Fourier space that remain after a square Fermi surface has been truncated around its corners. We refer to these line segments as {\em Fermi arcs}. Each Fermi arc is associated with one-particle degrees of freedom identified by flavor indices $(r,s)$ and momenta $\vk$. The truncated regions correspond to one-particle degrees of freedom that are assumed to be gapped by interactions and therefore are ignored. 

The fermions thus are represented by field operators $\hat\psi_{r,s}(\vk)$ in Fourier space, and the free part of the Mattis Hamiltonian has the form (see Section~\ref{The Mattis model} for the precise mathematical definition)
\begin{equation}
\label{IR limit of free part}
H_0 = \sum\limits_{r,s=\pm} \int{\ud^2k \; \epsilon_{r,s}(\vk):\!  \hat\psi^\dag_{r,s}(\vk) \hat\psi\pdag_{r,s}(\vk)\!: }
\end{equation}
with dispersion relations linear in $k_s$ (orthogonal to each Fermi arc) and constant in $k_{-s}$ (parallel to each Fermi arc), i.e.\ $\epsilon_{r,s}(\vk)=rv_Fk_s$ with a model parameter $v_F>0$ (referred to as {\em Fermi velocity}). The integration domain in \Ref{IR limit of free part} is unbounded in the $k_s$-direction, while $\left|k_{-s}\right|\leq \pi/\ta$ with $2\pi/\ta$ the length of each Fermi arc. 

The interaction part of the Mattis Hamiltonian consists of density-density interactions between different Fermi arc fermions. There are two different coupling parameters:\ $\gamma_1$ for interactions involving fermions on opposite arcs, and $\gamma_2$ for fermions on nearby arcs; see Figure~\ref{Truncated Fermi surface}. As seen in Section~\ref{The Mattis model}, the possible momentum exchange $\vp$ in these interactions are restricted such that $\left|p_\pm\right|\leq \pi/\ta$. Thus the length parameter $\ta$, which is proportional to the lattice constant in the original lattice fermion model, serves as a natural ultraviolet (UV) cutoff in the Mattis model. There is also a natural infrared (IR) cutoff provided by the linear size $L$ of the original square lattice.   

\subsection{Summary of results}
\label{Summary}
The Mattis model can be formally\footnote{We suppress details of the UV regularization here.} defined by the Hamiltonian
\begin{equation}
\label{Formal position space Hamiltonian}
\begin{split} 
H =& \sum_{r,s=\pm} \int \ud^2 x\, \Bigl(  rv_F :\! \psi^\dag_{r,s}(-\ii \partial_s)\psi_{r,s}\pdag\!: \\ & + \sum_{r',s'=\pm} g_{r,s,r',s'} :\!\psi^\dag_{r,s}\psi_{r,s}\pdag\!: :\!\psi^\dag_{r',s'}\psi_{r',s'}\pdag\!:  \Bigr)
\end{split} 
\end{equation}
with $\partial_{\pm}=\partial/\partial x_\pm$ and $x_\pm$ Cartesian coordinates of positions $\vx$. The $\psi_{r,s}^{(\dag)}=\psi_{r,s}^{(\dag)}(\vx)$ are fermion field operators obeying canonical anticommutator relations $\{\psi\pdag_{r,s}(\vx), \psi^{\dag}_{r',s'}(\vy)\} = \delta_{r,r'}\delta_{s,s'}\delta(\vx-\vy)$, etc.  As already mentioned, the model parameters include the Fermi velocity $v_F>0$, the coupling constants $g_{r,s,r',s'}$ parameterized as
\begin{equation}
\label{coupling constants}
g_{r,s,r',s'} = \ta\pi v_F\bigl(\delta_{s,s'}\delta_{r,-r'}\gamma_1+\delta_{s,-s'}\gamma_2/2\bigr) 
\end{equation} 
with dimensionless parameters $\gamma_{1,2}$, and the UV cutoff $\ta$. The colons indicate fermion normal ordering.

Our first aim of this paper is to give a rigorous definition of the Mattis model; see Section~\ref{The Mattis model}. For that we consider the regularized version obtained in \cite{EL1} (in the special case $\gamma_1=\gamma_2>0$). This regularized Mattis model describes an infinite number of fermion degrees of freedom, but it comes with particular short- and long distance regularizations controlled by the UV- and IR cutoff parameters $\ta$ and $L$, respectively. As we will see, this regularized Mattis model is well-defined if and only if
\begin{equation}
\label{Coupling restrictions}
|\gamma_1|<1,\qquad |\gamma_2|<|1+\gamma_1|. 
\end{equation} 
We emphasize that this parameter restriction does not mean that the Mattis model is weakly coupled:\ as shown in \cite{EL1}, the scaling of the coupling \Ref{coupling constants} naturally arises in a partial continuum limit of the above-mentioned lattice system, and $\gamma_1=\gamma_2=1/2$, for example, does not correspond to weak coupling in the latter system.

A key result (see Section~\ref{Bosonization}) is to make mathematically precise the equivalence of the Mattis model to a non-interacting boson model formally defined by the Hamiltonian \cite{EL1}
\begin{equation}
H = \frac{v_F}2 \sum_{s=\pm}\int \ud^2x\,:\! \Bigl((1-\gamma_1)\Pi_s^2 +(1+\gamma_1)\bigl( \partial_s\Phi_s\bigr)^2 +\gamma_2\bigl(\partial_s\Phi_s\bigr) \bigl(\partial_{-s}\Phi_{-s}\bigr) \Bigr)\!:   
\end{equation}
with 2D boson field operators $\Pi_s=\Pi_s(\vx)$ and $\Phi_s=\Phi_s(\vx)$ satisfying canonical commutator relations $[\Pi_s(\vx),\Phi_{s'}(\vy)]=-\ii\delta_{s,s'}\delta(\vx-\vy)$, etc. The boson operators are related to the fermions as 
\begin{equation}
\label{PiPhi from J}
\begin{split}
  \partial_s\Phi_s&=\sqrt{\pi\ta}\bigl(:\!  \psi^\dag_{+,s}\psi_{+,s}\pdag\!: + :\!  \psi^\dag_{-,s}\psi_{-,s}\pdag\!: \bigr) \\   \Pi_s &= -\sqrt{\pi\ta}\bigl(:\! \psi^\dag_{+,s}\psi_{+,s}\pdag\!: - :\!  \psi^\dag_{-,s}\psi_{-,s}\pdag\!: \bigr) 
\end{split} 
\end{equation}
at finite UV cutoff $\ta$. This relation, and the non-trivial scaling of the coupling constants $g_{r,s,r',s'}$ with $\ta$, make it clear that the UV cutoff in the Mattis model is essential. As we will see, the full continuum limit $\ta\to 0^+$ of the model, referred to as {\em the UV limit} in the following, makes sense only after non-trivial multiplicative renormalizations. Similarly, $L\to\infty$ is referred to as {\em the IR limit}. 

Since the regularized Mattis Hamiltonian can be written as a system of free bosons, it can be diagonalized by analytical methods. The result is given in Theorem~\ref{Solution}.  We also compute the free energy (Result~\ref{Result free energy}), the density-density correlation functions (Result~\ref{density correlation result}), and the fermion $N$-point correlation functions (Result~\ref{fermion correlation result}). We note that all these results are exact in the thermodynamic limit, and they depend on the UV cutoff $\ta$. Our results suggest that it is possible to remove the UV cutoff and, by appropriate renormalizations, obtain non-trivial results for all observables of the Mattis model. In the present paper, we give detailed proofs of this for the free energy and the fermion two-point functions (the latter at zero temperature); see Results~\ref{Result free energy 2}--\ref{QFT limit Fermion two-point function}.  We find that these two-point functions in this limit have an algebraic decay with exponents that depend non-trivially on both coupling parameters $\gamma_{1,2}$, which is one characteristic of Luttinger-liquid behavior \cite{Haldane,Anderson1}. We also discuss how our results can be used to substantiate the derivation of the Mattis model from the original lattice fermion system \cite{EL1}; see Section~\ref{Final remarks}, Remark~5.
   
\subsection{Related work}
The present paper is the third in a series \cite{EL1,dWL1} in which we propose and develop a method to do reliable, non-perturbative computations in models of strongly interacting fermions on a 2D square lattice \cite{EL0}. We have so far considered the simplest non-trivial case, the so-called 2D \ttpV model, describing spinless fermions with local interactions. In \cite{EL1}, a 2D analogue of the Luttinger model was derived from the 2D \ttpV model using a particular partial continuum limit.\footnote{The interested reader can find more details in Appendix~\ref{Lattice relation}.} This 2D Luttinger model is an extension of the Mattis model by so-called \textit{antinodal fermions} that cannot be bosonized. In \cite{dWL1} we showed that, within a mean field approximation, there is a non-trivial phase away from half-filling in which the antinodal fermions are gapped, and in this phase the 2D Luttinger model reduces to the Mattis model. One key ingredient of our approach is the notion of underlying Fermi surface arcs (not necessarily corresponding to a true Fermi surface) in the 2D \ttpV model away from half filling. Previous theoretical work predicting four disconnected Fermi surface arcs in 2D interacting lattice fermion systems include renormalization group studies in \cite{FRS,HSFR}. 

One of our main tenets is that progress in understanding these lattice models can be achieved if one clearly distinguishes between approximations justified by physical arguments and manipulations that are mathematically exact (rigorous); the present paper belongs to the latter category. We mention that the model derived in \cite{EL1} corresponds to the special case $\gamma\define \gamma_1=\gamma_2>0$; the generalization to different coupling parameters, allowing also for negative values, is natural from a mathematical point of view. We also note that, while an outline of how to diagonalize the Mattis Hamiltonian by bosonization was already given in \cite{EL1}, Section~6.2, the details and complete solution of the model are given here for the first time. 

We choose the name of the model studied in this paper to acknowledge a pioneering paper published by Mattis in 1987 in which he pointed out that a 2D interacting fermion Hamiltonian similar to ours can be mapped exactly to a non-interacting boson Hamiltonian \cite{Mattis}; the dispersion relations of this boson Hamiltonian were given by Mattis (see \cite{Mattis}, Equation (8)) but no details were provided. Mattis also argued that this model can arise from a tight-binding description of 2D lattice fermions at half-filling with a square Fermi surface and in the absence of nesting-type instabilities, but, as shown in \cite{EL1}, this latter proposal is doubtful (at half-filling, there are additional interaction terms that cannot be bosonized in a simple manner and which are likely to yield a gap, very different to what the Mattis model describes). 

Mattis' insight described above has received little attention up to now, and it was re-discovered in 1994 independently by Hlubina \cite{Hlubina} and Luther \cite{Luther2}. Hlubina presented a model similar to Mattis' that he claimed can be mapped to a non-interacting boson Hamiltonian, and he diagonalized the latter. He also gave phenomenological arguments suggesting that this model describes a two-dimensional Luttinger liquid. However,  from his discussion, it is not clear what the solvable model actually is (the model formulated in \cite{Hlubina}, Equation (1), is {\em not} exactly solvable but needs to be modified --- the necessary changes are similar to ones discussed in detail in \cite{EL1}). Luther studied spinfull fermions with a square Fermi surface and argued that the non-interacting part of the Hamiltonian can be simplified to one that can be bosonized exactly, and which is essentially the same as in Mattis' model. Luther diagonalized the bosonized Hamiltonian that one obtains by also including density-density interactions, initially restricting to interactions between opposite Fermi surface faces only \cite{Luther2} (this restriction corresponds to $\gamma_2=0$ in our notation). This was later extended to also include interactions between adjacent faces in \cite{SL}.

Our solution of the Mattis model not only includes the diagonalization of the bosonized Hamiltonian, but also the computation of all fermion correlation functions using the boson-fermion correspondence formula, which is an operator version of a formula that first appeared in \cite{SchotteSchotte}. In regards to this correspondence formula, we emphasize the role of the, at least in the condensed matter literature, sometimes overlooked {\em Klein factors}, i.e.\ the unitary operators that lower- or raise fermion excitation numbers when acting on states in the Fock space. For one-dimensional systems, the unitary Klein factors often combine to unity when computing physically interesting correlation functions, and they can then be safely ignored, as was done in the well-known work by Luther and Peschel \cite{LP}. However, the situation in higher dimension is different due to the profusion of flavor indices; previous work on computing correlation functions by Luther and collaborators \cite{FjaerestadSudboLuther1999,SL}  neglected to incorporate Klein factors, and their results are therefore not directly applicable to the model studied in this paper. In particular, as will be shown, Klein factors cannot be ignored in the solution of the Mattis model. In the condensed matter literature, the importance of Klein factors has been pointed out in papers by Haldane \cite{Haldane1979,Haldane} and Heidenreich {\em et al.} \cite{HSU} in the context of the Luttinger model; see also \cite{vonDelftSchoeller} for a review on constructive bosonization and Klein factors, and Remark~\ref{Klein factor remark} in Appendix~\ref{Bosonization identities}. Our formulation and solution of the Mattis model is based on mathematical formulations of these results proved in \cite{Frenkel,CH,CR,Kac,CL}, for example. 

An important question is whether the Mattis model describes a 2D Luttinger liquid \cite{Anderson1}. Different regularizations and treatments of Mattis-like models can give different results, as exemplified by Refs.\  \cite{VM} and \cite{ZYD}: these works are on such models that differ in regularization details of the interaction, they use different methods, and they obtain different results.  This suggests to us that it is important to be mathematically precise both in the definition of the model and of Luttinger-liquid behavior, as well as in the methods used to study them.

There exist many other papers studying 2D fermion systems by mapping them to bosonic models; see e.g.\ \cite{HKM} and references therein. However, these differ significantly in detail from our constructive bosonization methods. In particular, many of the various approaches, as indicated above, do not establish an operator correspondence between fermions and bosons, thus precluding the exact computation of fermion correlation functions.

\subsection{Notation and conventions}
We write ``$\define$'' to emphasize that an equation is a definition. For $z\in\mathbb{C}$, the complex conjugate is written $\overline z$, the real part is $\Re(z)$, the imaginary part is $\Im(z)$, and the argument (phase) is $\arg(z)$. We write a 2D vector as ${\bf u} = u_+\ve_+ + u_-\ve_-$, with $\ve_\pm\cdot\ve_\pm=1$ and $\ve_\pm\cdot\ve_\mp=0$. If $\mathcal{S}$ is a set, $x\in y\mathcal{S}$ and $x\in y+\mathcal{S}$ means $xy^{-1}\in \mathcal{S}$ and $x-y\in \mathcal{S}$, respectively. We identify $1$ with the identity operator. Unless otherwise stated, and when there is no risk of confusion, a free variable or index in an identity can take any allowed value. For example, the relations in \Ref{CAR} below are true for all $r,s,r',s'=\pm$ and $\vk,\vk'\in\Lambda^*_s$, whereas in equation \Ref{Highest Weight Condition}, the identity holds true for all $r,s=\pm$, but only those $\vk\in\Lambda^*_s$ satisfying the given condition.

\subsection{Plan of paper}
A detailed definition of the Mattis model, including all regularizations, is given in Section~\ref{Prerequisites}. This section also contains the mathematical results needed to bosonize the Mattis model, and a technical discussion on operator domains and the essential self-adjointness of the Mattis Hamiltonian (this latter part may be skipped without loss of continuity). In Section~\ref{Solution of the Mattis model}, we solve the Mattis model by diagonalizing the Hamiltonian and by computing all correlations functions. In Section~\ref{QFT limit}, we give the renormalized free energy and fermion two-point functions after all short- and long distance cutoffs have been removed. Section~\ref{Final remarks} contains final remarks. Technical details are deferred to Appendices~\ref{Lattice relation}-\ref{Index sets}. Appendix~\ref{Lattice relation} gives further details on the relation of the Mattis model to lattice fermions.  In Appendix~\ref{Bosonization identities}, we summarize well-known mathematical results on constructive bosonization in one dimension, and we explain how these are used to prove the results in Section~\ref{Prerequisites}. Appendix~\ref{Solution appendix} contains computational details of the solution of the Mattis model. Details on the removal of short- and long distance cutoffs in the Mattis model can be found in Appendix~\ref{Appendix QFT limit}. In Appendix~\ref{appX}, we discuss certain terms that were neglected in the derivation of the Mattis Hamiltonian from the lattice fermion Hamiltonian in \cite{EL1}. Appendix~\ref{Index sets} lists some index sets used throughout the paper. 

\newsection{Prerequisites}
\label{Prerequisites}
In this section, we define the regularized Mattis model in terms of fermion creation- and annihilation operators. We also summarize the mathematical results needed to bosonize the model.

\subsection{The Mattis model}
\label{The Mattis model}
We introduce fermion field operators $\hat\psi^{(\dag)}_{r,s}(\vk)$ labeled by flavor indices $r,s=\pm$ and momenta $\vk$ in the following sets\footnote{This and several other index sets are collected in Appendix~\ref{Index sets} for easy reference.}
\begin{equation}
\label{Continuum nodal momentum region}
\Lambda_{s}^* \define \left\{\vk\in \tPiL\left(\mathbb{Z} +\frac{1}{2}\right)^2 \; : \; -\frac{\pi}{\ta} \leq k_{-s}< \frac{\pi}{\ta} \right\}
\end{equation}
with parameters $\ta>0$ and $L>0$ such that $L/\ta$ is an odd integer. These operators obey the canonical anticommutator relations
\begin{equation}
\label{CAR}
\left\{\hat\psi\pdag_{r,s}(\vk),\hat\psi^\dag_{r',s'}(\vk')\right\} = \Bigl(\frac{L}{2\pi}\Bigr)^2 \delta_{r,r'}\delta_{s,s'}\delta_{\vk,\vk'},\qquad \left\{\hat\psi\pdag_{r,s}(\vk),\hat\psi\pdag_{r',s'}(\vk')\right\}= 0 
\end{equation}
and are defined on a fermion Fock space $\cF$ such that $\hat\psi\pdag_{r,s}(\vk)$ is the Hilbert space adjoint of $\hat\psi^\dag_{r,s}(\vk)$. The representation of these operators on $\cF$ is specified by the existence of a vacuum state (Dirac sea) $\Omega\in\cF$ satisfying
\begin{equation}
\label{Highest Weight Condition}
\hat\psi\pdag_{r,s}(\vk)\Omega = \hat\psi^\dag_{r,s}(-\vk)\Omega= 0, \quad \forall \vk\; \mbox{ such that }\; rk_s>0,
\end{equation} 
and $\langle\Omega,\Omega\rangle=1$, with $\langle\cdot,\cdot\rangle$ the inner product in $\cF$. We define normal ordering of fermion bilinears $\cO=\hat\psi^\dag_{r,s}(\vk) \hat\psi\pdag_{r',s'}(\vk')$ by 
\begin{equation}
\label{NormalOrder}
:\!\cO\!:\, \define \cO - \langle \Omega,\cO\Omega\rangle . 
\end{equation}
The same definition will be used for boson bilinears. 

The following characteristic (cutoff) function will be needed
\begin{equation}
\label{chis}
\chi(\vp)\define\begin{cases}
1& \text{if } - {\pi}/{\ta} \le p_{\pm} < {\pi}/{\ta}\\
0& \text{otherwise}
\end{cases}.
\end{equation}
We also introduce 
\begin{equation}
\label{Nodal density operators}
\hat{J}_{r,s}(\vp) \define  \sum_{\vk_1,\vk_2\in\Lambda^*_s}\Bigl(\tPiL\Bigr)^2:\!\hat\psi^\dag_{r,s}(\vk_1) \hat\psi\pdag_{r,s}(\vk_2)\!:\sum_{n\in\mathbb{Z}}\delta_{\vk_1+\vp,\vk_2+2\pi n\ta^{-1}\ve_{-s}}  
\end{equation}
for $\vp\in(2\pi/L)\mathbb{Z}^2$ and $r,s=\pm$. We refer to $\hat{J}_{r,s}(\vp)$ as {\em (fermion) densities}. We use the same name for the Fourier transform $J_{r,s}(\vx)$ of these operators (to be defined below). 

\begin{definition}
\label{Mattis model} 
Let $v_F>0$ and $g_{r,s,r',s'}$ as in \Ref{coupling constants}--\Ref{Coupling restrictions}. Then the regularized Mattis Hamiltonian is given by
\begin{equation}
\label{Regularized Mattis Hamiltonian}
H=H_0+H_1
\end{equation}
with the free part
\begin{equation}
\label{Free part of regularized Mattis Hamiltonian}
H_0=v_F\sum_{r,s=\pm}\sum_{\vk\in\Lambda_s^*}\Bigl(\tPiL\Bigr)^2 rk_s :\! \hat\psi^\dag_{r,s}(\vk) \hat\psi\pdag_{r,s}(\vk)\!: 
\end{equation}
and the interaction part 
\begin{equation}
\label{Interaction part of regularized Mattis Hamiltonian}
H_1= \sum_{r,r',s,s'=\pm}\sum_{\vp\in\frac{2\pi}{L}\mathbb{Z}^2}\Bigl(\frac{1}{L}\Bigr)^2 g_{r,s,r',s'}\chi(\vp)\hat{J}_{r,s}(-\vp)\hat{J}_{r',s'}(\vp).
\end{equation}
\end{definition}

Our results in Section~\ref{Solution of the Mattis model} prove that the regularized Mattis Hamiltonian is a self-adjoint operator bounded from below. 

\begin{remark} 
\label{Remark unbounded}
The densities in \Ref{Nodal density operators} and the Hamiltonians in \Ref{Regularized Mattis Hamiltonian}--\Ref{Interaction part of regularized Mattis Hamiltonian} are all unbounded operators, and a complete definition of the Mattis model thus require both a specification of their common domain of definition $\cD$, and a proof that $H_0$ and $H$ are essentially self-adjoint on $\cD$ \cite{ReedSimon1,ReedSimon2}. This can be done using general results on operator algebras arising in quasi-free boson- and fermion quantum field theories \cite{GL}; the interested reader can find the details in Section~\ref{Unboundedness}. An informal description of this goes as follows: In Lemma~\ref{Lemma Fock space basis}(b) below, we give a complete orthonormal 
set of exact eigenstates of $H_0$, and this proves essential self-adjointness of $H_0$ on the vector space $\cD$ of finite linear combinations of these eigenstates. Moreover, it is easy to see that $\cD$ is a common, invariant domain for the operators in \Ref{Nodal density operators}--\Ref{Interaction part of regularized Mattis Hamiltonian}, and for other operators to be introduced below.  For this reason, all identities involving commutators of unbounded operators that we write down are well-defined as identities on $\cD$.  Strictly speaking, one should distinguish between an unbounded operator $A$ defined on $\cD$ and its closure $\bar A$ (see e.g.\ \cite{ReedSimon1}, Section~VIII.1), but, as can be inferred from Section~\ref{Unboundedness}, it is safe to abuse notation and use the same notation for both operators. Finally, one can prove directly that the Mattis Hamiltonian $H$ is essentially self-adjoint on $\cD$ (Corollary~\ref{Mattis Hamiltonian is selfadjoint}), but, as already mentioned, this fact is also implied by our exact solution of the Mattis model in Section~\ref{Solution of the Mattis model} (in which a complete orthonormal set of exact eigenstates of $H$, all contained in $\cD$, is found).
\end{remark}

We can write the regularized Mattis model in position space, thereby making precise the formal Hamiltonian given in the introduction. Our conventions for the Fourier transform of fermion field operators are
\begin{equation}
\label{Position space operators}
\psi_{r,s}(\vx) \define \frac1{2\pi} \sum_{\vk\in\Lambda^*_{s}}\Bigl(\tPiL\Bigr)^2\hat\psi_{r,s}(\vk)\ee^{\ii\vk\cdot\vx}
\end{equation} 
with $\vx$ in the following ``position space'' corresponding to \Ref{Continuum nodal momentum region}, 
\begin{equation}
\label{Nodal position space}
\Lambda_s \define \left\{\vx\, : \, x_s\in\mathbb{R},\; x_{-s}\in \ta\mathbb{Z},\; -\frac{L}{2}\leq x_\pm < \frac{L}{2} \right\}, 
\end{equation}
which is different for $s=+$ and $s=-$. Note that this is not a subset of the lattice in the original fermion model \cite{EL1}: one component in $\vx\in\Lambda_s$ takes on continuous values, the other discrete ones, and the ``lattice constant'' $\ta$ differs from the original one. 

The operators in \Ref{Position space operators} satisfy the anticommutation relations 
\begin{equation}
\{\psi\pdag_{r,s}(\vx), \psi^{\dag}_{r',s'}(\vy)\} = \delta_{r,r'}\pdag\delta_{s,s'}\pdag\tilde\delta_s\pdag(\vx-\vy)
\end{equation}
with the partially regularized Dirac delta function 
\begin{equation}
\tilde\delta_s(\vx) \define \delta(x_s)\frac1{\ta}\delta_{x_{-s},0}. 
\end{equation}
Likewise, the Fourier transform of the densities in \Ref{Nodal density operators} are defined as\footnote{Only $J_{r,s}(\vx)$ is needed in this section, while $J_{r,s}(\vx;\epsilon)$ for $\epsilon>0$ is used in Section~\ref{Density correlation functions}.}
\begin{equation}
\label{Nodal densities in position space}
\begin{split}
J_{r,s}(\vx)\define&\lim_{\epsilon\to 0^+} J_{r,s}(\vx;\epsilon)\\ J_{r,s}(\vx;\epsilon) \define& \sum_{\vp\in \tilde\Lambda^*_{s}}\Bigl(\frac1{L}\Bigr)^2\hat{J}_{r,s}(\vp)\ee^{\ii\vp\cdot\vx}\ee^{-\epsilon |p_s|/2} \quad (\epsilon>0)
\end{split} 
\end{equation} 
with 
\begin{equation}
\label{Boson momentum set}
\tilde\Lambda_{s}^* \define \left\{\vp\in \tPiL\mathbb{Z}^2 \; : \; -\frac{\pi}{\ta} \leq p_{-s}< \frac{\pi}{\ta} \right\}. 
\end{equation}
One finds that 
\begin{equation}
\label{Jrs}
J_{r,s}(\vx) = \; :\!\psi^\dag_{r,s}(\vx)\psi\pdag_{r,s}(\vx)\!:, 
\end{equation} 
i.e.\ the fermion densities are local operators. We introduce the following notation
\begin{equation}
\label{tint}
\tint{s}\ud^2x\, \define \int_{-L/2}^{L/2}\ud x_s \sum_{x_{-s}\in\Lambda_{1\rm D}}\ta  
\end{equation}
with a 1D lattice 
\begin{equation}
\label{1D lattice}
\Lambda_{1\rm D} \define \left\{ x \in \ta\mathbb{Z}\; : \;  -\frac{L}{2}\leq x <\frac{L}{2} \right\}
\end{equation}
for the discrete component in \Ref{Nodal position space}. The regularized Mattis Hamiltonian can then be written in ``position space'' as (cf.\ \Ref{Formal position space Hamiltonian})
\begin{equation}
\label{Mattis Hamiltonian in position space} 
\begin{split}
H =& \sum_{r,s}\tint{s}\ud^2x\Bigl(r v_F  \!:\!\psi^\dag_{r,s}(\vx)(-\ii\partial_s)\psi\pdag_{r,s}(\vx)\!: 
\\
&+ \sum_{r',s'} g_{r,s,r',s'}\tint{s'} \ud^2y\,J_{r,s}(\vx)\tilde\delta(\vx-\vy)J_{r',s'}(\vy) \Bigr)
\end{split}
\end{equation}
with the fully regularized Dirac delta function
\begin{equation}
\tilde\delta(\vx) \define\sum_{\vp\in\frac{2\pi}{L}\mathbb{Z}^2} \Bigl(\frac1{L}\Bigr)^2\chi(\vp) \ee^{-\ii\vp\cdot\vx}.  
\end{equation}

\begin{remark}
For future reference, we note that $\psi_{r,s}(\vx)$ in \Ref{Position space operators} and $J_{r,s}(\vx)$ in \Ref{Nodal densities in position space} are well-defined for all $\vx\in\R^2$, although, to invert \Ref{Position space operators}, only $\psi_{r,s}(\vx)$ for $\vx\in\Lambda_s$ is needed: $\hat\psi_{r,s}(\vk)=\tint{s}\ud^2x\, \psi_{r,s}(\vx)\ee^{-\ii\vk\vx}/(2\pi)$, and similarly for $J_{r,s}(\vx)$. Later, when considering correlation functions in Section~\ref{Solution of the Mattis model} and \ref{QFT limit}, we find it convenient to consider the variable $\vx$ as continuous.
\end{remark}

\begin{remark}
\label{Relation Mattis model lattice fermions}
For simplicity in notation, we use in this paper a UV regularization that differs somewhat from the one obtained in \cite{EL1}. We expect that details of the UV regularization can be changed without significantly changing physically relevant results. Moreover, the generalization of our results to the more complicated UV regularization in \cite{EL1} is, in principle, straightforward. In any case, both regularizations coincide in the special case $\kappa=1/2$ (see Appendix~\ref{Lattice relation} for the definition of $\kappa$).
\end{remark}

\subsection{Bosonization}
\label{Bosonization}
In this section, we collect the mathematical results needed to bosonize the Mattis model. Some additional details can be found in Appendix~\ref{Bosonization identities}.

\begin{proposition}
\label{Proposition Density operators}
{\bf (a)} The fermion densities are well-defined\footnote{To be precise: These operators and the identities below are well-defined on a common, dense, invariant domain $\cD$ to be defined in Section~\ref{Unboundedness}; see also Remark~\ref{Remark unbounded}.} operators on $\cF$ obeying the commutation relations
\begin{equation} 
\label{Commutation relations of densities}
\Bigl[\hat{J}_{r,s}(\vp), \hat{J}_{r',s'}(\vp')\Bigr]= r \delta_{r,r'}\delta_{s,s'} \frac{2\pi p_s }{\ta}\Bigl(\frac{L}{2\pi}\Bigr)^2 \sum_{n\in\mathbb{Z}}\delta_{\vp+\vp',2\pi n\ta^{-1}\ve_{-s}}. 
\end{equation}
Moreover,
\begin{equation}
\hat{J}_{r,s}(\vp)^\dag=\hat{J}_{r,s}(-\vp)
\end{equation} 
and
\begin{equation}
\label{Nodal density annihilation}
\hat{J}_{r,s}(\vp)\Omega=0, \qquad \forall \vp\, \mbox{ such that }\, rp_s\geq 0. 
\end{equation}
\noindent {\bf (b)} The following operator identity  holds true, 
\begin{equation}
\label{Kronig relation}
\sum_{\vk\in\Lambda_s^*} \Bigl(\tPiL\Bigr)^2 rk_s :\!\hat\psi^\dag_{r,s}(\vk)\hat\psi\pdag_{r,s}(\vk)\!:\; = \ta\pi \sum_{\vp\in\tilde\Lambda_s^*} \Bigl(\frac{1}{L}\Bigr)^2 :\! \hat{J}_{r,s}(-\vp) \hat{J}_{r,s}(\vp)\!: 
\end{equation} 
with both sides defining self-adjoint operators on $\cF$ (the colons indicate normal ordering as in \Ref{NormalOrder}). 
\end{proposition}
\noindent (\textit{Proof:} See Appendix~\ref{Bosonization identities}.) 

\medskip

\begin{remark} 
It follows from equations \Ref{Nodal densities in position space} and \Ref{Commutation relations of densities} that
\begin{equation}
[{J}_{r,s}(\vx), {J}_{r',s'}(\vy)]=  \delta_{r,r'}\delta_{s,s'} r\frac1{2\pi\ta\ii}\frac{\partial}{\partial x_s} \tilde\delta_s(\vx-\vy).
\end{equation}
This and \Ref{Jrs} imply the identities in \Ref{PiPhi from J}. 
\end{remark} 

From Proposition \ref{Proposition Density operators}(a) follows that the rescaled densities
\begin{equation}
\label{Boson operators from nodal density operators}
b_s(\vp) \define
\begin{cases}
-\frac{\ii}L\sqrt{\frac{2\pi\ta}{|p_s|}} \hat{J}_{+,s}(\vp) & \text{if $p_s>0$}\\ \phantom{-}\frac{\ii}L\sqrt{\frac{2\pi\ta}{|p_s|}}\hat{J}_{-,s}(\vp) & \text{if $p_s<0$}
\end{cases} 
\end{equation} 
and $b^\dag_s(\vp)\define b\pdag_s(\vp)^\dag$, for $\vp$ in 
\begin{equation}
\label{Momentum set with zero-component}
\hat\Lambda_s^*\define\left\{\vp\in \tilde\Lambda_s^*\; :\; p_s\neq 0\right\}, 
\end{equation}
obey the defining relations of boson creation- and annihilation operators,
\begin{equation}
\label{ccr} 
[b\pdag_s(\vp),b^\dag_{s'}(\vp')]=\delta_{s,s'}\delta_{\vp,\vp'}, \qquad [b\pdag_s(\vp),b\pdag_{s'}(\vp')]=0, \qquad b\pdag_s(\vp)\Omega=0.
\end{equation}
Furthermore, \ref{Proposition Density operators}(b) implies that \Ref{Free part of regularized Mattis Hamiltonian} can be written as
\begin{equation}
\label{Kronig relation for Mattis Hamiltonian}
H_0= \ta\pi v_F \sum_{r,s=\pm}\sum_{\vp\in\tilde\Lambda_s^*} \Bigl(\frac{1}{L}\Bigr)^2 :\! \hat{J}_{r,s}(-\vp) \hat{J}_{r,s}(\vp)\!: 
\end{equation}
and thus the regularized Mattis Hamiltonian can be represented solely in terms of densities. 

It is important to note that not all Fourier modes of the fermion densities can be expressed as bosons: there are also the {\em zero modes} 
\begin{equation}
\label{hatNrs}
\hat{N}_{r,s}(p_{-s})\define\sqrt{\frac{\ta}{2\pi}} \left. \hat{J}_{r,s}(\vp)\right|_{p_s=0}
\end{equation}
with $p_{-s}$ in 
\begin{equation}
\label{1D Boson Fourier space}
\tilde\Lambda_{1\rm D}^* \define \left\{ p \in \tPiL\mathbb{Z}\, : \,  -\frac{\pi}{\ta} \leq p< \frac{\pi}{\ta} \right\},
\end{equation}
which require a different treatment. We find it convenient to consider the following Fourier transformed zero-mode operators 
\begin{equation}
\label{Zero-mode operator}
N_{r,s}(x) \define \sqrt{2\pi\ta} \sum_{p\in\tilde\Lambda^*_{1\mathrm{D}}}\frac{1}{L}   \hat{N}_{r,s}(p)\ee^{\ii px}, \quad x\in \Lambda_{1\mathrm{D}}.
\end{equation} 
It follows from \Ref{Nodal density annihilation} that
\begin{equation}
\label{Zero mode annihilation}
N_{r,s}(x)\Omega=0. 
\end{equation}
To fully formulate the Mattis model in bosonized language, we also need to introduce the  Klein factors $R_{r,s}(x)$ conjugate to the zero modes in \Ref{Zero-mode operator}. These, together with the boson operators, span the full fermion Fock space $\cF$:

\begin{lemma}
\label{Lemma Fock space basis}
{\bf (a)} There exist unitary operators $R_{r,s}(x)$ on the fermion Fock space $\cF$ commuting with all boson operators in \Ref{Boson operators from nodal density operators} and satisfying the commutation relations
\begin{equation}
\label{Klein factor commutator relations} 
\begin{split}
[N_{r,s}(x),R_{r',s'}(x')] &= r\delta_{r,r'}\delta_{s,s'}\delta_{x,x'}R_{r,s}(x),\\ \{ R_{r,s}(x),R_{r',s'}(x')^\dag \} &= 2\delta_{r,r'}\delta_{s,s'}\delta_{x,x'} .  
\end{split}
\end{equation} 
\noindent {\bf (b)} Let $\cQ$ be the set of all pairs $(\vn,\vnu)$ with 
\begin{equation}
\vn=\left\{ n_s(\vp) \right\}_{s=\pm,\, \vp\in \hat\Lambda_s^*},\qquad 
\vnu=\left\{ \nu_{r,s}(x) \right\}_{r,s=\pm, \, x\in\Lambda_{1\mathrm{D}}}
\end{equation} 
and integers $\nu_{r,s}(x)$ and $n_s(\vp)\geq 0$ such that
\begin{equation}
\label{Finite condition}
\sum_{r,s=\pm}\sum_{x\in\Lambda_{1\mathrm{D}}}\nu_{r,s}(x)^2 <\infty,\qquad \sum_{s=\pm}\sum_{\vp\in\hat\Lambda_s^*}|p_s| n_s(\vp)<\infty .
\end{equation} 
Then the states
\begin{equation}
\label{Boson basis}
\eta_{\vn,\vnu}\define \Bigl( \prod_{s=\pm}\prod_{\vp\in\hat\Lambda_s^*} \frac{b_s^\dag(\vp)^{n_s(\vp)}}{\sqrt{n_s(\vp)!}}\Bigr) \Bigl( \prod_{r,s=\pm}\prod_{x\in\Lambda_{1\mathrm{D}}} R_{r,s}(x)^{\nu_{r,s}(x)} \Bigr)\Omega,
\end{equation} 
with $(\vn,\vnu)\in\cQ$, provide a complete orthonormal basis in $\cF$.\\
\noindent {\bf (c)} The states $\eta_{\vn,\vnu}$ are common eigenstates of $N_{r,s}(x)$ and $b_{s}^\dag(\vp)b\pdag_s(\vp)$ with eigenvalues $\nu_{r,s}(x)$ and $n_{s}(\vp)$, respectively.
\end{lemma} 
\noindent (\textit{Proof:} See Appendix~\ref{Bosonization identities}.) 

\medskip

The above definitions allow to rewrite $H_0$ in \Ref{Kronig relation for Mattis Hamiltonian} as
\begin{equation}
H_0= v_F \sum_{s=\pm}\sum_{\vp\in\hat\Lambda_s^*} |p_s\pdag|b_s^\dag(\vp)b_s\pdag(\vp) + \frac{\pi v_F}{L} \sum_{r,s=\pm}\sum_{x\in\Lambda_{1\rm D}} N_{r,s}(x)^2
\end{equation}
where we used the Parseval-type relation 
\begin{equation}
\label{Parseval} 
\sum\limits_{p\in\tilde\Lambda^*_{1\rm D}}\tPiL\hat{N}_{r,s}(-p)\hat{N}_{r,s}(p) = \sum\limits_{x\in\Lambda_{1\rm D}} N_{r,s}(x)^2.
\end{equation}
This and Lemma~\ref{Lemma Fock space basis}(c) imply that the states in \Ref{Boson basis} are exact eigenstates of $H_0$ with eigenvalues $v_F\sum_{s,\vp}|p_s|n_s(\vp) + \pi v_F\sum_{r,s,x}\nu_{r,s}(x)^2/L$, and the latter are finite due to \Ref{Finite condition}. 

To compute fermion Green's functions, we need the following result expressing the fermion operators in terms of the bosons: 
\begin{proposition}
\label{Proposition Fermions from Bosons}
For $s=\pm$, $\vx\in\Lambda_s$, and $\epsilon>0$, the operators 
\begin{equation}
\label{psi and K 1}
\begin{split} 
\cV_{r,s}(\vx;\epsilon)\define& \frac1{\sqrt{2\pi \ta\epsilon}}\ee^{\ii r\pi x_sN_{r,s}(x_{-s})/L} R_{r,s}(x_{-s})^{-r} \ee^{\ii r\pi x_s N_{r,s}(x_{-s})/L} \\ & \times\exp\Bigl(r \frac{\ta}{2\pi} \sum_{\vp\in\hat\Lambda_{s}^*}\Big(\tPiL\Bigr)^2 \frac1{p_s} \hat{J}_{r,s}(\vp)\ee^{\ii \vp\cdot\vx}\ee^{-\epsilon |p_s|/2} \Bigr) 
\end{split}
\end{equation}
are proportional to unitary operators on the fermion Fock space $\cF$.\footnote{To be precise: $\sqrt{2\pi \ta\epsilon}\cV_{r,s}(\vx;\epsilon)$ is a unitary operator.} Moreover, 
\begin{equation}
\label{psi from K} 
\hat\psi_{r,s}(\vk) =\lim_{\epsilon\to 0^+} \frac1{2\pi} \tint{s} \ud^2 x \,  \cV_{r,s}(\vx;\epsilon)\ee^{-\ii\vk\cdot\vx}
\end{equation} 
using the notation in \Ref{tint}.\footnote{The limit is in the strong operator topology; see e.g.\ \cite{ReedSimon1}.} 
\end{proposition} 

\noindent (\textit{Proof:} See Appendix~\ref{Bosonization identities}.) 

\medskip

Note that, by taking the Hilbert space adjoint, \Ref{psi from K} implies a corresponding result for $\hat\psi^\dag_{r,s}(\vk)$.  

The operator $\cV_{r,s}(\vx;\epsilon)$ can be regarded as a regularized version of the fermion field operator $\psi_{r,s}(\vx)$. To be precise:\ $\psi_{r,s}(\vx)$ is an operator-valued distribution, and products of these need to be interpreted with care. However, for $\epsilon>0$, $\cV_{r,s}(\vx;\epsilon)$ is a well-defined bounded operator, and it converges to $\psi_{r,s}(\vx)$ in the limit $\epsilon\to 0^+$. This also suggests to write $\psi_{r,s}(\vx;\epsilon)$ instead of $\cV_{r,s}(\vx;\epsilon)$. 
In Section~\ref{Fermion correlation functions}, we compute expectation values of products of fermion field operators as the limit $\epsilon\to 0^+$ of the corresponding expectation values of products of these regularized operators. This circumvents potential difficulties with the distributional nature of the fermion fields. 

We find it convenient to introduce the notation
\begin{equation}
\begin{split} 
\label{Charges and chiral charges}
Q_{r,s}(x) \define & \frac1{\sqrt{2}}\Bigl( N_{+,s}(x)+r N_{-,s}(x) \Bigr) \\ \hat{Q}_{r,s}(p) \define & \frac1{\sqrt{2}}\Bigl( \hat{N}_{+,s}(p)+r \hat{N}_{-,s}(p) \Bigr) 
\end{split}. 
\end{equation}

In our computations in the next section we also need the following orthogonality
result implied by Lemma~\ref{Lemma Fock space basis}, 
\begin{equation}
\label{cR}
\langle\eta_{\vn,\vnu}, \prod_{r,s,x} R_{r,s}(x)^{m_{r,s}(x)}\eta_{\vn,\vnu} \rangle = \prod_{r,s,x}\delta_{m_{r,s}(x),0}
\end{equation} 
independent of $(\vn,\vnu)$, for all integers $m_{r,s}(x)$ such that $\sum_{r,s,x}m_{r,s}(x)^2<\infty$ and all states $\eta_{\vn,\vnu}$ in \Ref{Boson basis};  the products in \Ref{cR} are over all $r,s=\pm$ and $x\in\Lambda_{1\rm D}$. Note that the product on the left-hand side in \Ref{cR} is only well-defined with an ordering prescription (since operators $R_{r,s}(x)$ and $R_{r',s'}(x')$ anticommute for $(r,s,x)\neq (r',s',x')$). However, the result in \Ref{cR} holds true for any ordering, and we therefore do not specify the latter.

\subsection{Unboundedness of operators}
\label{Unboundedness}
In general, sums or products of unbounded Hilbert space operators need not be densely defined, and a symmetric unbounded operator need not define a self-adjoint operator. However, as elaborated in this section, such difficulties do not arise if a set of unbounded operators belongs to  what is called a {\em Op$^*$-algebra}, and this is the case for the unbounded operators introduced above. We also prove that the Mattis Hamiltonian is essentially self-adjoint on $\cD$. 

We recall the definition of a Op$^*$-algebra (see \cite{GL}, Section~II(c) and references therein). Let $\cD$ be a dense subset of some Hilbert space (in our case  $\cF$). Then $\cL_+(\cD)$ is the $*$-algebra of linear operators with $\cD$ as common, invariant domain of definition, and
with the involution $+$ defined as $A^+ \define  A^*|_{\cD}$, i.e.\ $A^+$ is such that 
\begin{equation} 
\label{involution}
\langle \eta_1,A\eta_2\rangle = \langle A^+\eta_1,\eta_2\rangle \qquad \forall \eta_1,\eta_2\in\cD
\end{equation} 
(in this section, we also use the symbol $*$ to denote the Hilbert space adjoint). It follows from this definition that all elements $A\in\cL_+(\cD)$ are closable \cite{ReedSimon1}. An Op*-algebra $\cA$ on $\cD$ is a $*$-subalgebra of $\cL_+(\cD)$ containing the identity. It is called {\em standard} if each symmetric element in $\cA$ is essentially self-adjoint.

We define $\cD$ as the space of all finite linear combinations of vectors in \Ref{Boson basis}. By definition, this set is dense in $\cF$  and a domain of essential self-adjointness of $H_0$. The densities $\hat J_{r,s}(\vp)$, the free fermion Hamiltonian $H_0$, and the regularized Mattis Hamiltonian $H$, are all unbounded operators defined on the common, invariant domain $\cD$. It follows from the above definition that these operators and the Klein factors $R_{r,s}(x)$, generate a Op$^*$-algebra containing the Mattis Hamiltonian $H$. This makes mathematically precise the identities involving unbounded operators given in the previous section; they are well-defined as Op$^*$-algebra relations on the domain $\cD$ if we interpret $\dag$ as involution 
 $+$  (rather than as Hilbert space adjoint).  A non-trivial question is if this Op$^*$-algebra is standard. However, this need not be addressed here in full generality. Instead, we restrict ourselves for the remainder of this section to a particular subspace of the Op$^*$-algebra, which contains the Mattis Hamiltonian, and for which the proof that all symmetric elements are essentially self-adjoint becomes simpler.

For $0<E<\infty$, let $P_E=\theta(E-H_0)$ with $\theta(x)=1$ for $x\geq 0$ and $0$ for $x<0$, i.e.\ $P_E$ is the projection to the Hilbert subspace generated by all eigenstates of $H_0$ with corresponding eigenvalues less or equal to $E$.  We define the set $\cO^\omega(H_0)$ of all linear operators $A$ on $\cF$ such that, for all $E<\infty$, $AP_E$ is bounded and 
\begin{equation}
\label{A conditions} 
||AP_E||\leq \alpha (E+E_0),\quad AP_E = P_{E+\Delta} AP_E
\end{equation} 
($||\cdot||$ is the operator norm) for some finite and non-negative integers $\alpha$, $E_0$ and $\Delta$ (all independent of $E$). Note that $\cO^\omega(H_0)$ is a vector space but not an algebra. 

\begin{lemma}
Symmetric operators $A\in\cO^\omega(H_0)$ are essentially self-adjoint on $\cD$. 
\end{lemma} 

\noindent {\em Proof:}  For all $\eta\in\cD$, there exists $E<\infty$ such that $\eta=P_E\eta$, and thus, using \Ref{A conditions},   
\begin{equation} 
\begin{split} 
||A^n\eta||=||AP_{E+(n-1)\Delta}\cdots AP_{E+\Delta} AP_E\eta||\leq \prod_{j=0}^{n-1}\alpha(E+ j\Delta + E_0)||\eta||
\end{split} 
\end{equation}  
implying $\sum_{n=1} ||A^n\eta|| t^n/n!<\infty$ for $0<t< 1/(\alpha \Delta)$ by the ratio test. Thus $\cD$ is a set of analytical vectors for any $A\in\cO^\omega(H_0)$, which gives the result by Nelson's analytic vector theorem (see e.g.\ \cite{ReedSimon2}, Section~X.6). \QED

\begin{remark}
More generally, $A\in\cO^\omega(H_0)$ implies $\overline{A^+}=A^*$; see \cite{GL}, Lemma~1. 
\end{remark} 

\begin{corollary}
\label{Mattis Hamiltonian is selfadjoint}
The Mattis Hamiltonian $H$ is essentially self-adjoint on $\cD$. 
\end{corollary} 

\noindent {\em Proof:} We only need to prove that $H\in\cO^\omega(H_0)$. The commutator relations 
\begin{equation}
[H_0,\hat J_{r,s}(\vp)] = -rv_F p_s \hat J_{r,s}(\vp) 
\end{equation} 
imply the second condition in \Ref{A conditions}  for $A=H$, with $\Delta=2\pi v_F/\tilde{a}$. 

To prove the first condition in \Ref{A conditions} we consider $A=\hat J_{r,s}(-\vp)\hat J_{r,s}(\vp)$. By definition of normal ordering, $A=c_1+\! :A:$ with $c_1= r p_s\theta(rp_s) L^2/(2\pi \tilde{a})$. Thus, for arbitrary $\eta\in\cD$ and $E<\infty$ such that $\eta=P_E\eta$, 
 \begin{equation}
 \begin{split}
 \langle\eta,A\eta\rangle = \langle\eta, (c_1 + \! :A:) \eta\rangle \leq  \langle\eta, (c_1 + c_2 H_0) \eta\rangle \leq \bigl(c_1+c_2 E)||\eta||^2 
 \end{split}
 \end{equation} 
 with $c_2= L^2/(\tilde{a}\pi v_F) $.  This and the Cauchy-Schwarz inequality imply the first condition in \Ref{A conditions} for all operators $A=\hat J_{r,s}(-\vp)\hat J_{r',s'}(\vp)$. Since the Mattis Hamiltonian is a sum of $H_0$ and a finite linear combination of such operators $A$, the triangle inequality and $||H_0P_E||\leq E$ imply the result.\QED

\newsection{Solution of the Mattis model}
\label{Solution of the Mattis model}
In this section, we show that the free energy and any correlation function of the Mattis model can be computed by analytical methods. Computational details are given in Appendix~\ref{Solution appendix}.

\subsection{Diagonalization of Hamiltonian} 
\label{Diagonalization of Mattis Hamiltonian}
Proposition~\ref{Proposition Density operators}(a) implies that, for $s=\pm$ and $\vp\in\hat\Lambda^*_s$,  the following linear combination of fermion densities 
\begin{equation}
\label{hatPihatPhi}
\begin{split}
\hat\Phi_s(\vp)\define&\sqrt{\frac{\ta}{4\pi}}\frac1{\ii p_s}\Bigl(\hat{J}_{+,s}(\vp) + \hat{J}_{-,s}(\vp) \Bigr)\\ \hat\Pi_s(\vp)\define&\sqrt{\frac{\ta}{4\pi}}\Bigl(-\hat{J}_{+,s}(\vp) + \hat{J}_{-,s}(\vp) \Bigr)
\end{split} \quad (p_s\neq 0)
\end{equation}
obey the canonical commutator relations of 2D neutral bosons, i.e.\
\begin{equation}
\begin{gathered}
\label{CCR}
[\hat\Phi\pdag_s(\vp),\hat\Pi^\dag_{s'}(\vp')]=\ii\delta_{s,s'}\Bigl(\frac{L}{2\pi}\Bigr)^2\delta_{\vp,\vp'}, \quad {[}\hat\Pi\pdag_s(\vp),\hat\Pi^\dag_{s'}(\vp'){]}={[}\hat\Phi\pdag_s(\vp),\hat\Phi^\dag_{s'}(\vp'){]}=0,\\ \hat\Pi^\dag_s(\vp)=\hat\Pi\pdag_s(-\vp),\quad \hat\Phi^\dag_s(\vp)=\hat\Phi\pdag_s(-\vp).   
\end{gathered} 
\end{equation} 
Moreover, Definition~\ref{Mattis model} and Proposition~\ref{Proposition Density operators}(b) imply that the Mattis Hamiltonian can be written in terms of these as\footnote{\label{Footnote Addition of zeros}Note the slight abuse of notation:\ the summation over momenta has been extended such that it superficially includes boson operators that are not defined. However, all these operators come with vanishing prefactors. The same remark applies to various expressions in Appendix~\ref{Solution appendix}.}
\begin{equation}
\label{Boson Hamiltonian}
\begin{split}
H = \frac{v_F}{2}\sum_{s=\pm}\sum_{\vp\in\hat\Lambda^*_s}\Bigl(\tPiL\Bigr)^2 :\!\Bigl(&(1-\gamma_1\chi)\hat\Pi^\dag_s\hat\Pi\pdag_s + (1+\gamma_1\chi)p_s^2\hat\Phi^\dag_s\hat\Phi\pdag_s \\ & +\gamma_2 p_+p_-\chi \hat\Phi^\dag_s\hat\Phi\pdag_{-s} + \hat\Xi^\dag_s\hat\Phi\pdag_s + \hat\Phi^\dag_s\hat\Xi\pdag_s\Bigr)\!:\,  + H^{(0)}_\zeromode 
\end{split} 
\end{equation} 
with
\begin{equation}
\label{Xi} 
\hat\Xi_s(\vp) = -\gamma_2\ii p_s\chi(\vp)\hat{Q}_{+,-s}(p_s)\delta_{p_{-s},0}, \qquad \hat\Xi^\dag_s(\vp)=\hat\Xi\pdag_s(-\vp)
\end{equation}
the zero modes that couple to the bosons, and
\begin{equation} 
\label{KK0} 
H^{(0)}_\zeromode = \frac{v_F}2\Bigl(\tPiL\Bigr)^2\Bigl(\sum_{r,s=\pm}\sum_{p\in\tilde\Lambda^*_{1\rm D} } (1+r\gamma_1)\hat{Q}_{r,s}(-p)\hat{Q}_{r,s}(p) + 2\gamma_2\hat{Q}_{+,+}(0)\hat{Q}_{+,-}(0) \Bigr)
\end{equation}
the terms containing only zero modes; we suppress common arguments $\vp$ in \Ref{Boson Hamiltonian}. Here and below, terms labeled by ``$B$'' and ``$\zeromode$'' generically correspond to boson- and zero-mode contributions, respectively. 

The zero-mode terms in \Ref{Xi} and \Ref{KK0} commute with all boson operators $\hat{\Pi}_s$ and $\hat{\Phi}_s$, and they can therefore be treated like $\C$-numbers. It is straightforward to find a Bogoliubov transformation that diagonalizes the Hamiltonian in \Ref{Boson Hamiltonian} into a sum of decoupled harmonic oscillators and zero-mode terms. In the following we summarize the main results from this diagonalization; see Appendix~\ref{Solution appendix} for details.
\begin{theorem}
\label{Solution}
There exists a unitary operator $\cU$ on the fermion Fock space $\cF$ diagonalizing the regularized Mattis Hamiltonian 
as follows:
\begin{equation}
\label{H2}
\begin{gathered}
\tilde{H}\define \cU^\dag H\cU = \tilde{H}_B+\tilde{H}_\zeromode+\cE_0\\  \tilde{H}_B =  \sum_{s=\pm}\sum_{\vp\in\hat\Lambda_s^*}\omega_s\pdag(\vp) b_s^\dag(\vp)b\pdag_s(\vp),\quad \tilde{H}_\zeromode=\tilde{H}_\zeromode\bigl(\{Q_{r,s}(x)\}\bigr)
\end{gathered}
\end{equation} 
with the boson operators $b_s^{(\dag)}(\vp)$ in \Ref{Boson operators from nodal density operators}--\Ref{ccr} and 
\begin{align}
\label{oms} 
\omega_\pm(\vp)&= \begin{cases} \tilde{v}_F\sqrt{\frac12\bigl( |\vp|^2 \pm \sqrt{|\vp|^4-A(2p_+p_-)^2 }\; \bigr)}& \mbox{ if }\; \gamma_2\chi(\vp)p_+p_-\neq 0
 \\v_F\sqrt{(1-\gamma_1^2\chi(\vp))}|p_\pm|& \mbox{ if }\;\gamma_2\chi(\vp)p_+p_-= 0  \end{cases}
\end{align} 
\begin{equation}
\label{AvF} 
A\define 1-\left(\frac{\gamma_2}{1+\gamma_1}\right)^2  ,\qquad \tilde{v}_F\define v_F \sqrt{1-\gamma_1^2}
\end{equation} 
the boson dispersion relations, 
\begin{equation} 
\begin{split}
\label{KK} 
&\tilde{H}_\zeromode\bigl(\{Q_{r,s}(x)\}\bigr) = \frac{v_F\pi}{L}\sum_{s=\pm}\sum_{x\in\Lambda_{1\rm D}} \Bigl((1+\gamma_1)A Q_{+,s}(x)^2+(1-\gamma_1)Q_{-,s}(x)^2\Bigr) \\ &+\frac{v_F\pi\ta}{L^2}\sum_{s=\pm} \Bigl(\frac{\gamma_2^2}{1+\gamma_1}\Bigl(\sum_{x\in\Lambda_{1\rm D}}Q_{+,s}(x)\Bigr)^2 + \gamma_2\Bigl(\sum_{x\in\Lambda_{1\rm D}}Q_{+,s}(x)\Bigr)\Bigl(\sum_{x\in\Lambda_{1\rm D}}Q_{+,-s}(x)\Bigr)\Bigr)
\end{split}
\end{equation} 
 the zero-mode contribution with the charges $Q_{r,s}(x)$ in \Ref{hatNrs}--\Ref{Zero-mode operator} and \Ref{Charges and chiral charges}, and 
\begin{equation}
\label{Ground state energy}
\cE_0 = \frac12\sum_{s=\pm} \sum_{\vp\in\hat\Lambda_s^*} (\omega_s(\vp)-v_F |p_s|)  
\end{equation} 
the groundstate energy. 
\end{theorem}

\noindent ({\em Proof:} See Appendix~\ref{Solution appendix}.)

\medskip

The explicit expression for the unitary operator $\cU$ is given in \Ref{cU1}, \Ref{cU full}, and \Ref{UhUdag2}. Note that the sum in \Ref{Ground state energy} has only a finite number of non-zero terms. Furthermore, the conditions in \Ref{Coupling restrictions} are necessary and sufficient for the dispersion relation in \Ref{oms}--\Ref{AvF} to be well-defined (the conditions imply that $0<A\leq1$). 

\begin{figure}[!ht]
\begin{center}
\includegraphics[width=1.0\textwidth]{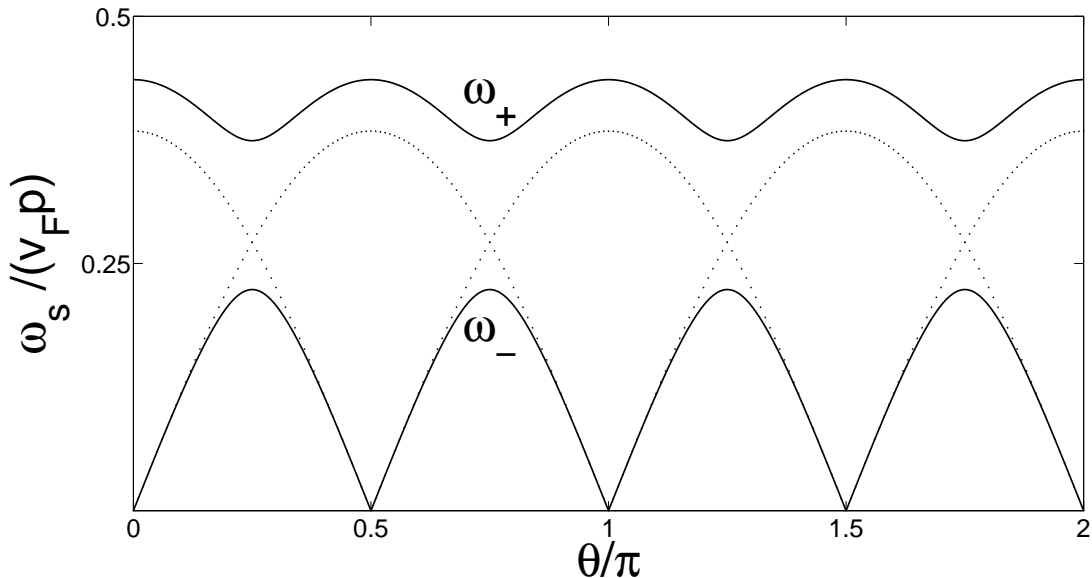} 
\end{center}
\caption{Solid line: the angular dependence of the boson dispersion relations $\omega_+(\vp)$ (top curve) and $\omega_-(\vp)$ (bottom curve) in \Ref{oms} for fixed $|\vp|>0$ and $\gamma_1=\gamma_2=0.9$. Dotted line: same angular dependence using instead the effective dispersion $\tilde\omega_\pm(\vp)$ defined in \Ref{Renormalized dispersion}.}
\label{Dispersion relation}
\end{figure}

When $\gamma_2\chi(\vp)p_+p_-\neq 0$, the boson dispersion relations $\omega_\pm(\vp)$ in \Ref{oms} are given by $v_F|\vp|$ times dimensionless functions depending only on the polar angle $\theta\define\arctan(p_-/p_+)$. The angular dependence is determined by the constant $A$ in \Ref{AvF}, and it is qualitatively similar for all values of $A$. For future reference, we note that both functions $\omega_\pm(\vp)/(v_F|\vp|)$ are invariant under $\theta\to \theta+\pi/2$ and $\theta\to -\theta$, and  $\omega_-(\vp)/(v_F|\vp|)$ vanishes like $const.|\theta|$ as $\theta \to 0$, for all $|\vp|>0$; see Figure~\ref{Dispersion relation}.    

Lemma~\ref{Lemma Fock space basis}(b) and Theorem~\ref{Solution} imply that the exact eigenstates of the regularized Mattis Hamiltonian are $\cU \eta_{\vnu,\vn}$, with the states $\eta_{\vnu,\vn}$ in \Ref{Boson basis}. The corresponding eigenvalues are
\begin{equation} 
\label{Emn} 
\cE_{\vn,\vnu} =\sum_{s=\pm}\sum_{\vp\in\hat{\Lambda}^*_s}\omega_s(\vp)n_s(\vp) + \tilde{H}_\zeromode(\{q_{r,s}(x)\}\bigr)+ \cE_0 
\end{equation}
with $\tilde{H}_\zeromode\bigl(\{q_{r,s}(x)\}\bigr)$ as in \Ref{KK}, but with the operator $Q_{r,s}(x)$ replaced by the eigenvalue $q_{r,s}(x)\define (\nu_{+,s}(x)+r\nu_{-,s}(x))/\sqrt{2}$. In particular, the Mattis model has a non-degenerate groundstate $\cU\Omega$ with energy $\cE_0$. 

Using these results, one can compute the partition function $\cZ_\beta\define\textrm{Tr}_{\cF}\bigl(\ee^{-\beta H}\bigr)$ and free energy $\Omega_\beta\define -\ln(\cZ_\beta)/\beta$ of the Mattis model, with $\beta$ the inverse temperature and the trace $\mathrm{Tr}_\cF$ in $\cF$. The additive form of the diagonalized Mattis Hamiltonian in \Ref{H2} implies $\cZ_\beta=\cZ_{B,\beta}\cZ_{\zeromode,\beta}\ee^{-\beta \cE_0}$, with $\cZ_{X,\beta}= \textrm{Tr}_{X}\bigl(\ee^{-\beta \tilde{H}_X}\bigr)$ for $X=B,\zeromode$, and with partial boson- and zero-mode traces
\begin{equation}
\label{TrBTrQ}
\begin{split}
\mathrm{Tr}_B(\tilde\cO_B)\define \sum_{\vn}\langle\eta_{\vn,\vzero},\tilde\cO_B\eta_{\vn,\vzero}\rangle\\ \mathrm{Tr}_\zeromode(\tilde\cO_\zeromode)\define \sum_{\vnu}\langle\eta_{\vzero,\vnu},\tilde\cO_\zeromode\eta_{\vzero,\vnu}\rangle
\end{split}
\end{equation} 
for boson operators $\tilde\cO_B$ (that only depend on the $b^{(\dag)}_s(\vp)$) and zero-mode operators $\tilde\cO_\zeromode$ (that only depend on the zero-mode operators $N_{r,s}(x)$ and $R_{r,s}(x)$), respectively, with $\eta_{\vn,\vnu}$, $\vn$ and $\vnu$ defined in Lemma~\ref{Lemma Fock space basis}(b). Furthermore, this implies that the free energy has the form $\Omega_\beta=\Omega_{B,\beta}+\Omega_{\zeromode,\beta}+\cE_0$ with the groundstate energy $\cE_0$ in \Ref{Ground state energy}. One finds by straightforward computations (see Appendix~\ref{Computation details} for details):

\begin{result} 
\label{Result free energy}
The boson- and zero-mode contributions to the free energy of the Mattis model are
\begin{align}
\Omega_{B,\beta}&= \sum_{s=\pm}\sum_{\vp\in\hat \Lambda^*_s}\frac1\beta \ln\bigl(1-\ee^{-\beta\omega_s(\vp)}\bigr)\label{OmB}\\ \Omega_{\zeromode,\beta}&=-\frac{2L}{\ta\beta}\ln\Bigl(\frac{L}{\beta\tilde{v}_F\sqrt{A}}\Bigr) -\frac1{2\beta}\ln(A) + O(L^3\ee^{-cL})\label{OmQ} 
\end{align}
for some $c>0$ that is independent of $L$, with $\omega_s(\vp)$ in \Ref{oms}, and $A$ and $\tilde{v}_F$ in \Ref{AvF}. 
\end{result} 

\noindent ({\em Proof:} The formula in \Ref{OmB} is a simple consequence of a well-known result stated in \Ref{cZboson}. The derivation of \Ref{OmQ} can be found in Appendix~\ref{Free energy appendix}.)

\medskip

\begin{remark} 
Note that the free energy density $\Omega_\beta/L^2$ has a well-defined IR limit (since the factor $1/L^2$ turns the $\vp$-sum in \Ref{Ground state energy} and \Ref{OmB} to Riemann sums converging to well-defined integrals), and the zero-mode contribution $\Omega_{\zeromode,\beta}/L^2$ vanishes in this limit. In Section~\ref{QFT limit}, we give simpler formulas for the free energy density obtained in the QFT limit $L\to\infty$ and $\ta\to 0^+$. 
\end{remark} 

For future reference, we summarize a few useful identities that are simple consequences of the above definitions and Theorem~\ref{Solution}.\footnote{Formulas like the ones given in this paragraph are also used in \cite{HSU}.} To compute thermal expectation values $\langle\cO\rangle_\beta\define\textrm{Tr}_{\cF}\bigl(\ee^{-\beta H}\cO\bigr)/\cZ_\beta$ of operators $\cO$ on the fermion Fock space $\cF$ one can use the (trivial) identity
\begin{equation}
\label{trivial 1}
\langle\cO\rangle_\beta = \frac1{\cZ_\beta}\mathrm{Tr}_{\cF}\bigl(\ee^{-\beta\tilde{H}}\tilde{\cO}),\qquad \tilde{\cO}\define \cU^\dag \cO\cU 
\end{equation} 
with $\tilde{H}$ in \Ref{H2}. Therefore, one only needs to know how to express $\tilde{\cO}$ in terms of boson- and zero-mode operators. Moreover, for operators such that $\tilde{\cO}$ can be factorized into a boson- and zero-mode part $\tilde\cO_B$ and $\tilde\cO_\zeromode$, respectively, the thermal expectation value factorizes accordingly:
\begin{equation}
\label{factorize} 
\begin{gathered}
\frac1{\cZ_\beta}\mathrm{Tr}_{\cF}\bigl(\ee^{-\beta\tilde{H}}\tilde{\cO}_B\tilde{\cO}_\zeromode)= \langle \tilde{\cO}_B\rangle_{B,\beta}\langle \tilde{\cO}_\zeromode\rangle_{\zeromode,\beta}\\ \langle \tilde{\cO}_{X}\rangle_{X,\beta}\define \frac1{\cZ_{X,\beta}}\mathrm{Tr}_{X}\bigl(\ee^{-\beta\tilde{H}_X}\tilde\cO_X), \quad X=B,\zeromode
\end{gathered}
\end{equation}
(this is a simple consequence of the definitions, including \Ref{TrBTrQ}). We are particularly interested in thermal expectation values of products of ``time'' evolved density- or fermion operators $\cA(t)\define\ee^{\ii Ht}\cA\ee^{-\ii Ht}$ with complex $t$ (in applications, $t$ is usually taken to be either real or purely imaginary, and our results allow for both). We use the trivial identities
\begin{equation}
\label{trivial 2} 
\begin{split}
&\langle\cA_1(t_1)\cdots \cA_N(t_N)\rangle_\beta \define\frac1{\cZ_\beta}\mathrm{Tr}_{\cF}\bigl(\ee^{-\beta \tilde{H}}\tilde\cA_1(t_1)\cdots \tilde\cA_N(t_N) \bigr), \quad N\in\N, \\ &\tilde\cA(t)=\cU^\dag \ee^{\ii Ht}\cA\ee^{-\ii Ht}\cU = \ee^{\ii t \tilde{H}}\cU^\dag \cA\cU\ee^{-\ii t \tilde{H}}, \quad t\in\mathbb{C}
\end{split}
\end{equation}
implied by \Ref{H2} and \Ref{trivial 1}. We refer to such thermal expectation values as {\em (thermodynamic) correlation functions}. \footnote{The extension of our results to include possible time orderings is straightforward.}

For the computation of correlation functions below, we will need $\tilde{\cA}(t)$ for the operators $\cA=N_{r,s}(x)$, $R_{r,s}(x)$ and $\hat J_{r,s}(\vp)$ introduced in Section~\ref{Bosonization}. The result is given by the following:
\begin{proposition}
\label{Observables}
Let $r,s=\pm$, $t\in\C$, and $\cU$ and $\tilde{H}$ as in Theorem~\ref{Solution}. Then, for all $x\in\Lambda_{1\rm D}$, 
\begin{equation}
\label{UNUdag}  
\ee^{\ii \tilde{H}t} \cU^\dag N_{r,s}(x)\cU \ee^{-\ii \tilde{H}t} = N_{r,s}(x)
\end{equation} 
and 
\begin{equation}
\label{URUdag}
\ee^{\ii \tilde{H}t} \cU^\dag R_{r,s}(x)\cU\ee^{-\ii \tilde{H}t} = 
\ee^{\ii r k_{r,s}(x) t/2}R_{r,s}(x)\ee^{\ii r k_{r,s}(x) t/2}\ee^{-\ii r [u_s(x,t)+u_s(x,t)^\dag]} 
\end{equation}
with
\begin{equation}
\label{krs} 
\begin{split}
k_{r,s}(x) = \frac{v_F\pi}{L}\sum_{r'=\pm}\Bigl[ \bigl( A(1+\gamma_1) +rr'(1-\gamma_1)\bigr) N_{r',s}(x) \\ +\frac{\ta}L \sum_{y\in\Lambda_{1\rm D}}\bigl( \frac{\gamma_2^2}{1+\gamma_1}N_{r',s}(y) +\gamma_2 N_{r',-s}(y) \bigr) \Bigr]
\end{split}
\end{equation} 
and 
\begin{equation}
\label{tkrs}
u_{s}(x,t) = \frac{\gamma_2}{1+\gamma_1}\sqrt{\frac{\pi}{2}} \frac{\sqrt{\ta}}{L}\sum_{p\in\hat\Lambda^*_{1\rm D}}\frac1p\sqrt{\frac{\omega_{-s}(p\ve_{-s})}{v_F}} b_{-s}(p\ve_{-s})\ee^{\ii px-\ii\omega_{-s}(p\ve_{-s})t}; 
\end{equation}
the index set in the last sum is defined as
\begin{equation}
\label{1D Boson Fourier space with no zero}
\hat\Lambda_{1\rm D}^* \define \left\{ p \in \tilde\Lambda_{1\rm D}^*\, : \,  p \neq 0 \right\}.
\end{equation}
Moreover, for all $\vp\in\hat\Lambda^*_s$,  
\begin{equation}
\begin{split}
\label{Jrs from b}
&\ee^{\ii \tilde{H}t}\cU^\dag\hat{J}_{r,s}(\vp)\cU\ee^{-\ii\tilde{H}t} = \frac{L}{\sqrt{\ta}}\sum_{s'=\pm} \Bigl( v^{s'}_{r,s}(\vp)b_{s'}(\vp)\ee^{-\ii\omega_s(\vp)t} 
\\ &+ \overline{v^{s'}_{r,s}(-\vp)}\, b^\dag_{s'}(-\vp)\ee^{\ii\omega_s(\vp)t}\Bigr) - \sqrt{\frac{\pi}{\ta}}\frac{\gamma_2}{1+\gamma_1}\chi(\vp)\hat{Q}_{+,-s}(p_s)\delta_{p_{-s},0}
\end{split} 
\end{equation} 
with 
\begin{equation}
\label{Mrssp}
v^{s'}_{r,s}(\vp) = \ii\sqrt{\frac{1}{8\pi}}U_{s,s'}(\vp)\Biggl(p_s\sqrt{\frac{v_F(1-\gamma_1\chi(\vp))}{\omega_{s'}(\vp)}} +r\sqrt{\frac{\omega_{s'}(\vp)}{v_F(1-\gamma_1\chi(\vp))}} \; \Biggr)
\end{equation}
and
\begin{equation}
U_{s,s'}(\vp)= \begin{cases} \bigl(\delta_{s,s'}\mp s\delta_{s,-s'}\bigr) \sqrt{\frac12\Bigl(1+\frac{ss'(p_+^2-p_-^2)}{\sqrt{|\vp|^4-A(2p_+p_-)^2}}\Bigr)} & \mbox{ if }\; \gamma_2\chi(\vp)p_+p_-\gtrless 0 \\ \delta_{s,s'} &\mbox{ if }\; \gamma_2\chi(\vp)p_+p_-= 0  \end{cases}.  \label{U}  
\end{equation}
\end{proposition}

\noindent ({\em Proof:} See Appendix~\ref{Diagonalization of the Mattis Hamiltonian}.)

\medskip

Note that in the special case $\gamma_1=\gamma_2=0$ and $t=0$, \Ref{Jrs from b}--\Ref{U} reduces to \Ref{Boson operators from nodal density operators}. 

\subsection{Density correlation functions} 
\label{Density correlation functions} 
We now consider density correlation functions or, more specifically, the two-point function
\begin{equation}
\label{cD}
\langle J_{r_1,s_1}(\vx_1,t_1)J_{r_2,s_2}(\vx_2,t_2)\rangle_\beta,\qquad J_{r,s}(\vx,t)=\ee^{\ii Ht}J_{r,s}(\vx)\ee^{-\ii H t}
\end{equation}
with the densities $J_{r,s}(\vx)$ in \Ref{Nodal densities in position space}.

Similar to the fermion field operators, the $J_{r,s}(\vx)$ are operator-valued distributions and thus do not automatically have well-defined products. When computing \Ref{cD}, it is therefore convenient to use the regularized densities $J_{r,s}(\vx;\epsilon)$ in \Ref{Nodal densities in position space}:\ for $\epsilon>0$, the latter are operators with well-defined products on the domain $\cD$ discussed in the paragraph after Lemma~\ref{Lemma Fock space basis}. Calculating $\tilde{J}_{r,s}(\vx,t;\epsilon)=\ee^{\ii\tilde{H}t}\cU^\dag J_{r,s}(\vx;\epsilon)\cU\ee^{-\ii\tilde{H}t}$ using \Ref{hatNrs}, \Ref{Zero-mode operator},  \Ref{Charges and chiral charges}, \Ref{UNUdag}, and \Ref{Jrs from b}, yields
\begin{equation}
\label{tildeJrs}
\begin{split}
\tilde{J}_{r,s}(\vx,t;\epsilon) = & \tilde{J}^+_{r,s}(\vx,t;\epsilon) + \tilde{J}^-_{r,s}(\vx,t;\epsilon)+\tilde{J}^0_{r,s}(\vx,t;\epsilon)\\ \tilde{J}^-_{r,s}(\vx,t;\epsilon) = & \frac1{L\sqrt{\ta}}\sum_{\vp\in\hat{\Lambda}^*_{s}}\sum_{s'=\pm} v^{s'}_{r,s}(\vp)b_{s'}(\vp)\ee^{\ii\vp\vx-\ii\omega_{s'}(\vp)t}\ee^{-\epsilon |p_s|/2}\\ \tilde{J}^+_{r,s}(\vx,t;\epsilon) = & \tilde{J}^-_{r,s}(\vx,t;\epsilon)^\dag\\ \tilde{J}^0_{r,s}(\vx,t;\epsilon) = &\frac1{L\ta}\Bigl( N_{r,s}(x_{-s}) 
-\sum_{r'=\pm}\sum_{y\in\Lambda_{1D}}f'(x_s-y;\epsilon)N_{r',-s}(y)\Bigr)
\end{split} 
\end{equation} 
with the function
\begin{equation}
\label{fx} 
f'(x;\epsilon)= \frac{\gamma_2}{2(1+\gamma_1)}\frac{\ta}{L}\sum_{p\in\hat\Lambda^*_{1\rm D}}\ee^{\ii px}\ee^{-\epsilon|p|/2} 
\end{equation} 
and superscripts $+$, $-$, and $0$ indicating the creation-, annihilation-, and zero-mode parts, respectively. Note that $f'(x;0^+)=\gamma_2[\delta_{x,0}-(\ta/L)]/[2(1+\gamma_1)]$. 

With that, one can use \Ref{trivial 2} to compute the correlation function in \Ref{cD}. The boson part of the thermal expectation values can be computed by a well-known result stated in Appendix~\ref{Neutral bosons}, Lemma~\ref{Boson results}(a). The zero-mode contribution is computed in Appendix~\ref{Zero-mode correlation functions appendix}. We summarize the result as follows. 

\begin{result}
\label{density correlation result}
For $j=1,2$, let $r_{j},s_{j}=\pm$, $\vx_j\in\Lambda_s$, and $t_j\in\C$. Then the density two-point function in \Ref{cD} is given by the limit $\epsilon\to 0^+$ of 
\begin{equation}
\label{cDresult}
\begin{split} 
\frac1{\ta}\sum_{\vp\in\hat\Lambda^*_{s_1}\cap\hat\Lambda^*_{s_2}} \Bigl(\frac1{L}\Bigr)^2 \sum_{s=\pm} \Bigl( \overline{v^{s}_{r_1,s_1}(\vp)}v^{s}_{r_2,s_2}(\vp)\frac{\ee^{\ii\omega_s(\vp)(t_1-t_2)}}{\ee^{\beta\omega_s(\vp)}-1} + v^s_{r_1,s_1}(-\vp)\overline{v^{s}_{r_2,s_2}(-\vp)} \\ \times \frac{\ee^{-\ii\omega_s(\vp)(t_1-t_2)}}{1-\ee^{-\beta\omega_s(\vp)}}\Bigr)\ee^{-\ii\vp(\vx_1-\vx_2)}\ee^{-\epsilon(|p_{s_1}|+|p_{s_2}|/2} +O(L^{-1}) 
\end{split} 
\end{equation}
with $\omega_s(\vp)$ in \Ref{oms} and $v^{s'}_{r,s}(\vp)$ in \Ref{Mrssp}.
\end{result}

As seen in Appendix~\ref{Zero-mode correlation functions appendix}, the zero-mode contribution to the correlation function $\langle \tilde J^{0}_{r_1,s_1}(\vx_1,t_1;0^+)\tilde J^0_{r_2,s_2}(\vx_2,t_2;0^+)\rangle_{\zeromode,\beta}$ vanishes like $1/L$ in the IR limit (represented by the term $O(L^{-1})$ in \Ref{cDresult}). Note that the $\vp$-sum in \Ref{cDresult} is a Riemann sum converging to an integral in the same limit.

Since the density operators are linear in the boson operators, it is straightforward to compute thermal expectation values of an arbitrary number of ``time'' evolved density operators using Wick's theorem (the zero-mode terms do not contribute in the IR limit; see Appendix~\ref{Computation details}). 

\subsection{Fermion correlation functions} 
\label{Fermion correlation functions}
A general $N$-point fermion correlation function for the Mattis model is given by 
\begin{equation}
\begin{split}
\label{cGN} 
\bigl\langle\psi^{q_1}_{r_1,s_1}(\vx_1,t_1) \cdots \psi^{q_N}_{r_N,s_N}(\vx_N,t_N)\bigr\rangle_\beta,\qquad  \psi^q_{r,s}(\vx,t)=\ee^{\ii Ht}\psi^q_{r,s}(\vx)\ee^{-\ii Ht}
\end{split} 
\end{equation}
for arbitrary $N\in\N$, with $q_j,r_j,s_j=\pm$, $\vx_j\in\Lambda_{s_j}$, and $t_j\in\C$; here and in the following, we use the convenient notation \cite{HSU} 
\begin{equation}
\label{psipm}
\psi^-_{r,s}(\vx)\define \psi\pdag_{r,s}(\vx),\qquad \psi^+_{r,s}(\vx)\define \psi^\dag_{r,s}(\vx).
\end{equation} 
Proposition~\ref{Proposition Fermions from Bosons} gives regularized fermion field operators $\psi^\pm_{r,s}(\vx,\epsilon)\define \cV_{r,s}(\vx;\epsilon)^{\pm 1}$ in terms of the operators $\hat J_{r,s}(\vp)$ and $R_{r,s}(x_{-s})$. This, Proposition~\ref{Observables}, and straightforward computations, yield the following expressions for the fermion fields $\tilde\psi^q_{r,s}(\vx,t;\epsilon)\define \ee^{\ii \tilde{H}t} \cU^\dag \psi^q_{r,s}(\vx,\epsilon)\cU \ee^{-\ii \tilde{H}t}$, $q=\pm$, 
\begin{equation}
\label{tpsi}
\begin{split}
\tilde\psi^q_{r,s}(\vx,t;\epsilon)=& \frac1{\sqrt{2\pi \ta \epsilon}}S^q_{r,s}(\vx,t;\epsilon)\ee^{-\ii q[K^{+}_{r,s}(\vx,t;\epsilon) + K^{-}_{r,s}(\vx,t;\epsilon)]} \\ S^q_{r,s}(\vx,t;\epsilon) = & \ee^{-\ii qK^{0}_{r,s}(\vx,t;\epsilon)/2} R_{r,s}(x_{-s})^{qr}\ee^{-\ii q K^{0}_{r,s}(\vx,t;\epsilon)/2}
\end{split} 
\end{equation}
with 
\begin{equation}
\label{Krs}
\begin{split}
K^{-}_{r,s}(\vx,t;\epsilon) =&  u_s(x_{-s},t)-\ii r\frac{2\pi\sqrt{\ta}}{L}\sum_{\vp\in\hat{\Lambda}^*_{s}}\sum_{s'=\pm} \frac{v^{s'}_{r,s}(\vp)}{p_s}b_{s'}(\vp)\ee^{\ii\vp\vx-\ii\omega_{s'}(\vp)t}\ee^{-\epsilon |p_s|/2}\\ K^{+}_{r,s}(\vx,t;\epsilon) =& K^-_{r,s}(\vx,t;\epsilon)^\dag\\ K^{0}_{r,s}(\vx,t;\epsilon) =&  r\frac{2\pi}{L}\Bigl( x_sN_{r,s}(x_{-s}) 
- \sum_{r'=\pm}\sum_{y\in\Lambda_{1\rm D}}f(x_s-y;\epsilon)N_{r',-s}(y)\Bigr) 
- k_{r,s}(x_{-s})t
\end{split}
\end{equation}
and 
\begin{equation}
\begin{split} 
f(x;\epsilon)= -\ii\frac{\gamma_2}{2(1+\gamma_1)}\frac{\ta}{L}\sum_{p\in\hat\Lambda^*_{1\rm D}} \frac{1}{p}\ee^{\ii px}\ee^{-\epsilon|p|/2}
\end{split} 
\end{equation}  
where, again, superscripts $+$, $-$, and $0$ indicate the creation-, annihilation-, and zero-mode parts of an operator $K_{r,s}(\vx,t;\epsilon)$ (cf. \Ref{tildeJrs}). Note that the operator $\tilde{J}_{r,s}(\vx,t;\epsilon)$ in \Ref{tildeJrs}, multiplied by a factor $2\pi\ta r$, is equal to the partial $x_s$-derivative of $K_{r,s}(\vx,t;\epsilon)$. 

With that, \Ref{trivial 2} provides a convenient recipe to compute the fermion Green's function in \Ref{cGN}:\ by \Ref{factorize} and \Ref{tpsi}, one has to compute the boson expectation value  
\begin{equation}
\label{B expectation value} 
\bigl\langle\ee^{-\ii q_1[K^{+}_{r_1,s_1}(\vx_1,t_1;\epsilon) + K^{-}_{r_1,s_1}(\vx_1,t_1;\epsilon)]}\cdots \ee^{-\ii q_N[K^{+}_{r_N,s_N}(\vx_N,t_N;\epsilon) + K^{-}_{r_N,s_N}(\vx_N,t_N;\epsilon)]}\bigr\rangle_{B,\beta}
\end{equation} 
and the zero-mode expectation value 
\begin{equation}
\label{Q expectation value} 
\bigl\langle S^{q_1}_{r_1,s_1}(\vx_1,t_1;\epsilon)\cdots S^{q_N}_{r_N,s_N}(\vx_N,t_N;\epsilon)\bigr\rangle_{\zeromode,\beta} .   
\end{equation} 

Since $K^\pm_{r,s}(\vx,t;\epsilon)$ is a linear combination of boson operators, the boson part in \Ref{B expectation value} can be computed using  well-known results stated in Appendix~\ref{Neutral bosons}, Lemma~\ref{Boson results}(b). We note that all terms involving $u_s(x_{-s},t)$ are $O(L^{-1})$ and thus do not contribute in the IR limit.

The zero-mode part in \Ref{Q expectation value} can be obtained by straightforward computations; see Appendix~\ref{Zero-mode correlation functions appendix} 
for details. One finds that, in the IR limit, this term simplifies to 
\begin{equation}
\label{RRR}
\langle\Omega,R_{r_1,s_1}(x_{1,-s_1})^{q_1r_1}\cdots R_{r_N,s_N}(x_{N,s_{-N}})^{q_Nr_N}\Omega\rangle.
\end{equation} 

We thus obtain the following:
\begin{result}
\label{fermion correlation result}
For $N\in\N$ and $j=1,2,\ldots,N$, let $q_j,r_j,s_j=\pm$, $\vx_j\in\Lambda_{s_j}$, and $t_j\in\C$. Then the fermion correlation function in \Ref{cGN} is given by the limit $\epsilon\to 0^+$ of 
\begin{equation}
\begin{split} 
\label{fermion correlation function}
\langle\Omega,R_{r_1,s_1}(x_{1,-s_1})^{r_1q_1}\cdots R_{r_N,s_N}(x_{N,-s_N})^{r_Nq_N}\Omega\rangle\Bigl( \prod_{j=1}^N g_{r_j,s_j}(\epsilon)^{-1/2} \Bigr) \\ \times\Bigl(\prod_{1\leq j<k\leq N} G_{r_j,s_j,r_k,s_k}(\vx_j-\vx_k,t_j-t_k;\epsilon)^{-q_jq_k} \Bigr)\Bigl(1 + O(L^{-1})\Bigr)
\end{split} 
\end{equation} 
where
\begin{equation}
\begin{split}
\label{G}
\ln G_{r_1,s_1,r_2,s_2}(\vx,t;\epsilon)= r_1r_2\ta\sum_{\vp\in\hat\Lambda^*_{s_1}\cap\hat\Lambda^*_{s_2}} \Bigl(\tPiL\Bigr)^2 \sum_{s=\pm} \frac1{p_{s_1}p_{s_2}}\Bigl( \overline{v^{s}_{r_1,s_1}(\vp)}v^{s}_{r_2,s_2}(\vp) \\ \times\frac{\ee^{\ii\omega_s(\vp)t}}{\ee^{\beta\omega_s(\vp)}-1} + v^s_{r_1,s_1}(-\vp)\overline{v^{s}_{r_2,s_2}(-\vp)}\frac{\ee^{-\ii\omega_s(\vp)t}}{1-\ee^{-\beta\omega_s(\vp)}}\Bigr)\ee^{-\ii\vp\vx}\ee^{-\epsilon (|p_{s_1}|+|p_{s_2}|)/2}
\end{split}
\end{equation}
\begin{equation}
\label{g}
g_{r,s}(\epsilon)= 2\pi\ta\epsilon G_{r,s,r,s}(\vzero,0;\epsilon)
\end{equation} 
with $\omega_s(\vp)$ in \Ref{oms} and $v^{s'}_{r,s}(\vp)$ in \Ref{Mrssp}.
\end{result}
Note that the $\vp$-sum in \Ref{G} is a Riemann sum converging to an integral in the IR limit. Even though this integral is singular, it is well-defined since its definition also specifies how to treat the singularity; see Section~\ref{Fermion two-point function for gamma2=0}. The terms $O(L^{-1})$ in \Ref{fermion correlation function} are mainly zero-mode contributions,\footnote{The other contribution comes from $u_s(x_{-s},t)$, as already mentioned.} and they are computed in Appendix~\ref{Computation details}.  

We emphasize that, in the IR limit, the effect of the zero-mode terms is negligible except for the factor in \Ref{RRR}, which is independent of temperature and coupling. This factor can be easily computed by ``moving Klein factors to the right'' using
\begin{equation}
\label{RR} 
R_{r,s}^{rq}(x)R_{r',s'}^{r'q'}(x')= \begin{cases} 1& \mbox{ if } (q,r,s,x)=(-q',r',s',x')\\ - R_{r',s'}^{r'q'}(x')R_{r,s}^{rq}(x)& \mbox{ otherwise} \end{cases} 
\end{equation} 
(this follows from \Ref{Klein factor commutator relations}) and \Ref{cR}. Thus, a fermion Green's function in \Ref{fermion correlation function} is non-zero only if $N$ is even and if, by using \Ref{RR} repeatedly, the product of Klein factors in \Ref{RRR} can be simplified to either $1$ or $-1$. In particular,\footnote{Below we write $x_j$ short for $x_{j,-s_j}$.} 
\begin{equation}
\label{N2} 
\langle\Omega,R_{r_1,s_1}(x_1)^{r_1q_1}R_{r_2,s_2}(x_2)^{r_2q_2}\Omega\rangle  = \delta_{q_1,-q_2}\delta_{r_1,r_2}\delta_{s_1,s_2}\delta_{x_1,x_2}
\end{equation}
\begin{equation}
\begin{split}
\label{N4} 
\langle  \Omega,R_{r_1,s_1}(x_1&)^{r_1q_1}\cdots R_{r_4,s_4}(x_4)^{r_4q_4}\Omega\rangle  = \\ &\delta_{q_1,-q_2}\delta_{r_1,r_2}\delta_{s_1,s_2}\delta_{x_1,x_2}\delta_{q_3,-q_4}\delta_{r_3,r_4}\delta_{s_3,s_4}\delta_{x_3,x_4} \\ &-\delta_{q_1,-q_3}\delta_{r_1,r_3}\delta_{s_1,s_3}\delta_{x_1,x_3}\delta_{q_2,-q_4}\delta_{r_2,r_4}\delta_{s_2,s_4}\delta_{x_2,x_4} \\ &+\delta_{q_1,-q_4}\delta_{r_1,r_4}\delta_{s_1,s_4}\delta_{x_1,x_4}\delta_{q_2,-q_3}\delta_{r_2,r_3}\delta_{s_2,s_3}\delta_{x_2,x_3}.
\end{split} 
\end{equation}

\newsection{Quantum field theory limit}
\label{QFT limit}  
The results of the previous section allow us to compute, in principle, any quantity of physical interest for the Mattis model. However, the given formulas are in general complicated, and it is not easy to discern from them the qualitative features of the model. For this reason, it is useful to only consider the physical behavior for length scales that are much larger than the UV cutoff $\ta$, but still much smaller than the IR cutoff $L$. Indeed, as will be shown in this section, one can obtain simplified formulas in the QFT limit $L\to \infty$ and $\ta\to 0^+$ (after suitable multiplicative renormalizations). As illustrative examples, we analyze in detail the free energy and fermion two-point function of the Mattis model. Computational details can be found in Appendix~\ref{Appendix QFT limit}.

\subsection{Free energy}
\label{QFT limit of Free energy} 
Computing the contributions to the free energy in leading order in $\ta$ and $1/L$ we obtain:

\begin{result}
\label{Result free energy 2}
The QFT limit of the (renormalized) free energy contribution $\Omega_\beta-\cE_0=\Omega_{B,\beta}+\Omega_{\zeromode,\beta}$ is
\begin{equation}
\label{F} 
\lim_{\ta\to 0^+} \lim_{L\to\infty} \ta L^{-2} \left(\Omega_\beta-\cE_0\right) = -\frac{\pi}{3\tilde{v}_F\sqrt{A}\beta^2}
\end{equation} 
with the parameters $A$ and $\tilde{v}_F$ in \Ref{AvF}. 
\end{result}

\noindent ({\em Proof:} See Appendix~\ref{QFT limit of free energy appendix}.) 

\medskip

This result is easy to derive when $\gamma_2=0$ ($A=1$) and, in this case, the corrections to the leading term of $\ta L^{-2} \left(\Omega_\beta-\cE_0\right)$ (at non-zero $\ta$) are exponentially suppressed (i.e.\ they are $O\bigl(\ee^{-\beta v_F/\ta}\bigr)$).\footnote{\label{Beta vs a-tilde}As seen in \cite{EL1}, the Fermi velocity $v_F$ is proportional to $\ta$ when the Mattis model is derived from the original lattice fermion model. Thus, in this case, the corrections to \Ref{F} do not vanish in the UV limit $\ta\to 0^+$, and Result~\ref{Result free energy 2} should be regarded as the low-temperature limit of the free energy.}

As seen in the proof in Appendix~\ref{QFT limit of free energy appendix}, the result for non-zero $\gamma_2$ can be obtained by replacing the original dispersion relation \Ref{oms} by the effective dispersion
\begin{equation}
\label{Renormalized dispersion}
\tilde\omega_\pm(\vp) \define \tilde{v}_F\sqrt{A}|p_\pm| \qquad \left(\gamma_2\chi(\vp)p_+p_-\neq 0\right)
\end{equation}
(corresponding to renormalizing $\gamma_1 \to\sqrt{1-A\left(1-\gamma_1^2\right)}$ and $\gamma_2\to0$ in \Ref{oms}) in the expression \Ref{OmB} for the boson free energy. This can be understood as follows. The dominating contributions to \Ref{OmB} come from terms for which the dispersion is minimized. A glance at Figure~\ref{Dispersion relation} shows that, in this regime, the dispersion relation is excellently approximated by \Ref{Renormalized dispersion}.

\subsection{Fermion two-point functions}
\label{Fermion two-point functions}
As discussed above, we are interested in the behavior of correlation functions for intermediate length and time scales defined by the regime
\begin{equation}
\label{QFT regime}
\ta \ll ||\vx|\pm v_F t| \ll L, \qquad \ta \ll ||\vx|\pm \tilde v_F \sqrt{A} t| \ll L. 
\end{equation}
In this section, we will focus solely on the case of fermion two-point functions in this regime and at zero temperature. 

To illustrate important features of the QFT limit, we first consider the special case $\gamma_2=0$ for which simple formulas can be obtained even before taking the limit $\ta\to 0^+$. We will need the following special function
\begin{equation}
\label{Fermion two-point function special function}
\sigma(z)\define \ee^{-E_1(z)}, \qquad E_1(z)\define \int_1^\infty (1/t)\ee^{-zt}\ud t \qquad (\arg(z)<\pi),
\end{equation}
whose asymptotic behavior is given by (see Appendix~\ref{Fermion two-point function for gamma2=0})
\begin{equation}
\label{Asymptotics of special function}
\sigma(z)=\ee^{\gamma}z\left(1+O(z)\right), \qquad \sigma(z)={1 - \frac{{{\ee^{ - z}}}}{z}\left( {1 + O({z^{ - 1}})} \right)},
\end{equation}
with Euler's constant $\gamma = 0.5772(1)$.
\begin{result}
\label{Result Fermion two-point function for gamma2=0}
Let $\gamma_2=0$, $q_j,r_j,s_j=\pm$ for $j=1,2$, $\vx\in\R^2$, and $t\in\C$. Then the fermion two-point functions at zero temperature and in the IR limit are
\begin{equation}
\label{IR limit fermion two-point function}
\begin{split}
&\lim\limits_{L \to \infty} \left\langle {\psi_{r_1,s_1}^{q_1}(\vx,t)\psi^{q_2}_{r_2,s_2} (\vzero,0)} \right\rangle_\infty = \delta_{q_1,-q_2}\delta_{r_1,r_2}\delta_{s_1,s_2} \frac{1}{\ta}\delta_{x_{-s_1},0} {\left(\ee^\gamma\pi\right)^{1 - K}} \\ &\times F_{r_1,s_1}(x_{s_1},t)\frac{1}{{2\pi }}\frac{1}{{{0^ + }  -\ii\left({r_1}{x_{{s_1}}} - {{\tilde v}_F}t\right)}} \left(\frac{\ta^2}{{{{ {(0^+ + \ii \tilde{v}_F t)^2 + (x_{s_1})^2} }}}}\right)^{\left( {K - 1} \right)/2} 
\end{split}
\end{equation}
with ${\tilde v}_F$ in \Ref{AvF}, 
\begin{equation}
\label{BK}
K=\frac12\bigl( B+B^{-1} \bigl),\qquad B=\sqrt{\frac{1-\gamma_1}{1+\gamma_1}}, 
\end{equation}
and
\begin{equation}
\label{Frsxt}
F_{r,s}(x,t) = \frac{{\sigma {{\left( {\frac{\pi }{{\ta}}({0^ + } - \ii({r}{x} -{{\tilde v}_F}t))} \right)}^{\left( {K + 1} \right)/2}}\sigma {{\left( {\frac{\pi }{{\ta}}({0^ + } + \ii({r}{x} + {{\tilde v}_F}t))} \right)}^{\left( {K - 1} \right)/2}}}}{{\sigma \left( {\frac{\pi }{{\ta}}({0^ + } - \ii({r}{x} -{v_F}t))} \right)}}.
\end{equation}
\end{result} 

\noindent ({\em Proof:} See Appendix~\ref{Fermion two-point function for gamma2=0}.) 

\medskip

It is instructive to compare the short- versus the intermediate length- and time scale structure of the correlation functions in \Ref{IR limit fermion two-point function}. Using the first relation in \Ref{Asymptotics of special function} one finds for short length- and time scales
\begin{equation}
\begin{split}
\lim\limits_{L \to \infty} \left\langle {\psi_{r,s}^{\dag}(\vx,t)\psi\pdag_{r,s} (\vzero,0)} \right\rangle_\infty =& \frac{1}{\ta}\delta_{x_{-s},0}\frac{1}{{2\pi }}\frac{1}{{{0^ + }  -\ii\left({r}{x_{{s}}} - {{v}_F}t\right)}}\\ &\times\left(1+O(x_s/\ta)+O(v_Ft/\ta)+O(\tilde v_Ft/\ta)\right),
\end{split}
\end{equation}
identical with the short-distance behavior in the non-interacting case. 

From the second relation in \Ref{Asymptotics of special function}, it follows that $F_{r,s}(x,t)\to 1$ in the regime \Ref{QFT regime} and in the QFT limit. The result in \Ref{IR limit fermion two-point function} thus shows that the fermion two-point functions in the Mattis model, for $\gamma_2=0$ and in the QFT limit, have an algebraic decay with exponents depending on the coupling parameter $\gamma_1$. We note that such behavior is one of the hallmarks of Luttinger-liquid behavior \cite{Haldane}. 

Note that $K>1$ for the interacting case, and therefore, to get a non-trivial UV limit in the regime \Ref{QFT regime}, one needs to multiply \Ref{IR limit fermion two-point function} by a factor $\propto \ta^{(1-K)}$ before taking $\ta\to 0^+$. We find it convenient to choose this multiplicative renormalization factor as $(\ee^{\gamma}\pi L_0/\ta)^{K-1}$ with a finite, arbitrary length scale parameter $L_0$. 

\begin{remark}
The computation above provides a generalization of the limit in which the 1D Luttinger model reduces to the massless Thirring model; see e.g.\ \cite{M1,GLR}. 
\end{remark} 

One of our main results is that, after a multiplicative renormalization similar to the one above, such an algebraic decay of fermion two-point functions holds true in the QFT limit even for non-zero $\gamma_2$:

\begin{result}
\label{QFT limit Fermion two-point function}
Let $q_j,r_j,s_j=\pm$ for $j=1,2$, $\vx\in\R^2$, and $t\in\C$. Then the (renormalized) fermion two-point functions at zero temperature and in the QFT limit are
\begin{equation}
\label{psipsi2} 
\begin{split}
\lim\limits_{\ta\to 0^+} \lim\limits_{L \to \infty} \Bigl(\frac{\ee^{\gamma}\pi L_0}{\ta}\Bigr)^{K-1} \left\langle {\psi_{r_1,s_1}^{q_1}(\vx,t)\psi^{q_2}_{r_2,s_2} (\vzero,0)} \right\rangle_\infty = \delta_{q_1,-q_2}\delta_{r_1,r_2}\delta_{s_1,s_2}\delta(x_{-s_1}) \\\times C(\gamma_1,\gamma_2)\frac{1}{{2\pi}}\frac{1}{{{0^+} - \ii({r_1}{x_{{s_1}}} - {{\tilde v}_F}\sqrt{A}t)}}{\left( {\frac{{{{\left( {{L_0}} \right)}^2}}}{ {(0^+ +\ii\tilde{v}_F \sqrt{A}t)^2 +(x_{s_1})^2}   }} \right)^{\left( {K - 1} \right)/2}}
\end{split}
\end{equation} 
with the dimensionless constant $C(\gamma_1,\gamma_2)$ given in Appendix~\ref{Fermion two-point function: general case},
\begin{equation}
\label{K2}
K=\frac12\Bigl( \frac{\sqrt{A}}{B}+\frac{B}{\sqrt{A}}\Bigr), \qquad B=\sqrt{\frac{1-\gamma_1}{1+\gamma_1}}, 
\end{equation}
and $A$, ${\tilde v}_F$ in \Ref{AvF}; $L_0>0$ is an arbitrary length scale parameter.
\end{result} 

\noindent ({\em Proof:} See Appendix~\ref{Fermion two-point function: general case}.)

\medskip

\begin{figure}[!ht]
\begin{center}
\includegraphics[width=0.8\textwidth]{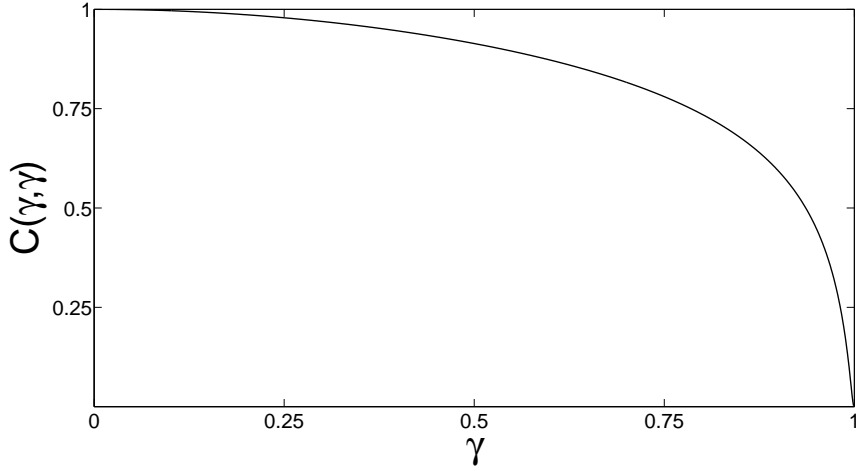} 
\end{center}
\caption{The dimensionless constant $C(\gamma,\gamma)$, occurring in the QFT limit of the fermion two-point functions, for the special case of $\gamma\define\gamma_1=\gamma_2$.}
\label{Multiplicative constant C}
\end{figure}

The constant $C(\gamma_1,\gamma_2)$ is defined by an integral of trigonometric functions (see \Ref{EE}--\Ref{C12}). The special case $\gamma\define\gamma_1=\gamma_2$ is plotted in Figure~\ref{Multiplicative constant C}.

The similarity between the correlation function in \Ref{psipsi2} and the corresponding result one would obtain for $\gamma_2=0$ ($A=1$), together with the  result derived for the free energy in section~\ref{QFT limit of Free energy}, might leave the reader with the impression that, in the QFT limit, no new physics is obtained by allowing $\gamma_2\neq 0$. We stress that this is not so. It is true that, apart from the overall multiplicative factor $C(\gamma_1,\gamma_2)$, Result~\ref{QFT limit Fermion two-point function} can be obtained from the corresponding result for $\gamma_2=0$ by effectively replacing ${\tilde v}_F$ by ${\tilde v}_F\sqrt{A}$ and $K$ in \Ref{BK} by that in \Ref{K2}. However, it is {\em not} possible to find an effective coupling parameter $\tilde\gamma_1$ such that the result in \Ref{psipsi2} is obtained from the one for $\gamma_2=0$ and 
by changing $\gamma_1$ to $\tilde\gamma_1$: the UV limit of the Mattis model depends non-trivially on both coupling parameters $\gamma_{1,2}$.

\subsection{Other correlation functions}
Our results in the previous section suggest that the following renormalized fermion operators  
\begin{equation}
\Psi^\pm_{r,s}(\vx)\define (\ee^{\gamma}\pi L_0/\ta)^{(K-1)/2} \psi^\pm_{r,s}(\vx), 
\end{equation}
with $K$ in \Ref{K2} and $L_0>0$ an arbitrary length scale parameter, have well-defined non-trivial correlation functions in the QFT limit: 
\begin{conjecture}
\label{QFT limit conjecture}
Let $q_j,r_j,s_j=\pm$, $\vx_j\in\R^2$, and $t_j\in\C$ for $j=1,2,\ldots,N$, and $N\in\N$. Then the following renormalized fermion correlation functions of the Mattis model in the QFT limit,  
\begin{equation}
\lim_{\ta\to 0^+}\lim_{L\to\infty}\bigl\langle  \Psi^{q_1}_{r_1,s_1}(\vx_1,t_1)\cdots \Psi^{q_N}_{r_N,s_N}(\vx_N,t_N)\bigr\rangle_\beta
\end{equation} 
are well-defined distributions for all $1/\beta\geq 0$. 
\end{conjecture} 
A proof of this conjecture, which would imply that the (constructive) QFT limit of the Mattis model exists and is non-trivial, involves taking the limit $\ta\to 0^+$ in Result~\ref{fermion correlation result} for arbitrary $N$. The calculation might be done in a similar way as the $N=2$ case presented in the previous section and Appendix~\ref{App:Fermion two-point functions}. However, due in part to the increasing complexity in the contribution from the Klein factors to higher order correlation functions (see for example Equation~\Ref{N4} corresponding to $N=4$), a formula for arbitrary $N$ will not be attempted in the present paper. 

It is also possible to compute the QFT limit of density correlation functions. Here we only give the result for the case $\gamma_2=0$ and in the IR limit (see also the remark at the end of Section~\ref{Density correlation functions}): 
\begin{result}
Let $\gamma_2=0$, $r_j,s_j=\pm$ for $j=1,2$, $\vx\in\R^2$, and $t\in\C$. Then the (renormalized) density two-point functions at zero temperature and in the IR limit are
\begin{equation}
\label{JJ1}
\begin{split}
&\lim\limits_{L \to \infty }{\ta \left\langle {J_{r_1,s_1} (\vx,t) J_{r_2,s_2}(\vzero,0)} \right\rangle_\infty } = \delta_{s_1,s_2}\frac{1}{\ta}\delta_{x_{-s_{1}},0}\frac1{4(2\pi)^2}\\&\times \Biggl( \frac{B+r_1r_2B^{-1}+(r_1+r_2)}{\left(0^+-\ii\left(x_{s_1}-\tilde v_F t\right)\right)^2} + \frac{B+r_1r_2B^{-1}-(r_1+r_2)}{\left(0^++\ii\left(x_{s_1}+\tilde v_F t\right)\right)^2} + e_{r_1,r_2}(x_{s_1},t)\Biggr) 
\end{split}
\end{equation}
with 
\begin{equation}
\begin{split}
e_{r,r'}(x,t) \define & {\left( {\frac{\pi }{{\ta}}} \right)^2}\left( {4{\delta _{r,r'}}{\alpha _1}\left( {\frac{\pi }{{\ta}}\left( {0^+  - \ii\left( {r{x} - {v_F}t} \right)} \right)} \right)} \right. \\  &\left. { - \sum\limits_{\tilde r =  \pm } {\left( {\left( {B + rr'{B^{ - 1}}} \right) + \tilde r\left( {r + r'} \right)} \right){\alpha _1}\left( {\frac{\pi }{{\ta}}\left( {0^+  - \ii\left( {\tilde r{x} - {{\tilde v}_F}t} \right)} \right)} \right)} } \right),
\end{split}
\end{equation}
the function ${\alpha _1}\left( z \right) \define z^{-2}\ee^{ - z} \left( {1 + z} \right)$, and the parameters ${\tilde v}_F$ and $B$ given in \Ref{AvF} and \Ref{K2} respectively.
\end{result}

\noindent ({\em Proof:} See Appendix~\ref{Density-density correlation function for gamma2=0}.)

\medskip

The corresponding QFT limit of \Ref{JJ1} should follow in a direct manner (we expect that $e_{r,r'}(x,t)\to 0$ in a distributional sense as $\ta\to 0^+$). Finding a simple analytical expression of the density two-point functions for general $\gamma_2$, and in the full QFT limit, appears more complicated and will not be attempted here.

\subsection{Positivity} 
\label{Positivity} 
As mentioned, the massless Thirring model can be obtained from the 1D Luttinger model as a limit that is similar to the full QFT limit of the Mattis model.  One can therefore regard the model obtained from the Mattis model in the full QFT limit (referred to as {\em Mattis QFT} for short in the following) as a 2+1 dimensional variant of the massless Thirring model. The Thirring model has been studied extensively in the context of axiomatic QFT (see e.g.\ \cite{CRW} and references therein), and it is therefore interesting to also discuss Mattis QFT from this point of view. 

Obviously, Mattis QFT is not Lorentz invariant, and it is therefore {\em not} a QFT theory in the sense of axiomatic QFT.\footnote{Note that axiomatic QFT has been mainly concerned with Lorentz invariant QFT models motivated by particle physics --- the fact that the full QFT limit of the 1D Luttinger model yields a Lorentz invariant QFT model is a coincidence.}  For this reason, many issues of axiomatic QFT do not apply to Mattis QFT. One important issue that does apply, however,  is positivity (in the sense of Osterwalder and Schrader \cite{OS}). We now sketch why positivity holds true for Mattis QFT.

Our results in the present paper show that the Mattis model is defined by a self-adjoint Hamiltonian on a separable Hilbert space, and this Hamiltonian is bounded from below. Moreover, the fermion field operators are well-defined operator valued distributions on this Hilbert space, i.e., smearing them by suitable testfunctions gives bounded operators.\footnote{We do not use test functions to deal with the distributional nature of the field operators but another method similar to the one in \cite{CRW}.} For these reasons, positivity is obvious for the Mattis model. Since positivity is preserved under limits (see e.g.\ \cite{CRW}), the result follows.

Our argument above is sketchy since the details are very similar to the ones in the proofs of positivity of the massless Thirring model in  \cite{M1,CRW}. 

\newsection{Final remarks}
\label{Final remarks}

\noindent {\bf 1.} In this paper, we presented a mathematically precise construction of the Mattis model formally defined by the regularized Hamiltonian in \Ref{Formal position space Hamiltonian}. As proposed in \cite{EL1,dWL1}, the Mattis model provides an effective description of 2D lattice fermions in a partially gapped phase close to half filling. We diagonalized this Hamiltonian, and we computed all thermodynamic correlation functions. We also proved that the fermion two-point functions have a non-trivial QFT limit.

One of our aims was to develop computational tools that can be straightforwardly generalized to more complicated models of similar type. One such model is a spinfull variant of the Mattis model that can be derived from the 2D Hubbard model by a partial continuum limit.\footnote{J. de Woul and E. Langmann (in preparation).} 

\medskip

\noindent {\bf 2.} Although the Mattis model is similar to models of parallel Luttinger chains, there is one important difference: the Mattis model describes {\em two} sets of parallel Luttinger chains that are orthogonal to each other. Apart from the limiting case $\gamma_2=0$ (for which the Mattis model is essentially equivalent to the model studied in \cite{VM}), the two sets are non-trivially coupled by density-density interactions. 

\medskip

\noindent {\bf 3.} As mentioned in the previous remark, the orthogonal Luttinger chains of the Mattis model are not coupled for $\gamma_2=0$. This explains the locality of the fermion correlation functions parallel with the Fermi surface directions in this case. We find that this locality holds true even at non-zero values of $\gamma_2$. This is due to the $\delta_{x_j,x_k}$-factors that appear in the Klein factor contributions in \Ref{N2} and \Ref{N4}. For density correlation functions, this locality seems to be restricted to the case $\gamma_2=0$, at least for length scales of order $\ta$. We suspect that this locality of the fermion correlations functions can be changed by adding correction terms to the Mattis Hamiltonian (see below). 

\medskip

\noindent {\bf 4.}  The results obtained in the present paper allow, in principle, to check the validity of the approximations used to obtain the 2D Luttinger model from the 2D \ttpV model \cite{EL1} (similarly as done in \cite{SL} for the model proposed by Luther \cite{Luther2}) and, if necessary, to improve upon these approximations. To be more specific we discuss one important approximation used in \cite{EL1}, namely replacing the full band relation by linearized ones (the approximation in \Ref{Approx 1}). As elaborated in Appendix~\ref{appX}, the leading corrections to this approximation are the following terms that can be added to the Mattis Hamiltonian, 
\begin{equation}
\label{Hcorr2} 
H_2 = \sum_{r,s=\pm} v_F \tint{s}\ud^2x :\!\Bigl( \alpha_{2,0}\ta\psi^\dag_{r,s}(-\ii\partial_s)^2\psi\pdag_{r,s} + \alpha_{3,0}\ta^2r\psi^\dag_{r,s}(-\ii\partial_s)^3\psi\pdag_{r,s}\Bigr)\!:  
\end{equation}
and
\begin{equation}
\label{Hcorr3} 
\begin{gathered} 
H_3 = -\sum_{r,s=\pm} v_F  \tint{s}\ud^2x :\!\Bigl( \alpha_{0,2}\ta \psi^\dag_{r,s}(\hat\partial^2_{-s}\psi\pdag_{r,s}) - \alpha_{1,2}\ta^2r\psi^\dag_{r,s}(\ii\partial_s)(\hat\partial^2_{-s}\psi\pdag_{r,s})\Bigr)\!: \\ (\hat\partial^2_{-s}\psi_{r,s})(\vx)\define \frac1{\ta^2}\Bigl( \psi_{r,s}(\vx+\ta\ve_{-s})+\psi_{r,s}(\vx-\ta\ve_{-s})-2\psi_{r,s}(\vx)\Bigr) 
\end{gathered}
\end{equation} 
with dimensionless constants $\alpha_{j,k}$ given in \Ref{alphajk}. The correction terms in  \Ref{Hcorr2} and \Ref{Hcorr3} take into account curvature effects of the band relations orthogonal to, and parallel with, the Fermi surface, respectively, up to terms $O(\ta^4\vk^4)$. The former can be bosonized as follows (the interested reader can find details in Appendix~\ref{AppX2}),
\begin{equation}
\label{Hcorr2_bosonized}
\begin{split} 
&H_2 =  \sum_{s=\pm} v_F \tint{s} \ud^2 x \xx\!\Bigl( \frac{\alpha_{2,0}\sqrt{\pi \ta^3}}3\Bigl[(\partial^{\phantom s}_s\Phi^{\phantom s}_s)^3+ 3\Pi_s^2\partial^{\phantom s}_s\Phi^{\phantom s}_s\Bigr] +\frac{\alpha_{3,0}\ta^2}4\Bigl[\pi\ta \\ &\times\Bigl( \Pi_s^4+6\Pi_s^2(\partial^{\phantom s}_s\Phi^{\phantom s}_s)^2+(\partial^{\phantom s}_s\Phi^{\phantom s}_s)^4\Bigr)  + (\partial^2_s\Phi^{\phantom s}_s)^2+(\partial^{\phantom s}_s\Pi^{\phantom s}_s)^2 \Bigr]+ O(L^{-1})\Bigr)\!\xx 
\end{split} 
\end{equation} 
(the corrections are zero-mode terms and terms given in Appendix~\ref{appX}). Thus $H_2$ corresponds to non-linear boson interactions. Note that the coupling constants in \Ref{Hcorr2_bosonized} are of higher order in the UV cutoff $\ta$ than in \Ref{Hcorr2}. This suggests that the corrections $H_2$ are less important than one might naively expect. 

It is also possible to bosonize $H_3$ in \Ref{Hcorr3} using Proposition~\ref{Proposition Fermions from Bosons}, but for this term the Klein factors $R_{r,s}(x_{-s})$ do not cancel. We believe that these Klein factor terms are important corrections to study since they can make fermion correlation functions non-local parallel with the Fermi surface directions. 

In a similar way, one can bosonize correction terms for all other approximations that lead from the \ttpV-model to the 2D Luttinger model (see Appendix~\ref{Lattice relation} for more details on these approximations). 
\medskip

\noindent {\bf 5.} The term proportional to $\alpha_{0,2}$ in $H_3$ describes hopping between nearest neighbor chains. Previous work on coupled Luttinger chains suggests that Luttinger-liquid behavior can be destroyed by arbitrarily small interchain hopping; see e.g.\ \cite{BBT,KMS,VC} and references therein. However, in our case the hopping parameter vanishes in the UV limit $\ta\to 0^+$, and it is therefore an open question if $H_3$ can change the qualitative behavior of the Mattis model in the UV limit. 

Note that there is a particular location of the Fermi surface of the 2D \ttpV model where $\alpha_{0,2}=0$ (see Appendix~\ref{AppX1}). In this case the leading correction in $H_3$ to the Mattis model is zero. 
 
\medskip

\noindent {\bf 6.} The Hamiltonian methods used in this paper have several merits, but renormalization group methods based on functional integrals \cite{Salmhofer,M} seem more powerful when it comes to determining the effect of correction terms that spoil exact solubility (like the ones in \Ref{Hcorr2}--\Ref{Hcorr3}). We believe that it would be fruitful to try to combine these methods with existing functional integral approaches to bosonization \cite{Fog,FGM,KHS}. 

\medskip

\noindent {\bf 7.} Anderson in an influential paper \cite{Anderson1} (appearing three years after Mattis' \cite{Mattis}) suggested that 2D fermions can have Luttinger-liquid behavior. Our Result~\ref{QFT limit Fermion two-point function} shows that the fermion two-point functions in the QFT limit have a non-trivial algebraic decay. This property is often regarded as a hallmark of a 1D Luttinger liquid \cite{Haldane}. However, 2D~Luttinger-liquid behavior is subtle (one difficulty being that different definitions seem to be used in the literature). 

A mathematical criterion for Luttinger-liquid behavior in 2D in the context of constructive QFT was given by Salmhofer \cite{MS}. Since our methods are different from the ones used in constructive QFT, it is not easy to directly apply the criterion in \cite{MS} to our results. However, Mastropietro \cite{VM} has shown Luttinger-liquid behavior in the sense of \cite{MS} in a model which is, essentially, the limiting case $\gamma_2=0$ of the Mattis model. His result, together with our Result~\ref{QFT limit Fermion two-point function}, suggest that the Mattis model has such behavior for all allowed values of the coupling parameters $\gamma_{1,2}$. This is consistent with earlier results suggesting that non-Fermi-liquid behavior is only possible in 2D models with Fermi surfaces that have flat parts; see \cite{Salmhofer1998,DR} and references therein (note that the latter results are restricted to weakly coupled fermion systems). 

We emphasize that, even if one proves Luttinger-liquid behavior for the Mattis model, this does not prove Luttinger-liquid behavior of the 2D \ttpV model. For that one would need to substantiate the physical arguments in \cite{EL1,dWL1} by mathematical proofs. 

\medskip

\noindent {\bf 8.} The most general form of the Mattis Hamiltonian that can be solved exactly using the methods developed in the present paper is 
\begin{equation} 
\label{generalized Mattis Hamiltonian} 
\begin{split} 
H = & v_F\sum_{r,s=\pm}\sum_{\vk\in\Lambda_s^*}\Bigl(\tPiL\Bigr)^2 rk_s :\! \hat\psi^\dag_{r,s}(\vk) \hat\psi\pdag_{r,s}(\vk)\!:  \\ & + \sum_{r,r',s,s'=\pm}\sum_{\vp\in\frac{2\pi}{L}\mathbb{Z}^2}\Bigl(\frac{1}{L}\Bigr)^2 V_{r,s,r',s'}(\vp):\! \hat{J}_{r,s}(-\vp)\hat{J}_{r',s'}(\vp)\! : 
\end{split} 
\end{equation} 
with suitable interaction potentials $V_{r,s,r',s'}(\vp) =\overline{V_{r',s',r,s}(-\vp)}$ (the notation here is the same as in Definition~\ref{Mattis model}). It is straightforward to write $H$ in \Ref{generalized Mattis Hamiltonian} as a non-interacting boson Hamiltonian, and by diagonalizing the latter one finds necessary and sufficient conditions on $V_{r,s,r',s'}(\vp)$ such that the generalized Mattis model is well-defined (these conditions are similar to those for the Luttinger model; see e.g.\ \cite{HSU}, Equations~(3.22)--(3.24)).

The model discussed by Mattis \cite{Mattis} and Hlubina \cite{Hlubina} correspond to the special case $V_{r,s,r',s'}(\vp)=V(\vp)$ (independent of $r,s,r',s'$), with the latter author also assuming that the interaction is long-ranged, i.e.\ $V(\vp)=0$ for $|\vp|>q_0$ and $q_0\ll \pi/\tilde{a}$. To not overburden the presentation, and due to the motivation provided by our previous work \cite{EL1,dWL1}, we restricted our discussion in this paper to the case for which the interaction is  short-ranged, i.e.\ $V_{r,s,r',s'}(\vp)$ is essentially independent of $\vp$. It can be expected that the Mattis model with long-range interactions has physical behavior that is qualitatively different. 

\noindent {\bf Acknowledgements:} 
This work was supported by the G\"oran Gustafsson Foundation, and the Swedish Research Council (VR) under contract no.\  621-2010-3708. We thank two referees for suggestions and critical remarks that helped us to improve our paper. We also thank Farrokh Atai for carefully reading the manuscript.

\appendix

\newsection{Relation to lattice fermions}
\label{Lattice relation}
As discussed in the introduction, one motivation for the Mattis model is that it can provide an effective description of  a 2D lattice fermion system, as proposed in \cite{EL1,dWL1}. Here we describe in more detail the lattice fermion system and the nature of the arguments on which our proposal is based. 

We first introduce some useful terminology: we define fermion models by giving the so-called {\em set of one-particle quantum numbers} $S$, the {\em normalization of the CAR} (canonical anticommutator relations) $z>0$, the dispersion relation $\epsilon_k$, and {\em interaction vertex} $v_{k_1,k_2,k_3,k_4}$. By this we mean a fermion model defined by fermion field operators $\psi^{(\dag)}_k$, $k\in S$, obeying CAR with $\{\psi_k,\psi_{k'}\}  = z\delta_{k,k'}$, and with the Hamiltonian  
\begin{equation} 
\sum_{k\in S} \epsilon_k \psi^\dag_k\psi^{\pdag}_k + \sum_{k_1,k_2,k_3,k_4\in S} v_{k_1,k_2,k_3,k_4}\psi^\dag_{k_1} \psi^{\pdag}_{k_2}\psi^\dag_{k_3}\psi^{\pdag}_{k_4} .
\end{equation} 
We note that this defines the model completely if (and only if) $S$ is a finite set. 

The lattice model describes spinless fermions on a square lattice with nearest-neighbor (nn)  and next-nearest neighbor hopping and nn density-density interactions. This so-called \ttpV-model can be defined as follows: The set of one-particle quantum numbers is equal to the Brillouin zone 
\begin{equation} 
\label{BZ} 
 \BZ \define \left\{ \vk=(k_1,k_2):\;   k_\pm\in \tPiL\left(\Z+\frac12\right),  -\frac{\pi} a \leq k_{1,2} <\frac{\pi} a \right\}
\end{equation} 
of the lattice, with $k_\pm \define (k_1\pm k_2)\sqrt{2}$; here $a>0$ is the lattice constant and $L>0$ the lattice size such that $L/(2\sqrt{2} a)$ is a positive integer. Choosing the normalization of the CAR as $[L/(2\pi)]^2$, the band relation of the model is 
\begin{equation}
\label{eps0} 
\epsilon(\vk) = -2t [\cos(ak_1) + \cos(ak_2) ] -  4t'\cos(ak_1)\cos(ak_2), 
\end{equation}
and the interaction vertex is  
\begin{equation}
\hat{v}(\vk_1,\vk_2,\vk_3,\vk_4) = \hat{u}(\vk_1-\vk_2) 
\sum_{\vn\in\Z^2} \left(\frac{L}{2\pi}\right)^2 
\delta_{\vk_1-\vk_2+\vk_3-\vk_4,2\pi\vn/a}    
\end{equation} 
with $ \hat{u}(\vp)= a^2 V[\cos(a p_1)+\cos(a p_2)]/(8\pi^2)$ the suitably scaled Fourier transform of the nn interaction potential; see Section~4.1 in \cite{EL1} for further details and explanations.  The box surrounding Figure~\ref{Truncated Fermi surface} shows the boundary of the Brillouin zone in \Ref{BZ}, with the horizontal axis corresponding to $k_1$ and the vertical axis to $k_2$. The vectors $\vk$ are interpreted as momenta of the fermions. 

One important parameter of the \ttpV-model is {\em filling} $\nu$, which is defined as the average number of fermions per lattice site; it is in the range $0\leq\nu\leq 1$. Various arguments (discussed in \cite{EL1}) suggest that, at half-filling (i.e.\ for $\nu=1/2$) and for sufficiently large values of $V/t$, this model describes an insulator, i.e.,  all fermion degrees of freedom are gapped. The challenge is to understand the model away from, but close to, half-filling. Experimental and theoretical results (discussed in \cite{EL1}) suggest that fermions associated with different regions in the Brillouin zone can have different physical properties for $\nu\neq 1/2$ and, in particular, fermions close to points
\begin{equation}
\label{Qrs} 
\mathbf{Q}_{r,s}=(rQ/a,rsQ/a), \quad r,s=\pm 
\end{equation} 
with $Q\approx \pi/2$ (the four points that are midpoints of the dashed line segments in Figure~\ref{Truncated Fermi surface}) are gapless, while fermions in other regions can still be gapped. This suggests to rewrite the fermion model by dividing the Brillouin zone in eight different subsets $\vQ_{r,s} + \Lambda^*_{r,s}$, $r=\pm$ and $s=0,\pm,2$, with $\vQ_{r,s}$ conveniently chosen points such that 
\begin{equation}
\sum_{\vk\in\BZ}f(\vk) = \sum_{r,s}\sum_{\vk\in\Lambda^*_{r,s}} f(\vQ_{r,s}+\vk)
\end{equation} 
for any function $f$ on the Brillouin zone; see Figure~2 in \cite{EL1} for how this division is defined. In particular, the regions $\Lambda^*_{r,s}$, $r,s=\pm$, associated with the points in \Ref{Qrs} are rectangles defined by $-\pi/\tilde{a}<k_{-s}<\pi/\tilde{a}$, and a similar constraint in the $k_s$-direction (see \cite{EL1}, Equation~(42)) with $\tilde a=\sqrt{2}a/(1-\kappa)$ and $0<\kappa<1$. This parameter $\kappa$ determines the size of the regions treated by bosonization and it also gives the UV cutoff of the Mattis model that we use in the present paper \cite{EL1,dWL1}. 

This division of the Brillouin zone allows to rewrite the \ttpV-model model as a model of eight flavors of fermions, distinguished  by labels $r=\pm$ and $s=0,\pm,2$, with each of them labeled by momenta $\vk$ in different sets $\Lambda^*_{r,s}$: the dispersion relations and the interaction vertices of these fermions are  $\epsilon_{r,s}(\vk)\define  \epsilon(\vQ_{r,s}+\vk)$ and 
\begin{equation}
\label{vv8}
\begin{split}
\hat{v}_{r_1,s_1,\ldots,r_4,s_4}(\vk_1,\ldots,\vk_4) =
\hat{u}(\vQ_{r_1,s_1} -\vQ_{r_2,s_2} + \vk_1-\vk_2)
\sum_{\vn\in\Z^2} \left(\frac{L}{2\pi}\right)^2
\\ 
\times \delta_{\vQ_{r_1,s_1}-\vQ_{r_2,s_2}+ \vQ_{r_3,s_3}
  -\vQ_{r_4,s_4} + \vk_1-\vk_2+\vk_3-\vk_4,2\pi /a\vn}. 
\end{split} 
\end{equation} 
We refer to the fermions with $r,s=\pm$ as {\em nodal}. The fermions with $r=\pm$, $s=0$ correspond to regions around the points $\vQ_{+,0}=(\pi/a,0)$ and $\vQ_{-,0}=(0,\pi/a)$ and are referred to as {\em antinodal}; the fermions with $r=\pm$, $s=2$ are included for completeness, but we expect that they can be safely ignored in the parameter regime of interest to us \cite{EL1}. The parameter $Q$ is chosen such that the points $\vQ_{r,s}$ lie on the Fermi arcs, and one expects that, for $\nu =  1/2$, $Q=\pi/2$ \cite{EL1,dWL1}.

To obtain a model that is better amenable to computations we proposed a series of approximations \cite{EL1}: 
First, to Taylor expand the dispersion relations $\epsilon_{r,s}(\vk)$, $r,s=\pm$, and keep only the leading non-trivial terms, e.g., to replace  
\begin{equation} 
\label{Approx 1}
\epsilon_{r,s}(\vk) = \epsilon(\vQ_{r,s}+\vk) \to \epsilon(\vQ_{r,s}) + r v_F k_s,\quad r,s=\pm
\end{equation} 
with $v_F$ in \cite{EL1}, Equation~(6) (Approximation~A1 in \cite{EL1}). Second, to  replace the interaction vertices of the model by
\begin{equation}
\label{Approx 2} 
\begin{split} 
\hat{v}_{r_1,s_1,\ldots,r_4,s_4}(\vk_1,\ldots,\vk_4) \to  \hat{u}(\vQ_{r_1,s_1}-\vQ_{r_2,s_2}) \left(\frac{L}{2\pi}\right)^2  \delta_{\vk_1-\vk_2+\vk_3-\vk_4,\vzero}  \\ \times  \sum_{\vn\in\Z^2}
\delta_{\vQ_{r_1,s_1}-\vQ_{r_2,s_2}+\vQ_{r_3,s_3}-\vQ_{r_4,s_4},2\pi\vn/a}   
\end{split} 
\end{equation} 
(A2 in \cite{EL1}). Third, to ignore the fermion degrees of freedom with $r=\pm$, $s=2$ (A3' in \cite{EL1}). One key result in \cite{EL1} is that, if $Q\neq \pi/2$, the approximations above lead to a model with density-density interactions only, i.e., all interactions can be written in terms of fermion densities
\begin{equation}
\label{hrho}
\hat\rho_{r,s}(\vp) \define \sum_{\vk_1,\vk_2\in\Lambda^*_{r,s}} \left(\tPiL\right)^2 
\hat\psi^\dag_{r,s}(\vk_1)\hat\psi^{\phantom\dag}_{r,s}(\vk_2) \delta_{\vk_1+\vp,\vk_2} .
\end{equation}
To obtain a model where the nodal fermions can be treated in a simple way by bosonization two further approximations are needed: First, to drop the restriction on $k_s$ in the nodal sets, i.e., to replace in \Ref{hrho} and in the free part of the Hamiltonian  
\begin{equation}
\label{Approx 3 and 4}  
\Lambda_{r,s}^*\to \Lambda^*_s
\end{equation} 
with the latter sets defined in \Ref{Continuum nodal momentum region}, which can be done after normal ordering (A3 and A4 in \cite{EL1}). Second, to replace the (normal ordered) nodal densities in \Ref{hrho} thus obtained by the ones in \Ref{Nodal density operators}:
\begin{equation} 
\label{Approx 5} 
:\!\hat{\rho}_{r,s}(\vp)\!: \,  \to \hat{J}^{\pdag}_{r,s}(\vp), \quad r,s=\pm  , 
\end{equation} 
 i.e., to insert  integer sums (A5 in \cite{EL1}).  The former of these approximations leads to a QFT, i.e., it replaces a model with a finite number of degrees of freedom by one with infinitely many ones. The model thus obtained was called  {\em 2D Luttinger model}  in \cite{EL1} since, first, it is related to 2D lattice fermions in a similar way as the 1D Luttinger model is related to a corresponding 1D lattice fermion system and, second, in this model the nodal fermion degrees of freedom can be treated using bosonization methods. In \cite{dWL1} we proposed to treat the antinodal fermions in the 2D Luttinger model by mean field theory, and we presented mean field results showing that there is a significant parameter regime where the antinodal fermions are gapped. Moreover, we showed that the latter result is robust, i.e., insensitive to changes in $\kappa$ and $Q$. We also proposed that, in this partially gapped phase, the low energy properties of the 2D Luttinger model is described (essentially\footnote{With $\gamma_1=\gamma_2$ and more general cutoff functions; see Remark~\ref{Relation Mattis model lattice fermions}.}) by the Mattis model. 
 
We note that all approximations above either only affect fermion degrees of freedom that are far away (in energy) from an assumed underlying Fermi surface (containing the points in \Ref{Qrs}), or they amount to changing the Hamiltonian by terms that are formally sub-leading in the lattice constant $a$. Physical arguments suggest that such changes do not (much) affect the low energy properties of the model. However, the validity of these arguments is far from obvious. It therefore is  important  to note that the results in the present paper provide a starting point to investigate the validity of the approximations in greater depth. In Section~\ref{Final remarks}, Remark~4 we explain this for the approximation in \Ref{Approx 1}. This is only one example, and the other approximations above certainly also deserve attention. In particular, one might be concerned about the approximation in \Ref{Approx 2}: this amounts to ignoring, in particular,  the fermion interaction terms associated with the interaction vertex
\begin{equation}
\label{corr} 
\begin{split} 
\Bigl(\hat{u}(\vQ_{r_1,s_1} -\vQ_{r_2,s_2} + \vk_1-\vk_2)- \hat{u}(\vQ_{r_1,s_1}-\vQ_{r_2,s_2})\Bigr)\qquad \\ \times   \left(\frac{L}{2\pi}\right)^2  \delta_{\vk_1-\vk_2+\vk_3-\vk_4,\vzero}  \sum_{\vn\in\Z^2}
\delta_{\vQ_{r_1,s_1}-\vQ_{r_2,s_2}+\vQ_{r_3,s_3}-\vQ_{r_4,s_4},2\pi\vn/a}   
\end{split} 
\end{equation} 
for $r_j,s_j=\pm$. While these terms are formally sub-leading in $a$, there are certainly scattering processes with momenta $\vk_1,\vk_2$ close to the underlying Fermi surface but where $|\vk_1-\vk_1|$ is of order $\pi/\tilde{a}$.  Furthermore, the approximation in \Ref{Approx 5} amounts to adding umklapp processes that are not present in the lattice model \cite{EL1}, and the role of these should be further investigated.

\newsection{Boson-fermion correspondence} 
\label{Bosonization identities}
In this appendix we summarize well-known mathematical results on constructive bosonization in one dimension, and we show how to obtain from these the results in Section~\ref{Bosonization}. 

\subsection{Bosonization in 1D} 
\label{Bosonization in 1D}
In this appendix, we write $r,r'=\pm$ for chirality indices, $A,A'\in\mathcal{I}$ for flavor indices with $\mathcal{I}$ some index set to be specified later, and $k,k'\in(2\pi/L)(\Z+1/2)$ for 1D Fourier modes. We consider fermion operators $c^{(\dag)}_{r,A}(k)$ defined on a fermion Fock space $\cF$ with normalized vacuum state $\Omega$, satisfying the usual relations of 1D relativistic fermions: 
\begin{equation}
\label{car} 
\{ c\pdag_{r,A}(k),c^\dag_{r',A'}(k')\} = \delta_{r,r'}\delta_{A,A'}\delta_{k,k'},\quad \{ c\pdag_{r,A}(k),c\pdag_{r',A'}(k')\}=0  
\end{equation}
and
\begin{equation}
\label{highest weight condition} 
c\pdag_{r,A}(k)\Omega = c^\dag_{r,A}(-k)\Omega  =0\quad \forall k\;\mbox{ such that }\; rk>0. 
\end{equation}
We also introduce so-called densities 
\begin{equation}
\label{hatj}
\hat\jmath_{r,A}(p)\define \sum_{k\in\frac{2\pi}{L}(\Z+\frac12)}
:\! c^\dag_{r,A}(k-p)c\pdag_{r,A}(k)\!:
\end{equation}
for $p\in(2\pi/L)\Z$; here, and in the following, the colons indicate normal ordering as in \Ref{NormalOrder}.
 
\begin{proposition}
\label{Proposition Bosonization 1D}
{\bf (a)} The densities in \Ref{hatj} are well-defined operators on $\cF$ and satisfy the commutator relations
\begin{equation*}
[\hat\jmath_{r,A}(p),\hat\jmath_{r',A'}(p')]=r\delta_{r,r'}\delta_{A,A'}\frac{Lp}{2\pi}\delta_{p+p',0}. 
\end{equation*} 
Moreover, $\hat\jmath_{r,A}(p)^\dag=\hat\jmath_{r,A}(-p)$ and
\begin{equation*}
 \hat\jmath_{r,A}(p)\Omega=0\quad \quad \forall p\;\mbox{ such that }\; rp\geq 0. 
\end{equation*} 
\noindent {\bf (b)} The following operator identity holds true 
\begin{equation*}
\sum_{k\in\frac{2\pi}L(\Z+\frac12)} r k:\! c^\dag_{r,A}(k)c\pdag_{r,A}(k)\!:\; = \frac{\pi}L\sum_{p\in\frac{2\pi}L\Z} :\!\hat\jmath_{r,A}(-p)\hat\jmath_{r,A}(p)\!:  
\end{equation*}
where both sides are well-defined operators on $\cF$. 

\noindent {\bf (c)} There exist unitary operators $R_{r,A}$ on $\cF$ satisfying the following relations, 
\begin{equation}
\label{R-relation 1}
[\hat\jmath_{r,A}(p),R_{r',A'}]=r\delta_{r,r'}\delta_{A,A'}\delta_{p,0}R_{r,A},\quad \{ R_{r,A}\pdag,R_{r',A'}^\dag\} = 2\delta_{r,r'}\delta_{A,A'}
\end{equation}
and
\begin{equation}
\label{R-relation 2}
\Bigl< \Omega,\prod_{r=\pm}\prod_{A\in\mathcal{I}}  R_{r,A}^{m_{r,A}}  \Omega \Bigr> = \prod_{r=\pm}\prod_{A\in\mathcal{I}}  \delta_{m_{r,A},0}  
\end{equation}
for all $m_{r,A}\in\Z$ such that $\sum_{r,A}|m_{r,A}|<\infty$.

\noindent {\bf (d)} The operators  $\hat\jmath_{r,A}(-rp)$ and $R_{r,A}$, with $p\in(2\pi/L)\N$,  $r=\pm$, and $A\in\mathcal{I}$, generate the full fermion Fock space $\cF$ from $\Omega$, i.e.\ the states 
\begin{equation}
\label{states} 
\left(\prod_{r=\pm}\prod_{A\in\mathcal{I}}\prod_{p\in\frac{2\pi}L\N}\hat\jmath_{r,A}(-rp)^{n_{r,A}(p)}\right)\left(\prod_{r=\pm}\prod_{A\in\mathcal{I}}  R_{r,A}^{m_{r,A}}\right) 
 \Omega
\end{equation}
with $m_{r,A}\in\Z$ and $n_{r,A}(p)\in\N_0$ such that $\sum_{r,A}m_{r,A}^2<\infty$ and $\sum_{p}|p|n_{r,A}(p)<\infty$, provide a complete orthogonal basis in the fermion Fock space $\cF$. 

\noindent {\bf (e)} For $\epsilon>0$ and $-L/2\leq x<L/2$, 
\begin{equation}
\label{V}
\begin{split} 
V_{r,A}(x;\epsilon)=& \frac1{\sqrt{2\pi\epsilon}}\ee^{r\pi\ii x\hat\jmath_{r,A}(0)/L}\left(R_{r,A}\right)^{-r} \ee^{r\pi\ii x\hat\jmath_{r,A}(0)/L} \\ &\times \exp\Bigl(r\frac{2\pi}L\sum_{p\in\frac{2\pi}L\Z\setminus\{ 0\}}\frac{\hat\jmath_{r,A}(p)}p\ee^{\ii px}\ee^{-\epsilon |p|/2}\Bigr)
\end{split} 
\end{equation}
define unitary operators on $\cF$. Moreover, 
\begin{equation} 
\label{c from V} 
c_{r,A}(k) = \lim_{\epsilon\to 0^+} \frac1{\sqrt{L}} \int_{-L/2}^{L/2}\ud x V_{r,A}(x;\epsilon)\ee^{-\ii kx}
\end{equation} 
where the limit is understood in the strong operator topology (see e.g.\ \cite{ReedSimon1}). 
\end{proposition} 

\noindent\textit{Proof:}  Proofs of (a)-(d) based on vertex operator algebras can be found in \cite{Kac}, Sections~5.1 and 5.2, for example (the details of the proof given there are only in the special case of a single fermion species, but the generalization to our case is trivial). An algebraic version of (e) is also stated and proved in  \cite{Kac}, Sections~5.2, but we find it convenient to use a different formulation proved in \cite{CH}, for example. \QED
\medskip

Even though the results summarized in Proposition~\ref{Proposition Bosonization 1D} are well-known, the mathematical literature in which rigorous proofs are provided is, to our opinion, not easily accessible for non-experts. We therefore give below a few remarks, where the first four and the last concern the mathematics- and the physics literature, respectively. 

\begin{remark}
Parts of Proposition~\ref{Proposition Bosonization 1D} are special cases of mathematical results stated and proved in \cite{Frenkel,CR}, for example. To be specific:\ (a) can be obtained from a wedge representation of an affine Kac-Moody algebra, (b) is a special case of the Sugawara construction,  and (c)-(e) can be deduced from standard vertex operator constructions. 
\end{remark} 

\begin{remark}
\label{Appendix Unboundedness}
Similar to what was discussed  in Remark~\ref{Remark unbounded}, issues related to the unboundedness of the operators appearing in Proposition~\ref{Proposition Bosonization 1D} can be dealt with using general mathematical results proved in \cite{GL}; see also Section~\ref{Unboundedness}. However, we will not elaborate on these results here since overly emphasizing this unboundedness issue would be misleading:  there exists an alternative rigorous approach based on vertex operator algebras that is purely algebraic \cite{Kac}.   
\end{remark} 

\begin{remark}
A delicate mathematical issue is the distributional nature of the fermion field operators, i.e., Part~(e) of Proposition~\ref{Proposition Bosonization 1D}. As discussed, our approach to this problem follows \cite{CH}: we consider regularized fermion operators, proportional to unitary operators, and which depend on a parameter $\epsilon>0$ such that, when $\epsilon\to 0^+$, they converge to the fermion field operators in a known distributional sense \cite{CH}. We find it convenient to use these regularized (unitary) operators when computing correlation functions, i.e.\ inside expectation values, since then the limit $\epsilon\to 0^+$ can be done in a fully controlled manner. This allows us to suppress some analytical details and thus simplify our presentation. Furthermore, overly emphasizing these details would be misleading since the essence of (e) is algebraic rather than analytic: as mentioned in Remark~\ref{Appendix Unboundedness}, 
there exists an alternative rigorous approach based on vertex operator algebras \cite{Kac}.   
\end{remark} 

\begin{remark}
\label{Klein factor remark} 
It is well-known that the Hilbert space obtained using just the boson operators is a proper subspace of the fermion Fock space, and to get equivalence one has to include different sectors in which the fermions are filled to different levels; see e.g.\ \cite{BenfattoGallavottiMastropietro1992}, end of Section~I, for a discussion of this point. Since fermion operators do not leave sectors invariant, while boson operators do, it is impossible to represent fermion operators in terms of bosons only. The Klein factors provide the remedy to this problem; see \cite{Kac}, Equations (5.2.1)--(5.2.4) where the Klein factor is denoted as $u$. As discussed in the introduction, in the physics literature, Klein factors can sometimes be ignored (and often are)  in 1D problems, but this is not true in higher dimensions. A constructive approach to Klein factors is discussed in \cite{CR}, Section~3.4, for example. 
\end{remark}

\begin{remark} 
A standard reference in the condensed matter literature in which the results above are derived is \cite{vonDelftSchoeller}. For the convenience of the reader, we therefore give the dictionary between our notation and that used in \cite{vonDelftSchoeller} (``$a\cong b$'' below means ``the symbol $a$ in \cite{vonDelftSchoeller} corresponds to our symbol $b$''). In the following, we suppress the irrelevant flavor index $\eta\cong A$.\footnote{Note that all fermions in \cite{vonDelftSchoeller} have chirality $r=+$ (at least in the first part of the text).}
\begin{equation*}
\delta_b\cong 1, \qquad |\vzero\rangle_0 \cong \Omega, \qquad c_{k}\cong c_{\pm}(\pm k),  \qquad \hat N\cong\hat\jmath_\pm(0),
\end{equation*}
\begin{equation}
F\cong (R_\pm)^{\mp 1}, \qquad {b(p) \cong  - \ii\sqrt {\frac{{2\pi}}{{Lp}}} \hat\jmath_{\pm} (\pm p)},\; p>0,
\end{equation}
\begin{equation*}
\psi(x) \cong \sqrt{2\pi}V_{\pm} ( \mp x;\epsilon ), \qquad a\cong \epsilon.
\end{equation*}
\end{remark} 

\subsection{Proofs}
The flavor index set in Section~\ref{Bosonization in 1D} is now identified with the set of pairs:
\begin{equation}
\label{Index set} 
\mathcal{I}\define \{ (s,x)\,:\, s=\pm,\, x\in\Lambda_{\mathrm{1D}} \}.
\end{equation} 

\noindent\textit{Proof of Proposition~\ref{Proposition Density operators}.} 
 It is easy to check that the relations in \Ref{car} and \Ref{highest weight condition} are equivalent to the operators
\begin{equation}
\label{psi from c} 
\hat\psi_{r,s}(\vk)=\frac{L}{2\pi}\sqrt{\frac{\ta}{L}}\sum_{x\in \Lambda_{\mathrm{1D}}} c_{r,s,x}(k_s)\ee^{-\ii k_{-s} x}
\end{equation} 
and $\hat\psi^\dag_{r,s}(\vk)=\hat\psi\pdag_{r,s}(\vk)^\dag$, $\vk=k_+\ve_{+}+k_{-}\ve_{-}$, satisfying the relations in \Ref{CAR} and \Ref{Highest Weight Condition}. Moreover, with this identification, the operators 
\begin{equation}
\label{J from j}
\hat{J}_{r,s}(\vp)=\sum_{x\in \Lambda_{\mathrm{1D}}}\hat\jmath_{r,s,x}(p_s)\ee^{-\ii p_{-s} x} \end{equation} 
with $\vp=p_+\ve_{+}+p_{-}\ve_{-}$, are identical with the nodal densities in \Ref{Nodal density operators}. Straightforward computations show that Parts~(a) and (b) of Proposition~\ref{Proposition Bosonization 1D}  are equivalent to Parts~(a) and (b) of
Proposition~\ref{Proposition Density operators}, respectively.\QED

\medskip

\noindent\textit{Proof of Lemma~\ref{Lemma Fock space basis}.} Using the identification in \Ref{Index set}, Part~(c) of Proposition \ref{Proposition Bosonization 1D} is obviously equivalent to Part~(a) of Lemma~\ref{Lemma Fock space basis}. Furthermore, the inverse of the relations in \Ref{J from j} is 
\begin{equation}
\label{j from J}
\hat\jmath_{r,s,x}(p_s) = \frac{\ta}{L} \sum_{p_{-s}\in\tilde\Lambda^*_{\mathrm{1D}}}\hat{J}_{r,s}(\vp)\ee^{\ii p_{-s} x} 
\end{equation} 
and allows to express any operator $\hat\jmath_{r,s,x}(-rp_s)$ with $p_s>0$ as a finite linear combination of the boson creation operators $b^\dag_s(\vp)$ in \Ref{Boson operators from nodal density operators}. Thus  Part~(d) of Proposition \ref{Proposition Bosonization 1D} implies that the states in \Ref{Boson basis} are a complete basis in $\cF$. Using the canonical commutator relations of the boson operators $b^{(\dag)}_s(\vp)$, and the relations in \Ref{R-relation 1} and \Ref{R-relation 2}, it is easy to check that the states in \Ref{Boson basis} are orthonormal, and this implies Part~(b) of Lemma~\ref{Lemma Fock space basis}. Part~(c) of this lemma is proved in a similar manner.\QED

\medskip

\noindent\textit{Proof of Proposition~\ref{Proposition Fermions from Bosons}.} Use the identification in \Ref{Index set}, insert \Ref{j from J} in \Ref{V}, and then use \Ref{c from V} in \Ref{psi from c}.\QED 

\newsection{Solution of the Mattis model: additional details}
\label{Solution appendix}
In this appendix, we give further details on how to obtain the results in Section~\ref{Solution of the Mattis model}. 

\subsection{Neutral bosons}
\label{Neutral bosons}
To fix our notation, and for easy reference, we collect some simple results on neutral bosons needed in Appendix~\ref{Diagonalization of the Mattis Hamiltonian}.

Consider the Hamiltonian
\begin{equation}
\label{h_0}
h_0 = \sum_{m=1}^N \lambda_{m}^0\bigl(b^\dag_m b\pdag_m + b^\dag_{-m}b\pdag_{-m} +1 \bigr) \qquad \left(\lambda_{m}^0>0\right)
\end{equation}
acting on a boson Fock space defined by annihilation (creation) operators $ b_m^{(\dag)}$, labeled by quantum numbers $m=\pm 1,\pm 2,\ldots,\pm N$, and a vacuum vector $\Omega_B$, with
\begin{equation}
\label{ccr for b and bdag}
[b_m\pdag,b_n\pdag]=0, \quad [b_m\pdag,b_n^\dag]=\delta_{m,n}\pdag, \quad b_m\Omega_B=0.
\end{equation}
Neutral boson operators $P_m$ and $Z_m$ are defined by
\begin{equation}
\label{PQ from bbdag}
P_m = -\ii\sqrt{\frac{\lambda_{|m|}^0}{2}} (b_m\pdag - b_{-m}^\dag), \quad Z_m = \frac{1}{\sqrt{2\lambda_{|m|}^0}} (b_m\pdag + b_{-m}^\dag)
\end{equation}
such that
\begin{equation}
\label{ccr1} 
[Z\pdag_m,P^\dag_n]=\ii\delta_{m,n},\quad [Z\pdag_m,Z\pdag_n]=[P\pdag_m,P\pdag_n]=0, \quad P_m^\dag=P\pdag_{-m},\quad Z_m^\dag=Z\pdag_{-m}.
\end{equation} 
In terms of these, the Hamiltonian in \Ref{h_0} becomes
\begin{equation}
\label{h_0 in PZ}
h_0 = \sum_{m=1}^N \Bigl(P_m^\dag P_m\pdag + (\lambda_{m}^0)^2 Z_m^\dag Z_m\pdag\Bigr).
\end{equation}

With these preliminary definitions at hand, we introduce a more general boson Hamiltonian of the form
\begin{equation}
\label{h}
h=\sum_{m,n=1}^N\Bigl( P_m^\dag A\pdag_{m,n}P\pdag_n + Z^\dag_m B\pdag_{m,n}Z\pdag_n\Bigr) + \sum_{m=1}^N\Bigl(Z^\dag_m K\pdag_m + K^\dag_m Z\pdag_m \Bigr) 
\end{equation} 
with ${\bf A}=(A_{m,n})$ and ${\bf B}=(B_{m,n})$ self-adjoint, positive definite, $N\times N$ matrices, and ${\bf K}=(K_m)$ a complex vector. It is useful to write this Hamiltonian in matrix form as follows, 
\begin{equation}
\label{General boson Hamiltonian}
h = {\bf P}^\dag{\bf A}{\bf P}+ {\bf Z}^\dag{\bf B}{\bf Z} + {\bf Z}^\dag{\bf K}+{\bf K}^\dag{\bf Z}. 
\end{equation} 
We define ${\bf C}\define {\bf A}^{1/2}{\bf B}{\bf A}^{1/2}$. Since the matrix ${\bf C}$ is self-adjoint, it can be diagonalized by a unitary matrix ${\bf U}$. We write $\vlambda^2 = {\bf U}^\dag {\bf C} {\bf U}$ with $\vlambda$ the (positive) diagonal matrix whose elements are the eigenvalues of ${\bf C}^{1/2}$. 

Our aim for the remainder of this section is to explicitly construct a unitary operator that transforms $h$ in \Ref{h} into a diagonal form similar to the one in \Ref{h_0}; see Equation \Ref{UhUdag} below. This is done in two steps stated in Lemmata~\ref{l1} and \ref{btilde to b} below.

In a first step, we construct a unitary operator $\cU'$ that transforms $h$ in \Ref{h} into a form similar to the one in \Ref{h_0 in PZ}.
\begin{lemma}
\label{l1} 
Assume that ${\bf U}=\ee^{\bf M}$ with ${\bf M}^\dag =-{\bf M}$. Then 
\begin{equation}
\label{cU} 
\cU' = \ee^{\ii ({\bf K}^\dag {\bf B}^{-1}{\bf P} + {\bf P}^\dag{\bf B}^{-1}{\bf K})}
\ee^{-\ii({\bf P}^\dag \ln({\bf A}){\bf Z} + {\bf Z}^\dag\ln({\bf A}){\bf P})/2}  
\ee^{-\ii({\bf P}^\dag {\bf M}{\bf Z} -{\bf Z}^\dag {\bf M}{\bf P})}
\end{equation}
is a unitary operator implementing a canonical transformation $({\bf Z},{\bf P})\to ({\bf \tilde{Z}},{\bf \tilde{P}})$, i.e.\ $\tilde{P}_m = \cU' P_m \cU'^\dag$ and $\tilde{Z}_m = \cU' Z_m \cU'^\dag$, and this canonical transformation is given by 
\begin{equation}
\label{PZ to tildePtildeZ}
{\bf P} = {\bf A}^{-1/2}{\bf U} {\bf \tilde{P}},\qquad {\bf Z} = {\bf A}^{1/2}{\bf U}  {\bf \tilde{Z}} - {\bf B}^{-1} {\bf K}.  
\end{equation} 
Moreover, this transformation applied to \Ref{General boson Hamiltonian} yields
\begin{equation}
\label{Diagonalized general boson Hamiltonian}
\cU'^\dag h\cU' = {\bf P}^\dag {\bf P} + {\bf Z}^\dag \vlambda^2 {\bf Z} - {\bf K}^\dag {\bf B}^{-1} {\bf K}. 
\end{equation}
\end{lemma} 

\noindent ({\em Proof:} See the end of this section.)

\medskip

The diagonalization of the Hamiltonian in \Ref{Diagonalized general boson Hamiltonian} is now straightforward. Introducing boson annihilation (creation) operators $ \tilde b_m^{(\dag)}$ through (cf. \Ref{PQ from bbdag})
\begin{equation}
\label{PQ from tilde bbdag}
P_m=-\ii\sqrt{\frac{\lambda_{|m|}}{2}}(\tilde b\pdag_m - \tilde b^\dag_{-m}),\qquad 
Z_m= \frac{1}{\sqrt{2\lambda_{|m|}}} (\tilde b\pdag_m + \tilde b^\dag_{-m}),
\end{equation}
with $\lambda_{m}$ the eigenvalues of ${\bf C}^{1/2}$ (= diagonal elements of $\vlambda$), one obtains
\begin{equation}
\label{Hamiltonian in terms of btilde}
\cU'^\dag h\cU' = \sum_{m=1}^N \lambda_{m}\pdag\bigl(\tilde b^\dag_m \tilde b\pdag_m + \tilde b^\dag_{-m}\tilde b\pdag_{-m}+ 1\bigr) - {\bf K}^\dag {\bf B}^{-1} {\bf K}.
\end{equation} 

In the second and final step, we define the unitary operator $\cU''$ that transforms between the boson operators $ b^{(\dag)}_m$ and $\tilde b^{(\dag)}_m$:
\begin{lemma}
\label{btilde to b}
Introduce the unitary operator
\begin{equation}
\cU'' = \exp\Bigl(-\sum_{m=1}^N \mu_m\pdag( b_m^\dag b_{-m}^\dag - b_{-m}\pdag b_{m}\pdag)\Bigr), \qquad \tanh(\mu_m) = \frac{\lambda_{m}\pdag -\lambda_{m}^0}{\lambda_{m}\pdag +\lambda_{m}^0},
\end{equation}
and set $\cU=\cU'\cU''$. Then
\begin{equation}
\label{UhUdag}
\cU^\dag h\cU = \sum_{m=1}^N \lambda_{m}\pdag\bigl( b^\dag_m b\pdag_m + b^\dag_{-m} b\pdag_{-m}+ 1\bigr) - {\bf K}^\dag {\bf B}^{-1} {\bf K}.
\end{equation} 
\end{lemma}

\noindent ({\em Proof:} See the end of this section.)

\medskip

Using this lemma, one can construct all exact eigenstates and corresponding eigenvalues of the Hamiltonian in \Ref{h}.

From the last two relations in \Ref{ccr1} follows that \Ref{h} can equivalently be written as
\begin{equation}
\label{h2}
h=\sum_{m,n=1}^N\Bigl( P^\dag_{-m} A\pdag_{n,m}P_{-n}\pdag + Z^\dag_{-m} B\pdag_{n,m}Z_{-n}\pdag \Bigr) + \sum_{m=1}^N\Bigl( Z^\dag_{-m} K\pdag_{-m}  + K^\dag_{-m} Z\pdag_{-m}  \Bigr)
\end{equation} 
with $K^\dag_{-m}=K\pdag_m$ etc.

As noted in Section~\ref{Diagonalization of the Mattis Hamiltonian}, the boson part of the Mattis Hamiltonian can be expressed as an average of the right-hand sides of \Ref{h} and \Ref{h2}. One can thus diagonalize the boson part by a unitary operator $\cU$, using the results above. As explained at the end of Section~\ref{Diagonalization of Mattis Hamiltonian}, this reduces correlation functions of the Mattis model to expectation values of products of simple boson operators for a system of decoupled harmonic oscillators (up to the zero-mode terms which require a separate treatment; see Section~\ref{Computation details}). 

\medskip

\noindent\textit{Proof of Lemma~\ref{l1}.} The Hamiltonian in \Ref{h} can be written as $h={\bf \tilde{P}}^\dag {\bf \tilde{P}} + {\bf \tilde{Z}}^\dag \vlambda^2{\bf \tilde{Z}} - {\bf K}^\dag {\bf B}^{-1} {\bf K}$ with
\begin{equation}
\label{tPtQ from PQ}
{\bf \tilde{P}} = {\bf U}^\dag{\bf A}^{1/2}{\bf P} , \qquad {\bf \tilde{Z}} = {\bf U}^\dag{\bf A}^{-1/2}({\bf Z} + {\bf B}^{-1}{\bf K})
\end{equation}
equivalent to \Ref{PZ to tildePtildeZ}. 

We consider the unitary operators 
\begin{equation}
\label{cU1cU2def} 
\cU_1(\valpha)\define \ee^{\ii (\valpha^\dag{\bf P} + {\bf P}^\dag\valpha)},\qquad \cU_2({\bf L})\define \ee^{{\bf P}^\dag {\bf L}{\bf Z} -{\bf Z}^\dag {\bf L}^\dag{\bf P}}
\end{equation} 
for a complex $N$-vector $\valpha=(\alpha_m)$ and a $N\times N$-matrix ${\bf L}=(L_{m,n})$. These operators implement canonical transformations as follows, 
\begin{gather}
\begin{aligned}
\label{cU1cU2} 
\cU_1(\valpha)P_m \cU_1(\valpha)^\dag &=P_m, & \qquad \cU_2({\bf L})P_m \cU_2({\bf L})^\dag &=(\ee^{-\ii{\bf L}^\dag}{\bf P})_m, \\ \cU_1(\valpha)Z_m \cU_1(\valpha)^\dag &=Z_m + \alpha_m, & \qquad \cU_2({\bf L})Z_m \cU_2({\bf L})^\dag &=(\ee^{-\ii{\bf L}}{\bf Z})_m.
\end{aligned} 
\end{gather} 
(For the bottom left relation in \Ref{cU1cU2} consider $Z_m(t)\define \cU_1(t\valpha)Z_m \cU_1(t\valpha)^\dag$ with a real parameter $t$. We compute $dZ_m(t)/dt = \cU_1(t\valpha)\ii[\valpha^\dag{\bf P} + {\bf P}^\dag\valpha,Z_m]\cU_1(t\valpha)^\dag = \alpha_m$ using \Ref{ccr1}. This together with $Z_m(0)=Z_m$ implies $Z_m(t)=Z_m + t\alpha_m$ which, for $t=1$, gives the result. The other three relations in \Ref{cU1cU2} can be proved in a similar manner.) 

By inspection, we see that the canonical transformation $({\bf Z},{\bf P})\to ({\bf \tilde{Z}},{\bf \tilde{P}})$ in \Ref{tPtQ from PQ} corresponds to a sequence of three canonical transformations implemented by the unitary operators $\cU_2(-\ii{\bf M})$, $\cU_2(-\ii\ln({\bf A})/2)$, and $\cU_1({\bf B}^{-1}{\bf K})$; see \Ref{cU1cU2def}--\Ref{cU1cU2}. Thus, the full transformation \Ref{tPtQ from PQ} is implemented by the operator $\cU'\define \cU_1({\bf B}^{-1}{\bf K}) \cU_2(-\ii\ln({\bf A})/2)\cU_2(-\ii{\bf M})$, equal to \Ref{cU}.\QED

\medskip

\noindent\textit{Proof of Lemma~\ref{btilde to b}.} Computations similar to those in the proof of Lemma~\ref{l1} give  
\begin{equation}
\label{Unitary transform of boson operators}
\cU''b_m\cU''^\dag = \cosh(\mu_{|m|})b_m\pdag + \sinh(\mu_{|m|})b_{-m}^\dag.
\end{equation}
Combining \Ref{PQ from bbdag} with \Ref{PQ from tilde bbdag}, and using the fact that
\begin{equation}
\cosh(\mu_{|m|}) = \frac{1}{2}\Bigl(\sqrt{\frac{\lambda_{|m|}}{\lambda_{|m|}^0}}+\sqrt{\frac{\lambda_{|m|}^0}{\lambda_{|m|}}}\Bigr), \qquad \sinh(\mu_{|m|}) = \frac{1}{2}\Bigl(\sqrt{\frac{\lambda_{|m|}}{\lambda_{|m|}^0}}-\sqrt{\frac{\lambda_{|m|}^0}{\lambda_{|m|}}}\Bigr),
\end{equation}
one sees that the right-hand side of \Ref{Unitary transform of boson operators} is equal to $\tilde b_m $. Equation \Ref{UhUdag} follows from \Ref{Hamiltonian in terms of btilde}.\QED

\subsubsection{Correlation functions}
We give some well-known results on (free) boson correlation functions needed in Section~\ref{Solution of the Mattis model}.
 
Let $b_m^{(\dag)}$ be boson operators satisfying \Ref{ccr for b and bdag} and $\tilde{H}_B=\sum_m\omega\pdag_mb^\dag_mb\pdag_m$; unlike before, in this subsection $\sum_m$ means the sum over $m=\pm 1,\pm 2,\ldots,\pm N$, and similarly for products $\prod_m$. Denote by $\langle A\rangle_{B,\beta}$ the usual thermal expectation value of a boson operator $A$ with respect to that free boson Hamiltonian, i.e.\ $\langle A\rangle_{B,\beta}=\mathrm{Tr}_B\bigl(\ee^{-\beta\tilde{H}_B} A\bigr)/\cZ_{B,\beta}$ with $\mathrm{Tr}_B$ the trace over the boson Fock space and
\begin{equation} 
\label{cZboson}
\cZ_{B,\beta}=\mathrm{Tr}_B\bigl(\ee^{-\beta\tilde{H}_B}\bigr) = \prod_m\frac1{1-\ee^{-\beta\omega_m}};
\end{equation} 
see e.g.\ Appendix~C in \cite{vonDelftSchoeller} for more details. 
\begin{lemma}
\label{Boson results}
\noindent {\bf (a)} For $j=1,2$, let 
\begin{equation}
J\pdag_j=J^0_j+\sum_m \bigl(J^+_{j,m}b^\dag_m+J^-_{j,m}b\pdag_m\bigr)  
\end{equation}
with complex numbers $J^0_j$ and $J^\pm_{j,m}$. Then
\begin{equation}
\label{Boson result 1} 
\langle J_1 J_2 \rangle_{B,\beta} =J^0_1J^0_2+\sum_m\Bigl( J^-_{1,m}J^+_{2,m}\frac1{1-\ee^{-\beta\omega_m}} + J^+_{1,m}J^-_{2,m}\frac1{\ee^{\beta\omega_m}-1} \Bigr). 
\end{equation}

\noindent {\bf (b)} For $j=1,2,\ldots,N$, let 
\begin{equation}
K^+_j=  \sum_m K^+_{j,m}b^\dag_m,\qquad K^-_j=\sum_m K^-_{j,m}b\pdag_m
\end{equation}
with complex numbers $K^\pm_{j,m}$. Then
\begin{equation}
\label{Boson result 2}
\begin{split} 
\bigl\langle \ee^{K^+_1+K^-_1}\ee^{K^+_2+K^-_2}\cdots\ee^{K^+_N+K^-_N} \bigr\rangle_{B,\beta} =\Biggl( \prod_{j=1}^N G_{j,j}^{1/2} \Biggr)\Biggl( \prod_{1\leq j<k\leq N} G_{j,k} \Biggr) \\ G_{j,k}=\exp{\Biggl(\sum_m\Bigl( K^-_{j,m}K^+_{k,m}\frac1{1-\ee^{-\beta\omega_m}} + K^+_{j,m}K^-_{k,m}\frac1{\ee^{\beta\omega_m}-1} \Bigr) \Biggr)}  
\end{split} .
\end{equation} 
\end{lemma}

\noindent\textit{Outline of proof.}  Part~(a) is a simple consequence of the following well-known results, $\langle b\pdag_nb^\dag_m\rangle_{B,\beta} = \delta_{n,m}/(1-\ee^{-\beta\omega_m})$ and $\langle b\pdag_nb\pdag_m\rangle_{B,\beta} = \langle b\pdag_m\rangle_{B,\beta}  =0$. 
Part~(b) can be proved using well-known identities collected and proved in \cite{vonDelftSchoeller}, Appendix~C, for example. To be specific, writing $K_j=K^+_j + K^-_j$, one finds 
\begin{equation}
\begin{split}
{\ee^{{K_1}}}{\ee^{{K_2}}}\ldots{\ee^{{K_N}}} = {\ee^{\sum_{j} {{K_j}} }}{\ee^{\frac{1}{2}\sum_{j < k} {\left[ {{K_j},{K_k}} \right]} }} = \\ {\ee^{\sum_{j} {{K_j}} }}{\ee^{\sum_{j < k} {\sum_m {( { - K_{j,m}^ + K_{k,m}^ -  + K_{j,m}^ - K_{k,m}^ + } )/2} } }}
\end{split}
\end{equation}
using Theorem~2, Equation~(C4) in \cite{vonDelftSchoeller}, and \Ref{ccr for b and bdag}, respectively. The result follows from the relation
\begin{equation}
\left\langle {{\ee^{\sum_j {{K_j}} }}} \right\rangle  = {\ee^{\sum_{j,k} {\left\langle {{K_j}{K_k}} \right\rangle/2 } }}
\end{equation}
(see Theorem~4, Equation~(C10) in \cite{vonDelftSchoeller}), applying \Ref{Boson result 1}, and by collecting terms.\QED

\begin{remark}
In the proof above we suppressed technical details concerning the unboundedness of the operators $K_j^\pm$. It is well-known how to take care of these details using Weyl operators; see e.g.\ \cite{CRW}, Proposition 2.1{\em ff}.
\end{remark} 

\subsection{Proofs}
\label{Diagonalization of the Mattis Hamiltonian}
We give the proofs of Theorem~\ref{Solution} and Proposition~\ref{Observables} stated in Section~\ref{Solution of the Mattis model}.

The Mattis Hamiltonian \Ref{Boson Hamiltonian} can be written as a sum of Fourier modes $H(\vp)$ that mutually commute. In order to diagonalize the Hamiltonian, one can either treat each mode $H(\vp)$ separately using the results in Appendix~\ref{Neutral bosons} for $N=2$ or $N=1$ (the latter for the cases $\vp=p_\pm\ve_\pm$, the former otherwise), or treat all the modes in one go. Although we have opted to present the latter approach here for conciseness, the careful reader might prefer to write out the explicit details of the former approach. Note that the matrix ${\bf B}$ in Appendix~\ref{Neutral bosons} is non-trivial only in the cases when $\gamma_2\chi(\vp)p_+p_-\neq 0$. Moreover, ${\bf K}$ is non-zero only if $\vp=p_\pm\ve_\pm$.

\medskip

\noindent\textit{Proof of Theorem~\ref{Solution}.} From \Ref{Boson operators from nodal density operators} and \Ref{hatPihatPhi} follows that 
\begin{equation}
\label{PZ from b} 
\hat{\Pi}_{s}(\vp)=-\ii\frac{L}{2\pi}\sqrt{\frac{|p_s|}{2}}\bigl( b\pdag_{s}(\vp) - b^\dag_{s}(-\vp)\bigr),\quad \hat{\Phi}_{s}(\vp)= \frac{L}{2\pi}\sqrt{\frac{1}{2|p_s|}} \bigl( b\pdag_s(\vp) + b^\dag_{s}(-\vp)\bigr)
\end{equation}
analogous to \Ref{PQ from bbdag}.

The Hamiltonian in \Ref{Boson Hamiltonian}--\Ref{KK0} is of the form $H = v_F :\!h\!:  + H^{(0)}_\zeromode$ with $h$ the average of the Hamiltonians in \Ref{h} and \Ref{h2} if we identify $m$ with $(s,\vp)$, $s=\pm$, $\vp\in\hat\Lambda^*_s$, and set
\begin{equation}
\label{identification}
P_{s,\vp} = \frac{2\pi}L\hat\Pi_s(\vp),\qquad Z_{s,\vp} = \frac{2\pi}L \hat\Phi_s(\vp) ,\qquad K_{s,\vp} = \frac{2\pi}L\hat\Xi_s(\vp)
\end{equation} 
with $-m$ corresponding to $(s,-\vp)$. The matrices ${\bf A}$ and ${\bf B}$ have block diagonal form, i.e.\ $A_{s,\vp,s',\vp'}=A_{s,s'}(\vp)\delta_{\vp,\vp'}$ and similarly for ${\bf B}$, with
\begin{equation}
\label{AB}
A_{s,s'}(\vp)= ( 1-\gamma_1\chi(\vp) )\delta_{s,s'},\quad B_{s,s'}(\vp) = \bigl( (1+\chi(\vp)\gamma_1)\delta_{s,s'}+\chi(\vp)\gamma_2\delta_{s,-s'}\bigr)p_sp_{s'}.
\end{equation}
It follows that the matrix elements of ${\bf U}$ in Lemma~\ref{l1} also have the form $U_{s,s'}(\vp)\delta_{\vp,\vp'}$, and similarly for ${\bf M}$. We note that ${\bf A}$ and ${\bf B}$ commute, and thus ${\bf C} = {\bf A}{\bf B}$ (see Section~\ref{Neutral bosons}). When $\gamma_2\chi(\vp)p_+p_-\neq 0$, the diagonalization of ${\bf C}$ reduces to that of $2\times 2$ matrices. When $\gamma_2\chi(\vp)p_+p_-=0$, the block $B_{s,s'}(\vp)$ of the matrix ${\bf B}$ has only diagonal components $s=s'=\pm$, as is the case for all other matrices ${\bf A}$, ${\bf C}$, ${\bf U}$, and ${\bf M}$. Furthermore, we can treat ${\bf K}$ as a complex vector since its components commute with those of ${\bf P}$ and ${\bf Z}$. With the identification in \Ref{identification}, the commutator relations in \Ref{CCR} are equivalent to the ones in \Ref{ccr1}. 

By straightforward computations we find that the eigenvalues $\lambda_{s,\vp}$ of ${\bf C}^{1/2}=({\bf A}{\bf B})^{1/2}$, multiplied by $v_F$, are equal to $\omega_s(\vp)$ in \Ref{oms}, and the result for $U_{s,s'}(\vp)$ is given in \Ref{U}. Moreover, 
\begin{equation}
\label{BinvK}
({\bf B}^{-1}{\bf K})_{s,\vp} = \frac{2\pi}{L}\frac{\hat\Xi_s(\vp)}{(1+\gamma_1\chi(\vp))p_s^2}= -\frac{2\pi}L\frac{\gamma_2}{1+\gamma_1}\frac{\ii}{p_s}\chi(\vp)\hat{Q}_{+,-s}(p_s)\delta_{p_{-s},0}
\end{equation} 
for $\vp\in\hat\Lambda^*_s$, and
\begin{equation}
\label{KBinvK}
{\bf K}^\dag{\bf B}^{-1}{\bf K} = \frac{\gamma_2^2}{1+\gamma_1}\frac12\sum_{s=\pm}\sum_{p \in {\frac{2\pi}{L}\mathbb{Z}\setminus \{0\} }}
\Bigl(\tPiL\Bigr)^2 \chi(p\ve_{-s})\hat{Q}_{+,s}(-p) \hat{Q}_{+,s}(p) ,
\end{equation} 
and for the unitary operator in \Ref{cU} we obtain (see also Footnote~\ref{Footnote Addition of zeros} on page~\pageref{Footnote Addition of zeros})  
\begin{equation}
\label{cU1}
\begin{split} 
\cU' =& \exp\Biggl(\ii\sum_{s=\pm}\sum_{\vp\in\hat\Lambda^*_s}\Bigl(\tPiL\Bigr)^2\bigl(\hat\Xi^\dag_{s}(\vp)\hat\Pi\pdag_{s}(\vp)+\hat\Pi^\dag_{s}(\vp)\hat\Xi\pdag_{s}(\vp)\bigr)/[2(1+\gamma_1)p_s^2]\Biggr)\\ & \times\exp\Biggl(-\ii\sum_{s=\pm}\sum_{\vp\in\hat\Lambda^*_s}\Bigl(\tPiL\Bigr)^2\ln\bigl( 1-\gamma_1\chi(\vp)\bigr)\bigl(\hat\Pi^\dag_s(\vp)\hat\Phi_s\pdag(\vp)+\hat\Phi^\dag_s(\vp)\hat\Pi_s\pdag(\vp)\bigr)/4\Biggr) \\ & \times\exp\Biggl(-\ii\sum_{s=\pm}\sum_{\vp\in\hat\Lambda^*_s}\Bigl(\tPiL\Bigr)^2M_{s,-s}(\vp)\bigl(\hat\Pi^\dag_{s}(\vp)\hat\Phi\pdag_{-s}(\vp)-\hat\Phi^\dag_{s}(\vp)\hat\Pi\pdag_{-s}(\vp)\bigr)/2\Biggr)
\end{split} 
\end{equation} 
with
\begin{equation} 
M_{s,s'}(\vp)=-s\delta_{s,-s'}\arctan(U_{-+}(\vp)/U_{++}(\vp))
\end{equation}
and $U_{s,s'}(\vp)$ in \Ref{U}. Thus \Ref{Diagonalized general boson Hamiltonian} and \Ref{Hamiltonian in terms of btilde} give 
\begin{equation}
\label{UhUdag1}
\cU'^\dag\!:\!h\!:\!\cU' = \sum_{s=\pm}\sum_{\vp\in\hat\Lambda^*_s} \Bigl(\frac{\omega_{s}(\vp)}{v_F}[\tilde b^\dag_s(\vp)\tilde b\pdag_s(\vp)+1/2]-|p_s|/2\Bigr)  - {\bf K}^\dag {\bf B}^{-1} {\bf K}
\end{equation}
where we write $\tilde b^{(\dag)}_s(\vp)\define \tilde b^{(\dag)}_{s,\vp}$; we used $\omega_{s}(\vp)=\omega_{s}(-\vp)$ and 
\begin{equation}
\label{PQ from btilde} 
\begin{split}
\hat{\Pi}_{s}(\vp)=&-\ii\frac{L}{2\pi}\sqrt{\frac{\omega_{s}(\vp)}{2v_F}}\bigl(\tilde b\pdag_{s}(\vp) - \tilde b^\dag_{s}(-\vp)\bigr) \\ \hat{\Phi}_{s}(\vp)=& \frac{L}{2\pi}\sqrt{\frac{v_F}{2\omega_{s}(\vp)}} \bigl( \tilde b\pdag_s(\vp) + \tilde b^\dag_{s}(-\vp)\bigr)
\end{split}
\end{equation}
implied by \Ref{PQ from tilde bbdag}. The additional constants $-|p_s|/2$ in \Ref{UhUdag1} are due to the normal ordering. 

Following Lemma~\ref{btilde to b}, we set
\begin{equation}
\label{cU2}
\cU'' = \exp\Biggl(\ii\sum_{s=\pm}\sum_{\vp\in\hat\Lambda^*_s}\Bigl(\tPiL\Bigr)^2 \mu_s(\vp) \bigl(\hat\Pi^\dag_{s}(\vp)\hat\Phi\pdag_{s}(\vp)+\hat\Phi^\dag_{s}(\vp)\hat\Pi\pdag_{s}(\vp)\bigr)/2\Biggr)
\end{equation}
with
\begin{equation}
\tanh(\mu_s(\vp)) = \frac{\omega_s(\vp)-v_F|p_s|}{\omega_s(\vp)+v_F|p_s|}
\end{equation}
and where we have used \Ref{PZ from b}. Finally, we define the unitary operator that diagonalizes the Mattis Hamiltonian as the product of \Ref{cU1} and \Ref{cU2}, 
\begin{equation}
\label{cU full}
\cU=\cU'\cU'', 
\end{equation}
such that
\begin{equation}
\label{UhUdag2}
\cU^\dag\!:\!h\!:\!\cU = \sum_{s=\pm}\sum_{\vp\in\hat\Lambda^*_s} \Bigl(\frac{\omega_{s}(\vp)}{v_F}[b^\dag_s(\vp)b\pdag_s(\vp)+1/2]-|p_s|/2\Bigr)  - {\bf K}^\dag {\bf B}^{-1} {\bf K}.
\end{equation}
Since $H = v_F :\!h\!: + H^{(0)}_\zeromode$, \Ref{UhUdag2} implies \Ref{H2}{\em ff} with $\tilde H_\zeromode=H^{(0)}_\zeromode-v_F{\bf K}^\dag{\bf B}^{-1}{\bf K}$ and $\cE_0$ in \Ref{Ground state energy}. Using \Ref{KK0} and \Ref{KBinvK}, together with the Parseval relation \Ref{Parseval}, \Ref{Charges and chiral charges}, and $\hat{Q}_{\pm,s}(0)=\sqrt{\ta/(2\pi)}\sum_{x\in\Lambda_{1\rm D}}Q_{\pm,s}(x)$, one obtains \Ref{KK}.\QED 

\medskip

\noindent\textit{Proof of Proposition~\ref{Observables}.} Since $N_{r,s}(x)$ obviously commutes with $\cU$ in \Ref{cU full} and $\tilde{H}$ in \Ref{H2}, \Ref{UNUdag} is trivial. 

For the Klein factor, note that $\cU'^\dag R_{r,s}(x)\cU'=\ee^{-\ii D}R_{r,s}(x)\ee^{\ii D}$ (recall $\cU'$ in \Ref{cU1}) with 
\begin{equation}
D =\sum_{s=\pm}\sum_{\vp\in\hat\Lambda^*_s}\Bigl(\tPiL\Bigr)^2\frac1{2(1+\gamma_1)p_s^2}\bigl(\hat\Xi^\dag_{s}(\vp)\hat\Pi\pdag_{s}(\vp)+\hat\Pi^\dag_{s}(\vp)\hat\Xi\pdag_{s}(\vp)\bigr) 
\end{equation}
and $\hat\Xi_s(\vp)$ in \Ref{Xi}. To compute this we define $f(z)=\ee^{-\ii zD}R_{r,s}(x)\ee^{\ii z D}$, $z\in\C$, and find $df(z)/dz$ using $[\hat{Q}_{+,s}(p),R_{r,s}(x)]=r\sqrt{\ta/\pi}R_{r,s}(x)\ee^{-\ii px}/2$ (the latter follows from \Ref{hatNrs}, \Ref{Zero-mode operator}, \Ref{Klein factor commutator relations}, and \Ref{Charges and chiral charges}). The resulting differential equation for $f(z)$ can be solved with the condition $f(0)=R_{r,s}(x)$, and setting $z=1$ gives 
\begin{equation}
\cU'^\dag R_{r,s}(x)\cU' = R_{r,s}(x)\exp\Bigl(-r\frac{\gamma_2}{2(1+\gamma_1)}\sqrt{\frac{\ta}{\pi}} \sum_{p\in\tilde\Lambda^*_{1\rm D}}\Bigl(\tPiL\Bigr)^2 \frac1{p}\hat\Pi^\dag_{-s}(p\ve_{-s})\ee^{-\ii px}\Bigr).
\end{equation} 
Using \Ref{PZ from b} and \Ref{PQ from btilde}, one sees that
\begin{equation}
\label{U'' on PZ}
\cU''^\dag  \hat \Pi_s(\vp)\cU'' = \sqrt {\frac{{\omega_s(\vp)}}{{v_F \left| {p_s}\right|}}} \hat\Pi _s (\vp), \qquad \cU''^\dag  \hat \Phi_s(\vp)\cU'' = \sqrt {\frac{{v_F \left| {p_s}\right|}}{{\omega_s(\vp)}}} \hat\Phi _s (\vp),
\end{equation}
from which $\cU^\dag R_{r,s}(x)\cU$ then follows. In a similar manner we compute
\begin{equation}
\ee^{\ii \tilde{H} t}R_{r,s}(x) \ee^{-\ii \tilde{H} t} = \ee^{\ii k_{r,s}(x) t/2} R_{r,s}(x)\ee^{\ii k_{r,s}(x) t/2}
\end{equation} 
with $k_{r,s}(x)$ in \Ref{krs}. We recall \Ref{PZ from b} and obtain \Ref{URUdag}--\Ref{tkrs} by straightforward computations. 

Lemma~\ref{l1}, \Ref{identification}, and \Ref{BinvK} imply  
\begin{equation}
\label{UPiPhiUdag} 
\begin{split} 
\cU'^\dag\hat\Pi_s(\vp)\cU' =& (1-\gamma_1\chi(\vp))^{-1/2}\sum_{s'=\pm} U_{s,s'}(\vp)\hat\Pi_{s'}(\vp)\\ \cU'^\dag\hat\Phi_s(\vp)\cU' =& (1-\gamma_1\chi(\vp))^{1/2}\sum_{s'=\pm} U_{s,s'}(\vp)\hat\Phi_{s'}(\vp) \\&+ \frac{\gamma_2}{1+\gamma_1}\chi(\vp)\frac{\ii}{p_s}\hat{Q}_{+,-s}(p_s)\delta_{p_{-s},0}
\end{split} 
\end{equation}
for $s=\pm$. Using \Ref{U'' on PZ}, $\hat{J}_{r,s}(\vp)=\sqrt{\pi/\ta}\bigl(\ii p_s\hat\Phi_s(\vp)-r\hat\Pi_s(\vp)\bigr)$ for $p_s\neq 0$ (this follows from \Ref{hatPihatPhi}), \Ref{UNUdag}, \Ref{U}, and 
\begin{equation}
\ee^{\ii\tilde H t} b\pdag_s(\vp) \ee^{-\ii\tilde H t} = b\pdag_s(\vp)\ee^{-\ii \omega_s(\vp) t}, 
\end{equation}
we obtain the result in \Ref{Jrs from b}--\Ref{Mrssp}.\QED

\subsection{Zero modes}
\label{Computation details} 
We show that the contribution from zero modes to the free energy density and any correlation function is irrelevant in the IR limit. Much of what is stated in this section has its origin in the following result.

\begin{lemma}
\label{Lemma Bound on generating function}
Let ${\bf H}={\bf H}^T$ be a real positive definite $M\times M$ matrix, $\vm\in\mathbb{R}^M$, and\footnote{Note that $Z$ is (a variant of) the Riemann Theta function, albeit using an unconventional notation.}
\begin{equation}
\label{Riemann Theta function}
Z=\sum\limits_{\vnu \in\mathbb{Z}^M } {{\ee^{ - \vnu^T{\bf H}\vnu + \ii\vm^T\vnu}}}.
\end{equation}
Assume
\begin{equation}
\label{Inverse eigenvalue}
\lambda\define\min\limits_{\vx\in\mathbb{R}^M}\frac{\vx^T{\bf H}^{-1}\vx}{|\vx|^2}>\frac{2}{\pi^2}, \qquad \mu\define\max\limits_{1\leq k\leq M}\left|\left({\bf H}^{-1}\vm \right)_k\right| \leq 2\pi \lambda.
\end{equation}
Then
\begin{equation}
\label{Bound on generating function}
1 \leq \frac{Z}{J} \leq 1 + O\left(M\lambda^2\ee^{-\lambda\pi^2}\right)
\end{equation}
with
\begin{equation}
\label{Gaussian integral}
J = \int\limits_{\mathbb{R}^M } {\ee^{ - \vx^T{\bf H}\vx + \ii\vm^T\vx}\ud^M x}  =\sqrt {\frac{{\pi ^M }}{{\det ({\bf H})}}} \ee^{-\frac{1}{4}\vm^T {\bf H}^{ - 1} \vm}.
\end{equation}
\end{lemma}
Note that ${\bf H}^{-1}$ exists since ${\bf H}$ is positive definite, and that $1/\lambda$ is equal to the largest eigenvalue of ${\bf H}$. Furthermore, the exponential damping in \Ref{Bound on generating function} shows that \Ref{Riemann Theta function} can be estimated, to leading order in $\lambda$, by replacing the sum by a Gaussian integral.

\medskip

\noindent\textit{Outline of proof.} Using the Poisson summation formula and \Ref{Gaussian integral}, the sum \Ref{Riemann Theta function} can be rewritten as
\begin{equation}
Z = \sqrt {\frac{{\pi ^M }}{{\det ({\bf H})}}} \sum\limits_{\vn \in \mathbb{Z}^M } \ee^{ - \frac{1}{4}\left( {\vm + 2\pi \vn} \right)^T {\bf H}^{ - 1} \left( {\vm + 2\pi \vn} \right)} = J
\sum\limits_{\vn \in \mathbb{Z}^M} \ee^{-{\pi^2}{\vn^T}{{\bf H}^{-1}}\vn - \pi {\vm^T}{{\bf H}^{ - 1}}\vn}.
\end{equation}
The last sum is bounded by
\begin{equation}
1 \leq \sum\limits_{\vn \in \mathbb{Z}^M} \ee^{-{\pi^2}{\vn^T}{{\bf H}^{-1}}\vn - \pi {\vm^T}{{\bf H}^{ - 1}}\vn} \leq {\left({\sum\limits_{n \in \mathbb{Z}} {{\ee^{ -\lambda\pi^2 {n^2} + \mu\pi\left|n\right|}}} } \right)^M}
\end{equation}
with $\lambda$ and $\mu$ in \Ref{Inverse eigenvalue}. For $\mu\leq 2\pi \lambda$, it holds that
\begin{equation}
\sum\limits_{n \in \mathbb{Z}} {\ee^{ - \lambda \pi ^2 n^2  + \mu \pi \left| n \right|} }  \leq 1 + 2\left( {\ee^{ - \lambda \pi ^2  + \mu \pi }  + \ee^{\mu ^2 /(4\lambda )} \sum\limits_{n = 1}^\infty  {\ee^{ - \lambda \pi ^2 n^2 } } } \right)
\end{equation}
and, if $\lambda>2/\pi^2$, one finds (see for example \cite{Kahn2004}, p.153-156)
\begin{equation}
\sum\limits_{n = 1}^\infty  {{\ee^{ - \lambda\pi^2 {n^2}}}}  = O\left( {\lambda ^2}\ee^{ - \lambda\pi^2 } \right)
\end{equation}
from which \Ref{Bound on generating function} straightforwardly follows.\QED

\subsubsection{Free energy density}
\label{Free energy appendix}
We derive the right-hand side of \Ref{OmQ}. The zero-mode contribution to the partition function is (cf. \Ref{TrBTrQ})
\begin{equation}
\label{Zero mode partition function}
\cZ_{\zeromode,\beta} = \sum_{\vnu}\langle\eta_{\vzero,\vnu},\ee^{-\beta \tilde{H}_\zeromode}\eta_{\vzero,\vnu}\rangle
\end{equation}
with $\tilde{H}_\zeromode$ in \Ref{KK}. This can be written in the form \Ref{Riemann Theta function} with $\vm = \vzero$ and $M=4L/\ta$. From \Ref{KK}, one sees that $\lambda=O(L)$, and it is immediately clear from Lemma~\ref{Lemma Bound on generating function} that the leading order contribution to \Ref{Zero mode partition function} is obtained by the corresponding Gaussian integral. A straightforward calculation yields (using \Ref{Bound on generating function})
\begin{equation}
\cZ_{\zeromode,\beta }  = \sqrt A \left( {\frac{L}{{\beta \tilde v_F \sqrt A }}} \right)^{2L/\ta} \left(1 + O\left(L^3\ee^{-cL}\right)\right)
\end{equation}
with $A$ and $\tilde v_F$ in \Ref{AvF}, and $c>0$ some constant independent of $L$. This also implies that the IR limit of the zero-mode contribution to the free energy density vanishes
\begin{equation}
\label{IR limit of zero-mode free energy}
\frac{\Omega_{\zeromode,\beta}}{L^2} = -\frac{1}{\beta L^2}\ln(\cZ_{\zeromode,\beta}) = O\left(L^{-1}\ln L\right).
\end{equation}

\subsubsection{Correlation functions}
\label{Zero-mode correlation functions appendix}
The irrelevance of contributions from zero modes to thermal expectation values in the IR limit is a simple consequence of the following:
\begin{lemma}
Let $m_{r,s}(x)\in\mathbb{R}$, with $r,s=\pm$ and $x\in\Lambda_{1D}$. Then
\begin{equation}
\label{Generating function for zero-mode correlators}
\begin{split}
&\Bigl\langle {{\ee^{\frac{\ii}{L}\sum\limits_{r,s =  \pm } {\sum\limits_{x \in {\Lambda _{1D}}} {{m_{r,s}}(x){N_{r,s}}(x)} } }}} \Bigr\rangle_{\zeromode,\beta}  =
\\ &\ee^{ - \frac{1}{{8\pi \beta v_F L}}\sum\limits_{r,r' =  \pm } {\sum\limits_{s =  \pm } {\sum\limits_{x \in \Lambda _{1D} } {\left( {\frac{1}{{A\left( {1 + \gamma _1 } \right)}} + \frac{{rr'}}{{1 - \gamma _1 }}} \right)m_{r,s} (x)m_{r',s} (x)} } } }\\ &\times\ee^{\frac{1}{{8\pi \beta v_F L}}\frac{{\tilde a}}{L}\sum\limits_{r,r' =  \pm } {\sum\limits_{s =  \pm } {\sum\limits_{x,x' \in \Lambda _{1D} } {\frac{{\gamma _2 }}{{A\left( {1 + \gamma _1 } \right)^2 }}m_{r,s} (x)m_{r, - s} (x')} } } }
\left(1 + O\left(L^3\ee^{-cL}\right)\right)
\end{split} 
\end{equation}
with $c>0$ some constant.
\end{lemma}

\noindent\textit{Proof.}
By definition,
\begin{equation}
\Bigl\langle {{\ee^{\frac{\ii}{L}\sum\limits_{r,s =  \pm } {\sum\limits_{x \in {\Lambda _{1D}}} {{m_{r,s}}(x){N_{r,s}}(x)} } }}} \Bigr\rangle_{\zeromode,\beta}  = \frac{1}{\cZ_{\zeromode,\beta}}\sum_{\vnu} \ee^{- \beta {{\tilde H}_\zeromode} +\frac{\ii}{L}\sum\limits_{r,s =\pm} {\sum\limits_{x \in {\Lambda_{1D}}} {m_{r,s}(x){\nu_{r,s}}(x)} } }.
\end{equation}
As in Section~\ref{Free energy appendix}, we can apply Lemma~\ref{Lemma Bound on generating function}. The result follows by replacing sums by Gaussian integrals.\QED

\medskip

All $N$-point correlation functions involving zero modes can be generated from \Ref{Generating function for zero-mode correlators} by differentiation with respect to the $m_{r,s}(x)$. We have, for instance, the important special case
\begin{equation}
\label{NN}
\begin{split}
\frac{1}{{L^2 }}\left\langle {N_{r_1 ,s_1 } (x_1 )N_{r_2 ,s_2 } (x_2 )} \right\rangle_{\zeromode,\beta}  =&
\frac{1}{{4\pi \beta v_F L}}\left( {\left( {\frac{1}{{A\left( {1 + \gamma _1 } \right)}} + \frac{{r_1 r_2 }}{{1 - \gamma _1 }}} \right)\delta _{s_1 ,s_2 } \delta _{x_1 ,x_2 }  } \right. 
\\ &\left.{ -\frac{{\ta}}{L}\frac{{\gamma _2 }}{{A\left( {1 + \gamma _1 } \right)^2 }}\delta _{ - s_1 ,s_2 } } \right) \left(1 + O\left(L^3\ee^{-cL}\right)\right)
\end{split}
\end{equation}
whose IR limit clearly vanishes. We also note from the above lemma
\begin{equation}
\label{IR limit of generating function}
\lim_{L\to\infty} \Bigl\langle {{\ee^{\frac{\ii}{L}\sum\limits_{r,s =  \pm } {\sum\limits_{x \in {\Lambda _{1D}}} {{m_{r,s}}(x){N_{r,s}}(x)} } }}} \Bigr\rangle_{\zeromode,\beta} = 1
\end{equation}
if all but a few of the $m_{r,s}(x)$ are non-zero (i.e.\ the number of non-zero terms is $O(L^0)$).

\subsubsection{Computation details}
We end this section by verifying the claims made regarding zero-mode contributions in connection with Results~\ref{density correlation result} and \ref{fermion correlation result}, respectively.

\medskip

\noindent\textit{On Result~\ref{density correlation result}.} From the scaling with $L$ in \Ref{tildeJrs} and comparing with \Ref{NN}, it is clear that $\left\langle {\tilde{J}^0_{r_1,s_1}(\vx_1,t_1;\epsilon)\tilde{J}^0_{r_2,s_2}(\vx_2,t_2;\epsilon)} \right\rangle_{\zeromode,\beta}=O(L^{-1})$. Thus, the zero modes do not contribute to the density-density correlation functions in the IR limit. The same holds true for higher-order density correlation functions, as seen by successive differentiation of the generating function \Ref{Generating function for zero-mode correlators}.\QED

\medskip

\noindent\textit{On Result~\ref{fermion correlation result}.} We want to compute the zero-mode expectation value \Ref{Q expectation value} for the operators $S^q_{r,s}(\vx,t;\epsilon)$ in \Ref{tpsi}. One can move all terms of the form $\ee^{-\ii qK^{0}_{r,s}(\vx,t;\epsilon)/2}$ to the right by successively applying 
\begin{equation}
\label{B part}
{\ee^{-\ii qK^{0}_{r,s}(\vx,t;\epsilon)/2}R_{r',s'}(y)^{r'q'}=R_{r',s'}(y)^{r'q'}\ee^{-\ii qK^{0}_{r,s}(\vx,t;\epsilon)/2}\ee^{\ii qq'\pi \ell_{r,s,r',s'}(\vx,y,t;\epsilon)/L}}
\end{equation} 
where
\begin{equation}
\begin{split} 
\ell_{r,s,r',s'}(\vx,y,t;\epsilon)=\frac{v_F t}2\Bigl( \bigl[(1+\gamma_1)A + rr'(1-\gamma_1)\bigr]\delta_{x_{-s},y}\delta_{s,s'} +\frac{\ta}{L}\gamma_2\\\times \bigl[\frac{\gamma_2}{1+\gamma_1}\delta_{s,s'}+\delta_{s,-s'}\bigr]  \Bigr) -rx_s\delta_{r,r'}\delta_{s,s'}\delta_{x_{-s},y} + r\delta_{s,-s'}f(x_s-y;\epsilon)
\end{split} 
\end{equation}
(this follows from \Ref{Krs} and the commutation relations in \Ref{Klein factor commutator relations}). Then \Ref{Q expectation value} becomes
\begin{equation}
\begin{split} 
\label{Q part 2} 
&\langle\Omega,R_{r_1,s_1}(x_{1,-s_1})^{q_1r_1}\cdots R_{r_N,s_N}(x_{N,s_{-N}})^{q_Nr_N}\Omega\rangle \bigl\langle\ee^{-\ii\sum_{j=1}^Nq_j K^0_{r_j,s_j}(\vx_j,t_j;\epsilon)}\bigr\rangle_{\zeromode,\beta} \\& \times\ee^{\ii\frac{\pi}L\sum_{j=1}^N\ell_{r_j,s_j,r_j,s_j}(\vx_j,x_{j,-s_j},t_j;\epsilon) + 2\ii\frac{\pi}L\sum_{1\leq j<k\leq N}q_jq_k\ell_{r_j,s_j,r_k,s_k}(\vx_j,x_{k,-s_k},t_j;\epsilon)}.
\end{split} 
\end{equation} 
The second factor becomes $1$ in the IR limit (cf. \Ref{IR limit of generating function} and \Ref{Krs}). The same holds true for the third factor, leaving us with \Ref{RRR}.\QED

\newsection{Quantum field theory limit: additional details}
\label{Appendix QFT limit} 
\subsection{Free energy}
\label{QFT limit of free energy appendix}
We outline the computations leading to Result~\ref{Result free energy 2}. The zero modes do not contribute in the IR limit by \Ref{IR limit of zero-mode free energy}, and it suffices to only consider the boson part given in \Ref{OmB}. We separate the latter into terms with quantitatively different momentum dependence for the dispersion in \Ref{oms}, writing $\Omega_{B,\beta} = \Omega_{B,\beta}^< + \Omega_{B,\beta}^>$ with (in the IR limit)
\begin{equation}
\label{Division of Omega}
\begin{split}
\lim\limits_{L\to\infty}L^{-2}\Omega_{B,\beta}^< &= P.V.\int\limits_{-\pi/\ta}^{\pi/\ta}\int\limits_{-\pi/\ta}^{\pi/\ta}\frac{\ud^2p}{(2\pi)^2} \sum_{s=\pm}\frac1\beta \ln\bigl(1-\ee^{-\beta\omega_s(\vp)}\bigr),\\ \lim\limits_{L\to\infty}L^{-2}\Omega_{B,\beta}^> &= 4\frac{1}{\ta}\int\limits_{\pi/\ta}^{\infty}\frac{\ud p}{2\pi}\frac1\beta \ln\bigl(1-\ee^{-\beta v_Fp}\bigr),
\end{split}
\end{equation}
where the Cauchy principal value for the improper double integral is defined as
\begin{equation}
\label{Cauchy principal value}
P.V.\int\limits_{-\pi/\ta}^{\pi/\ta}\ud p_\pm\left(\cdots\right)\define\lim\limits_{L\to\infty}\left[\int\limits_{-\pi/\ta}^{-2\pi/L}\ud p_\pm\left(\cdots\right) +\int\limits_{2\pi/L}^{\pi/\ta}\ud p_\pm\left(\cdots\right) \right].
\end{equation}
The second term in \Ref{Division of Omega} is exponentially suppressed $O\bigl(\ee^{-\beta v_F/\ta}\bigr)$ and thus vanishes in the UV limit (although see the footnote on page \pageref{Beta vs a-tilde}). For the first term, we consider initially the simpler case of $\gamma_2=0$. By inserting \Ref{oms}, using the symmetry of the integrand, and rescaling the integration variable, we get
\begin{equation}
\label{IR limit of free energy with gamma2=0}
\lim\limits_{L\to\infty}L^{-2}\Omega_{B,\beta}^< =\frac{2}{{\pi \tilde v_F \ta\beta ^2 }} \int\limits_{0}^{\beta \tilde v_F \pi /\ta} {\ln \left( {1 - \ee^{ - x}}\right) \ud x}, \qquad (\gamma_2=0)
\end{equation}
which, in the (properly scaled) UV limit, becomes
\begin{equation}
\label{UV limit of free energy with gamma2=0}
\lim\limits_{\ta\to 0^+}\lim\limits_{L\to\infty}\ta L^{-2}\Omega_{B,\beta}^< = - \frac{\pi}{3{\tilde v_F }\beta^2}, \qquad (\gamma_2=0),
\end{equation}
using that
\begin{equation}
\label{Gamma-zeta integral relation}
\int\limits_0^\infty  {x^z \ln \left( {1 - e^{ - x} } \right)\ud x}  =  - \Gamma (z + 1)\zeta(z + 2)\qquad (\Re(z)>-1)
\end{equation}
and $\zeta(2)=\pi^2/6$. For the computation with non-zero $\gamma_2$, it is useful to introduce the effective dispersion in \Ref{Renormalized dispersion} and a corresponding free energy contribution $\tilde \Omega_{B,\beta}^<$. The QFT limit of $\tilde \Omega_{B,\beta}^<$ is computed as in \Ref{IR limit of free energy with gamma2=0}--\Ref{UV limit of free energy with gamma2=0}, and this gives the right-hand side of \Ref{F}. In order to prove Result~\ref{Result free energy 2}, it is enough to show that the remainder term $\Omega_{B,\beta}^<-\tilde \Omega_{B,\beta}^<$ is subleading in $\ta$ and therefore does not contribute to \Ref{F}. In the IR limit,
\begin{equation}
\lim\limits_{L \to \infty }L^{-2}\left( {\Omega _{B,\beta }^ <   - \tilde \Omega _{B,\beta }^ <  }\right) = P.V.\int\limits_{ - \pi /\ta}^{\pi /\ta} {\int\limits_{ - \pi /\ta}^{\pi /\ta} {\frac{{\ud^2 p}}{{\left( {2\pi } \right)^2 }}\frac{1}{\beta }\ln \left( \prod_{s=\pm}{\frac{{1 - e^{ - \beta \omega _s (\vp)} }}{{1 - e^{ - \beta \tilde \omega _s (\vp)} }}} \right)} } .
\end{equation}
Symmetry of the integrand allows us to restrict the integration domain to the following triangular region: $0< p_+\leq \pi/\ta$ and $p_- \leq p_+$. Changing to polar coordinates gives
\begin{equation}
\begin{split}
\label{Free energy error}
\lim\limits_{L \to \infty }L^{-2}&\left( {\Omega _{B,\beta }^ <   - \tilde \Omega _{B,\beta }^ <  }\right) = \\&\frac{8}{{\left( {2\pi } \right)^2 }} \lim\limits_{\varepsilon\to 0^+}\int\limits_\varepsilon^{\pi /4} {\int\limits_\varepsilon^{\pi /(\ta\cos (\theta) )} {\frac{1}{\beta }\ln \left( \prod_{s=\pm}{\frac{{1 - e^{ - \beta \omega _s (\vp)} }}{{1 - e^{ - \beta \tilde \omega _s (\vp)} }}} \right)p\ud p \ud\theta } }
\end{split} 
\end{equation}
with
\begin{equation}
\label{Dispersion in polar coordinates}
\omega _\pm (\vp) =\tilde v_F p g_\pm(\theta),\qquad g_\pm(\theta) \define \sqrt {\frac{1}{2}\left( {1 \pm\sqrt {1 - A\sin ^2 (2\theta )} } \right)}
\end{equation}
and similarly for $\tilde \omega_\pm (\vp)$. We note that both the numerator and the denominator in the argument of the logarithm in \Ref{Free energy error} become zero as $p\to 0$ (for $s=\pm$) or $\theta\to 0$ (for $s=-$). However, their ratio is non-zero and finite in these limits (this is the reason for the specific choice of \Ref{Renormalized dispersion}), and the double integral therefore has a well-defined limit as $\varepsilon\to 0^+$.

To finish the proof of Result~\ref{Result free energy 2}, we show that the right-hand side of \Ref{Free energy error} remains finite as $\ta\to 0^+$. The expression \Ref{Free energy error} is of the form $\int_0^{\pi /4}f(\theta)\ud\theta$, with $f$ given as an integral over $p$. By the mean value theorem, there is a $\theta^*$ in $\left(0,\pi/4\right)$ (possibly depending on $\ta$) such that $\int_0^{\pi /4}f(\theta)\ud\theta=f(\theta^*)\pi /4$, i.e.\
\begin{equation}
\lim\limits_{L \to \infty }L^{-2}\left( {\Omega _{B,\beta }^ <   - \tilde \Omega _{B,\beta }^ <  }\right) = \frac{1}{{2\pi  }} {\int\limits_0^{\pi /(\ta\cos (\theta^*) )} {\frac{1}{\beta }\left.\ln \left( \prod_{s=\pm}{\frac{{1 - e^{ - \beta \omega _s (\vp)} }}{{1 - e^{ - \beta \tilde \omega _s (\vp)} }}} \right)\right|_{\theta=\theta^*}p\ud p } },
\end{equation}
which, after changing integration variables, can be written as
\begin{equation}
\begin{split}
\lim\limits_{L \to \infty }L^{-2}\left( {\Omega _{B,\beta }^ <   - \tilde \Omega _{B,\beta }^ <  }\right) = 
 \frac{1}{{2\pi }}\frac{1}{{{{\tilde v}_F}^2}}\frac{1}{{{\beta^3}}}  \left[ {\sum\limits_{s =  \pm } {\frac{1}{{{{ {{g_s}({\theta ^ * })} }^2}}}\int\limits_0^{\beta {{\tilde v}_F}{g_s}({\theta ^ * })\pi /(\ta\cos ({\theta ^ * }))} {\ud x} } } \right. \\ \left. { - \frac{1}{{A{{\cos }^2}({\theta ^ * })}}\int\limits_0^{\beta {{\tilde v}_F}\sqrt A \pi /\ta} {\ud x}  - \frac{1}{{A{{\sin }^2}({\theta ^ * })}}\int\limits_0^{\beta {{\tilde v}_F}\sqrt A \tan ({\theta ^ * })\pi /\ta} {\ud x} } \right]x\ln \left( {1 - {e^{ - x}}} \right).
\end{split}
\end{equation}
If $\theta^*/\ta$ remains finite in the UV limit, it follows that
\begin{equation}
\lim\limits_{\ta\to 0^+}\lim\limits_{L \to \infty }L^{-2}\left( {\Omega _{B,\beta }^ <   - \tilde \Omega _{B,\beta }^ <  }\right) =  - \frac{1}{{2\pi }}\zeta(3)\frac{1}{{{{\tilde v}_F}^2}}\left( {1 - \frac{1}{A}} \right)\frac{1}{{{\beta ^3}}},
\end{equation}
where we used \Ref{Gamma-zeta integral relation} to compute the integrals. Otherwise, one finds
\begin{equation}
\begin{split}
&\lim\limits_{\ta\to 0^+}\lim\limits_{L \to \infty }L^{-2}\left( {\Omega _{B,\beta }^ <   - \tilde \Omega _{B,\beta }^ <  }\right) = -\frac{1}{2\pi }  \zeta(3)\frac{1}{{\tilde v_F ^2 }} \frac{1}{{\beta ^3 }}\\ &\times  \lim\limits_{\ta\to 0^+} {\left( \sum\limits_{s =  \pm }{\frac{2}{{1 + s\sqrt {1 - A\sin ^2 (2\theta^* )} }} - \frac{1}{{A{{\cos }^2}({\theta ^ * })}} - \frac{1}{{A{{\sin }^2}({\theta ^ * })}}} \right)}.
\end{split}
\end{equation}
Using trigonometric identities, one shows that the terms in parenthesis on the right-hand side above add identically to zero for all $\theta^*$. This finishes the proof that the remainder is subleading in $\ta$.

\subsection{Fermion two-point functions}
\label{App:Fermion two-point functions}
Result~\ref{fermion correlation result} and Equation \Ref{RR} imply that the IR limit of the fermion two-point functions at zero-temperature are 
\begin{equation}
\label{IR limit fermion two-point function Appendix}
\begin{split}
\lim\limits_{L \to \infty} \left\langle {\psi_{r,s}^{q}(\vx,t)\psi^{q'}_{r',s'} (\vzero,0)} \right\rangle_\infty =  &\delta_{q,-q'}\delta_{r,r'}\delta_{s,s'}\delta_{x_{-s},0}\\ &\times \lim\limits_{L \to \infty}\lim\limits_{\epsilon \to 0^ +  }\frac1{2\pi\ta\epsilon}\frac{G_{r,s,r,s}(x_s\ve_s,t;\epsilon)}{G_{r,s,r,s}(\vzero,0;\epsilon)}
\end{split}
\end{equation} 
with (cf. \Ref{G})
\begin{equation}
\label{G for zero temperature}
\begin{split}
G_{r,s,r,s}(\vx,t;\epsilon ) & = G_{r,s,r,s}^< (\vx,t;\epsilon ) G_{r,s,r,s}^> (\vx,t;\epsilon )
\\ \ln G_{r,s,r,s}^\lessgtr (\vx,t;\epsilon ) & \define \ta\sum\limits_{\genfrac{}{}{0cm}{1}{\vp \in \hat \Lambda_s^*}{\left| {p_s} \right| \lessgtr \frac{\pi}{\ta}}  } {\left( {\frac{{2\pi }}{L}} \right)^2 \sum\limits_{s' =  \pm } {\frac{{\left| {v_{r,s}^{s'} (\vp)} \right|^2 }}{{\left| {p_s } \right|^2 }}\ee^{\ii \vp\cdot\vx} \ee^{ - \ii\omega _{s'} (\vp)t} \ee^{ - \epsilon \left| {p_s } \right|} } }.
\end{split}
\end{equation} 

\subsubsection{Proof of Result~\ref{Result Fermion two-point function for gamma2=0}}
\label{Fermion two-point function for gamma2=0}
For $\gamma_2=0$ and $|p_{\pm}|\leq \pi/\ta$ the formulas in \Ref{oms} and \Ref{Mrssp} simplify to
\begin{equation}
\label{omsvrss}
\omega_s(\vp)=\tilde v_F|p_s|,\qquad v^{s'}_{r,s}(\vp) = \ii\sqrt{\frac{1}{8\pi}}\delta_{s,s'}\sqrt{|p_s|}\Bigl(\sgn(p_s)\sqrt{B}+r/\sqrt{B}\Bigr)
\end{equation}
with $\tilde v_F$ in \Ref{AvF} and $B=v_F(1-\gamma_1)/\tilde v_F$ as in \Ref{BK}. Inserting this into \Ref{G for zero temperature} gives
\begin{equation}
\label{lnG1}
\begin{split}
&\ln G_{r,s,r,s}(x_s\ve_s,t;\epsilon) = \frac{\ta}{L}\sum_{|p_{-s}|\leq {\pi}/\ta} \Biggl(\frac12\sum_{p_s\neq 0}\tPiL  \frac1{|p_{s}|}\bigl(K+r\sgn(p_{s})\bigr) \\ &\times\ee^{-\ii \tilde v_F|p_{s}| t}\ee^{\ii p_{s}x_{s}}\ee^{-\epsilon |p_{s}|} +\frac12\sum_{|p_{s}|>{\pi}/{\ta}}\tPiL \frac1{|p_{s}|}\Bigl(\bigl(1+r\sgn(p_{s})\bigr)\ee^{-\ii v_F|p_{s}| t} \\ &- \bigl(K+r\sgn(p_{s})\bigr)\ee^{-\ii \tilde v_F|p_{s}| t} \Bigr) \ee^{\ii p_{s}x_{s}}\ee^{-\epsilon |p_{s}|}\Biggr)
\end{split}
\end{equation} 
with $K$ in \Ref{BK} and $p_\pm\in (2\pi/L)\Z$; the second $p_s$-sum in \Ref{lnG1} is to account for the fact that \Ref{omsvrss} holds true only for $|p_s|\leq\pi/\ta$, and  $\omega_s(\vp)=v_F|p_s|$ and $|v^{s'}_{r,s}(\vp)|^2 = \delta_{s,s'}|p_s|\bigl(1+r\sgn(p_s)\bigr)/(4\pi)$ for $|p_s|>\pi/\ta$. The sum over $p_{-s}$ in \Ref{lnG1} cancels the overall factor $\ta/L$. The first sum over $p_s$ can be computed using the series $\sum_{n=1}^\infty \ee^{-zn}/n=-\ln(1-\ee^{-z})=-\ln(z)+O(z)$ for $\Re(z)>0$, leading to
\begin{equation}
-\ln \left[ {{{\left( {\frac{{2\pi }}{L}} \right)}^{K}}{{\left( {\epsilon  - \ii(r{x_s} -{{\tilde v}_F}t)} \right)}}{{\left({(\epsilon+\ii {\tilde v}_F t)^2 + (x_s)^2} \right)}^{\left( {K - 1} \right)/2}}} \right]
\end{equation}
up to $O({L^{ - 1}})$ corrections, while the second sum over $p_s$ can be approximated, up to corrections of $O(L^{-1})$, by the integral
\begin{equation}
\begin{split} 
&\frac12\int_{\frac{\pi}{\ta}}^\infty \frac{\ud p}{p}\Bigl(2\ee^{-p(\epsilon+\ii v_F t-\ii rx_s)}- \sum_{r'=\pm}(K+r')\ee^{-p(\epsilon+\ii\tilde v_F t-\ii r'rx_s)}\Bigr) = \\ &E_1\bigl(\frac{\pi}{\ta}(\epsilon+\ii v_F t-\ii rx_s)\bigr)-\frac12\sum_{r'=\pm} (K+r')E_1\bigl(\frac{\pi}{\ta}(\epsilon+\ii\tilde v_F t-\ii r'rx_s)\bigr)
\end{split} 
\end{equation}
with $E_1$ defined in \Ref{Fermion two-point function special function}. Putting all this together, we arrive at
\begin{equation}
\label{IR limit of fermion Green's function}
\begin{split}
G_{r,s,r,s}(x_s\ve_s,t;\epsilon) =  {\left( {\frac{L}{{2\pi}}}\right)^K}{{\frac{1}{{\epsilon  - \ii(r{x_s} - {{\tilde v}_F}t)}}}}{\left( {\frac{1}{{(\epsilon+\ii\tilde{v}_F t)^2  +({x_s})^2}}} \right)^{\left( {K - 1} \right)/2}} \\ \times \frac{{\sigma {{\left( {\frac{\pi }{{\ta}}\left( {\epsilon  - \ii(r{x_s} - {{\tilde v}_F}t)} \right)} \right)}^{\left( {K + 1} \right)/2}}\sigma {{\left( {\frac{\pi }{{\ta}}\left( {\epsilon  + \ii(r{x_s} + {{\tilde v}_F}t)} \right)} \right)}^{\left( {K - 1} \right)/2}}}}{{\sigma \left( {\frac{\pi }{{\ta}}\left( {\epsilon  - \ii(r{x_s} - {v_F}t)} \right)} \right)}}
\left( {1 + O({L^{ - 1}})} \right)
\end{split}
\end{equation}
with $\sigma(z)$ defined in \Ref{Fermion two-point function special function}. The function $E_1$ has the asymptotic behavior (see e.g.\ \cite{AS}, Equations 5.1.11 and 5.1.52)
\begin{equation}
E_1(z)=-\gamma-\ln(z)+O(z),\qquad E_1(z)=\frac{\ee^{-z}}{z+1}\Bigl(1 + O\left((z+1)^{-2}\right)\Bigr), 
\end{equation} 
implying \Ref{Asymptotics of special function}. It follows that, for $\epsilon\ll\ta$,
\begin{equation}
\label{Green's function at x=t=0}
G_{r,s,r,s}(\vzero,0;\epsilon) = {\left( {\frac{L}{{2\pi }}} \right)^K}{\left( {{\ee^\gamma }\frac{\pi }{{\ta}}} \right)^{K - 1}}\frac{1}{\epsilon }\left( {1 + O\left( {\epsilon /\ta} \right) + O({L^{ - 1}})} \right). 
\end{equation}
Inserting this and \Ref{IR limit of fermion Green's function} into \Ref{IR limit fermion two-point function Appendix}, we obtain \Ref{IR limit fermion two-point function}--\Ref{Frsxt}. 

\subsubsection{Proof of Result~\ref{QFT limit Fermion two-point function}}
\label{Fermion two-point function: general case}
The computation of \Ref{IR limit fermion two-point function Appendix}--\Ref{G for zero temperature} for non-zero $\gamma_2$ mimics the computation of the free energy in Appendix~\ref{QFT limit of free energy appendix}. We add and subtract to the non-trivial part (i.e.\ $\ln G_{r,s,r,s}^< $) of \Ref{G for zero temperature} a term
\begin{equation}
\ln \tilde G_{r,s,r,s}^< (x_s\ve_s,t;\epsilon ) \define \ta\sum\limits_{\genfrac{}{}{0cm}{1}{\vp \in \hat \Lambda_s^*}{\left| {p_s} \right| < {\pi}/{\ta}}  } {\left( {\frac{{2\pi }}{L}} \right)^2 \frac{{K\left| {p_s } \right| + rp_s }}{{4\pi\left| {p_s } \right|^2 }}\ee^{\ii p_s x_s} \ee^{ - \ii\tilde \omega _{s} (\vp)t} \ee^{ - \epsilon \left| {p_s } \right|} } 
\end{equation}
with $\tilde \omega _{s} (\vp)$ in \Ref{Renormalized dispersion} and $K$ in \Ref{K2}. We note that this expression follows from that of $\ln G_{r,s,r,s}^< $ by everywhere replacing the original dispersion relation $\omega_\pm (\vp)$ with the effective dispersion $\tilde \omega_\pm (\vp)$ (i.e.\ making the replacement in the coefficients \Ref{Mrssp}--\Ref{U} for the case $\gamma_2=0$, and in the time-dependent part of \Ref{G for zero temperature}). Result~\ref{QFT limit Fermion two-point function} can then be obtained by analyzing the QFT limits of $\tilde G_{r,s,r,s}\define\tilde G_{r,s,r,s}^<  G_{r,s,r,s}^>$ and $E_{r,s,r,s}\define G_{r,s,r,s}^< / \tilde G_{r,s,r,s}^< $ separately. The former can be computed as in Appendix~\ref{Fermion two-point function for gamma2=0}, and one finds expressions identical to \Ref{IR limit of fermion Green's function} and \Ref{Green's function at x=t=0}, except with ${\tilde v}_F$ replaced by ${\tilde v}_F\sqrt{A}$ and $K$ now given in \Ref{K2}. In the regime \Ref{QFT regime}, we obtain 
\begin{equation}
\begin{split}
&\tilde G_{r,s,r,s}(x_s\ve_s,t;\epsilon) = 
 {\left( {\frac{L}{{2\pi }}} \right)^K}{ {\frac{1}{{\epsilon  - \ii(r{x_s} - {{\tilde v}_F}\sqrt{A}t)}}}}\\ &\times {\left( {\frac{1}{{(\epsilon+\ii\tilde{v}_F\sqrt{A}t)^2+(x_s)^2} }} \right)^{\left( {K - 1} \right)/2}}
\left( 1 + O({\ta})\right) \left(1+O({L^{ - 1})} \right)
\end{split}
\end{equation}
using (the equivalent of) \Ref{IR limit of fermion Green's function} and the second asymptotic relation in \Ref{Asymptotics of special function}. 

The error term in the IR limit is given by
\begin{equation}
\label{IR limit of error term}
\begin{split}
\lim\limits_{L \to \infty} \ln E_{r,s,r,s} (x_s\ve_s,t;\epsilon ) = & P.V. \int\limits_{ - \pi /\ta}^{\pi /\ta} {\int\limits_{ - \pi /\ta}^{\pi /\ta} {\ud^2 p\; \ta\left( {\sum\limits_{s' = \pm } {\frac{{\left| {v_{r,s}^{s'} (\vp)} \right|^2 }}{{\left| {p_s } \right|^2 }}\ee^{ - \ii\omega _{s'} (\vp)t} }} \right.}} \\& \left. {- \frac{1}{{4\pi \left| {p_s } \right|^2 }}\left( {K\left| {p_s } \right| + rp_s } \right)e^{ - \ii\tilde \omega _s (\vp)t} } \right)\ee^{\ii p_s x_s } \ee^{ - \epsilon \left| {p_s } \right|}
\end{split}
\end{equation}
with the Cauchy principal value defined in \Ref{Cauchy principal value}. We will see that, in the limit $\ta\to 0^+$, this error term yields a finite constant (depending only on the coupling parameters) if $x_s = t = 0$, while it vanishes in the regime \Ref{QFT regime}. Transforming to polar coordinates and using the independence of $E_{r,s,r,s} (\vzero,0;\epsilon ) $ on $r$ and $s$, we find by integrating over the radial coordinate and using the rotational symmetry of the integrand (we set, without loss of generality, $s=+$ on the right-hand side of \Ref{IR limit of error term})
\begin{equation}
\label{EE} 
\begin{split}
&\lim\limits_{L \to \infty } \ln E_{r, s ;r, s } (\vzero,0;0^+) = \mathop {\lim }\limits_{\varepsilon  \to 0^ +  } \left[ {\int\limits_\varepsilon ^{\pi /4} { \frac{\ud\theta}{{\cos (\theta )}}}  + \int\limits_{\pi /4}^{\pi /2 - \varepsilon } { \frac{\ud\theta}{{\sin (\theta )}}} } \right]\frac{1}{{\cos ^2 (\theta )}} \\ & \times \left( {\sum\limits_{s' =  \pm } {\frac{1}{4}\left( {1 + \frac{{s'\cos  (2\theta)}}{{\sqrt {1 - A\sin ^2 (2\theta )} }}} \right)\left( {\frac{{B\cos ^2 (\theta )}}{{g_{s'} (\theta )}} + \frac{{g_{s'} (\theta )}}{B}} \right) - K\left|\cos (\theta )\right|} } \right)  
\end{split}
\end{equation}
with $g_{s'} (\theta )$ in \Ref{Dispersion in polar coordinates}. The limit $\varepsilon  \to 0^ +$ of the right-hand side exists (the integrand remains bounded as $\theta  \to 0^ +$ and $\theta  \to (\pi/2)^-$)   and it is conveniently denoted by $ - \ln \left( {C(\gamma _1 ,\gamma _2 )} \right)$,  i.e.\
\begin{equation}
\label{C12} 
C(\gamma_1,\gamma_2) \define \exp\Bigl( - \lim\limits_{L \to \infty } \ln E_{r, s ;r, s } (\vzero,0;0^+)\Bigr) . 
\end{equation} 
We observe that $C(\gamma_1,0)=1$.

We proceed by showing that \Ref{IR limit of error term} vanishes in the regime \Ref{QFT regime} in the limit $\ta\to 0^+$. We only discuss the case $s=-$ (since the case $s=+$ is nearly identical). Furthermore, it is enough to restrict the integration domain on the right-hand side of \Ref{IR limit of error term} to the following triangular region, $0< p_+\leq \pi/\ta$ and $p_- \leq p_+$ (the other seven triangles are analyzed in the same way). Transforming to polar coordinates and integrating over the radial coordinate, we find 
{\small\begin{equation}
\begin{split}
\lim\limits_{\varepsilon  \to 0^ +  } \int\limits_\varepsilon ^{\pi /4} {\ud\theta \frac{{\tilde a}}{{4\pi \ii\sin ^2 (\theta )}}} \left( {\sum\limits_{s' =  \pm } {\frac{1}{4}\left( {1 - \frac{{s'\cos  (2\theta )}}{{\sqrt {1 - A\sin ^2 (2\theta )} }}} \right)\left( {\frac{{B\sin ^2 (\theta )}}{{g_{s'} (\theta )}} + \frac{{g_{s'} (\theta )}}{B} + 2r\sin (\theta )} \right)} } \right. \\ \left. { \times \frac{{\ee^{\ii\pi \left( {\sin (\theta )x_s  - \tilde v_F g_{s'} (\theta )t} \right)/\left( {\tilde a\cos (\theta )} \right)}  - 1}}{{\sin (\theta )x_s  - \tilde v_F g_{s'} (\theta )t}} - \left( {K + r} \right)\frac{{\ee^{\ii\pi \tan (\theta )\left( {x_s  - \tilde v_F \sqrt A t} \right)/\tilde a}  - 1}}{{x_s  - \tilde v_F \sqrt A t}}} \right)
\end{split}.
\end{equation} }
The integrand remains bounded in the whole integration domain as $\varepsilon  \to 0^ +$. We can therefore apply the mean value theorem (cf. Appendix~\Ref{QFT limit of free energy appendix}) to its real and imaginary parts separately. The end result is $\ta$ multiplied by a bounded complex number (the bound is independent of $\ta$). It follows that the UV limit of the above expression vanishes, and, more generally
\begin{equation}
\lim\limits_{\ta \to 0^ +  } \lim\limits_{L \to \infty} \ln E_{r,s,r,s} (x_s\ve_s,t;\epsilon ) = 0. 
\end{equation}
Inserting all results above into  
\begin{equation}
\begin{split}
&\lim\limits_{\ta \to 0^ +  }\lim\limits_{L \to \infty} \Bigl(\frac{\ee^{\gamma}\pi L_0}{\ta}\Bigr)^{K-1}\left\langle {\psi_{r,s}^{q}(\vx,t)\psi^{q'}_{r',s'} (\vzero,0)} \right\rangle_\infty = \delta_{q,-q'}\delta_{r,r'}\delta_{s,s'}\\ &\times\lim\limits_{\ta \to 0^ +  }\lim\limits_{L \to \infty}\lim\limits_{\epsilon \to 0^ +  }\delta_{x_{-s},0} \bigl(\frac{\ee^{\gamma}\pi L_0}{\ta}\bigr)^{K-1}\frac1{2\pi\ta\epsilon}\frac{\tilde G_{r,s,r,s}(x_s\ve_s,t;\epsilon)}{\tilde G_{r,s,r,s}(\vzero,0;\epsilon)}\frac{E_{r,s,r,s}(x_s\ve_s,t;\epsilon)}{E_{r,s,r,s}(\vzero,0;\epsilon)}
\end{split}
\end{equation} 
and using that $\delta_{x_{-s},0}/\ta\to \delta(x_{-s})$ as $\ta\to 0^+$ yields Result~\ref{QFT limit Fermion two-point function}.

\subsection{Density-density correlation function for $\gamma_2=0$}
\label{Density-density correlation function for gamma2=0}
Inserting \Ref{omsvrss} in Result~\ref{density correlation result}, we find after a trivial integration and by adding and subtracting terms
{\small
\begin{equation}
\begin{split}
&\lim\limits_{L \to \infty }{\ta \left\langle {J_{r,s} (\vx,t;\epsilon) J_{r',s'}(\vzero,0;\epsilon)} \right\rangle_\infty } = {\delta _{s,s'}}\frac{1}{{\ta}}{\delta _{{x_{ - s}},0}}\frac{1}{{4{{\left( {2\pi } \right)}^2}}} \\ &\times \Biggl( {\int\limits_{ - \infty }^\infty  {\ud p\left| p \right|\left( {B + rr'{B^{ - 1}} + {\mathop{\rm sgn}} (p)\left( {r + r'} \right)} \right){\ee^{ - \ii{{\tilde v}_F}\left| p \right|t}}{\ee^{ - \epsilon \left| p \right|}}{\ee^{\ii p{x_s}}}}  + \int\limits_{\left| p \right| > \pi /\ta} {\ud p\left| p \right|} } \Biggr. \\ &\Biggl. { \times \left( {{\delta _{r,r'}}2\left( {1 + {\mathop{\rm sgn}} (p)r} \right){\ee^{ - \ii{v_F}\left| p \right|t}} - \left( {B + rr'{B^{ - 1}} + {\mathop{\rm sgn}} (p)\left( {r + r'} \right)} \right){\ee^{ - \ii{{\tilde v}_F}\left| p \right|t}}} \right){\ee^{\ii p{x_s}- \epsilon \left| p \right|} }} \Biggr)
\end{split}.
\end{equation} 
}
Computing the first integral above gives the first two terms within the parenthesis in \Ref{JJ1}. For the second integral, we scale integration variables and use 
\begin{equation}
{{\alpha _1}\left( z \right) \define \int\limits_1^\infty  {t{\ee^{ -zt}}\ud t}  = \frac{{{\ee^{ - z}}}}{{{z^2}}}\left( {1 + z} \right)} \qquad (\Re(z) > 0)
\end{equation}
(see \cite{AS}, Equations 5.1.5 and 5.1.8). This gives the last term within the parenthesis in \Ref{JJ1}.

\newsection{Correction terms}
\label{appX} 
\subsection{Derivation} 
\label{AppX1} 
We consider the tight binding band relations $\epsilon(\vk)$ in \Ref{eps0} and the nodal points $\vQ_{r,s}$ in \Ref{Qrs}. We Taylor expand
\begin{equation}
\label{expand} 
\epsilon(\vQ_{r,s}+\vk) =   \epsilon(\vQ_{r,s}) + v_F\sum_{j=1}^3\sum_{k=0}^j \alpha_{j-k,k}\ta^{j-1} (rk_s)^{j-k}(k_{-s})^k + O(\ta^4|\vk|^4) 
\end{equation} 
with $v_F$, given in Equation (6) in \cite{EL1}, defined such that $\alpha_{1,0}=1$. By straightforward computations we find the following other non-zero coefficients, 
\begin{gather}
\begin{aligned}
\label{alphajk}
\alpha_{2,0} &= \frac{(1-\kappa)[t\cos(Q) + 2t'\cos(2Q)]}{4\sin(Q)[t+2t'\cos(Q)]}, & \qquad \alpha_{0,2} &= \frac{(1-\kappa)[t\cos(Q) +2t']}{4\sin(Q)[t+2t'\cos(Q)]} \\ \alpha_{3,0} &= -\frac{(1-\kappa)^2[t+8t'\cos(Q)]}{24[t+2t'\cos(Q)]}, & \qquad \alpha_{1,2} &= -\frac{(1-\kappa)^2 t}{8[t+2t'\cos(Q)]}
\end{aligned} 
\end{gather} 
with parameters $t$, $t'$, $Q$ and $\kappa$ defined in \cite{EL1} (we used (11) in \cite{EL1}). The first non-trivial term $v_Frk_s$ on the right-hand side in \Ref{expand} corresponds to $H_0$ in \Ref{Free part of regularized Mattis Hamiltonian}. The terms $v_F(\alpha_{2,0}\ta k_s^2+\alpha_{3,0}\ta^2rk_s^3)$ correspond to $H_2$ in \Ref{Hcorr2}. Finally, the terms $(\alpha_{0,2}\ta + \alpha_{1,2}\ta^2 r k_s)k_{-s}^2$ correspond to $H_3$ in \Ref{Hcorr3}; we inserted 
\begin{equation}
\ta^2 k_{-s}^2 = 2-2\cos(k_{-s}\ta) +O(\ta^4k_{-s}^4)
\end{equation}
and ignored the $O(\ta^4k_{-s}^4)$-terms to obtain an expression local in the ``position space'' defined in \Ref{Nodal position space} (this corresponds to using a lattice second derivative in the $x_{-s}$-direction).  

Note that $\alpha_{0,2}=0$ at the special parameter value $Q=\arccos(-2t'/t)$, as mentioned in Section~\ref{Final remarks}, Remark~5. 

\subsection{Bosonization} 
\label{AppX2} 
We introduce the notation $\xxa \cdots \xxe$ for normal ordering of general boson multilinears, defined as usual:
\begin{equation}
\label{xx} 
\xxa\hat\jmath_{r_1,A_1}(p_1)\cdots\hat\jmath_{r_N,A_N}(p_N)\xxe \;  \define \hat\jmath_{r_{\sigma(1)},A_{\sigma(1)}}(p_{\sigma(1)})\cdots\hat\jmath_{r_{\sigma(N)},A_{\sigma(N)}}(p_{\sigma(N)})
\end{equation}
with the permutation $\sigma$ of $N$ objects such that $r_{\sigma(1)}p_{\sigma(1)}\leq r_{\sigma(2)}p_{\sigma(2)}\leq \cdots \leq r_{\sigma(N)}p_{\sigma(N)}$. Note that, for boson bilinears, this normal ordering is equivalent to the one in \Ref{NormalOrder}.

The following is a straightforward generalization of Proposition~\ref{Proposition Bosonization 1D}(b).
\begin{proposition}
For $b\in\R$,
\begin{equation}
\label{Generalized Kronig identity}
\begin{split}
\lim_{\epsilon\to 0^+} :\! V_{r,A}(x;\epsilon)^\dag V_{r,A}(x + b;\epsilon) \!:\; = \left(2\ii L \sin\left(\frac{rb\pi}{L}\right)\right)^{-1}  \\  \times\Big[\xxa\exp\Bigl({{r2\pi \ii\sum\limits_{m = 0}^\infty  {\frac{{b^{m + 1} }}{{(m + 1)!}}(\partial _x)^m j_{r,A} (x)} }}\Bigr) \xxe - 1 \Bigr]
\end{split}
\end{equation}
with $V_{r,A}(x;\epsilon)$ in  \Ref{V} and
\begin{equation}
j_{r,A} (x) = \frac{1}{L}\sum\limits_{p\in\frac{2\pi}L\Z} {\hat\jmath_{r,A} (p)\ee^{\ii px} }.
\end{equation}
\end{proposition}
\noindent ({\em Proof:} See e.g.\, \cite{CL}.)

\medskip

A formal expansion in the parameter $b$ of the left- and right-hand sides of \Ref{Generalized Kronig identity}, together with an integration of both sides with respect to $x$, give
{\small \begin{equation}
\begin{split}
&\lim_{\epsilon\to 0^+} \sum\limits_{n=0}^\infty\frac{b^n}{n!} \int\limits_{-L/2}^{L/2}\ud x:\! V_{r,A}(x;\epsilon)^\dag (\partial_x)^n V_{r,A}(x;\epsilon) \!:\; = 
\int\limits_{-L/2}^{L/2}\ud x \xxa j_{r,A}\xxe \\ &+b \int\limits_{-L/2}^{L/2} {\ud x  \frac{1}{2} \xxa r 2\pi\ii ({ j_{r,A}})^2 \xxe } + \frac{b^2}{2!} \int\limits_{-L/2}^{L/2} {\ud x \frac{1}{3} \xxa \left( {{{\left( {r2\pi \ii} \right)}^2}{({ {{j_{r,A}}} })^3} + {{\left( {\frac{\pi}{L}} \right)}^2}{j_{r,A}}} \right) \xxe}  \\ &+\frac{b^3}{3!} \int\limits_{-L/2}^{L/2} {\ud x \frac{1}{4} \xxa \left( {{{\left( {r2\pi \ii} \right)}^3}({j_{r,A}})^4 - r2\pi \ii{{\left( {{\partial _x}{j_{r,A}}} \right)}^2}}+ 2{{\left( {\frac{\pi }{L}} \right)}^2} {  r2\pi \ii{ ({{j_{r,A}}})^2}} \right) \xxe} 
+ O(b^4)
\end{split}
\end{equation}}
where we suppress the argument in $j_{r,A}(x)$. Using that, we can bosonize the correction term in \Ref{Hcorr2} and obtain
\begin{equation}
\begin{split} 
H_2 =& \sum_{r,s=\pm} v_F \tint{s} \ud^2 x \xx\!\Biggl(\ta\alpha_{2,0}\Bigl[ \frac43(\pi\ta)^2 (J_{r,s})^3 -\frac13\Bigl(\frac{\pi}{L}\Bigr)^2J_{r,s}\Bigr] \\ &+ \ta^2\alpha_{3,0}\Bigl[ 2(\pi\ta)^3 (J_{r,s})^4  +\frac12(\pi\ta)(\partial_sJ_{r,s})^2 -(\pi\ta)\Bigl(\frac{\pi}{L}\Bigr)^2 (J_{r,s})^2 \Bigr]\Biggr)\!\xx   
\end{split} 
\end{equation} 
where we use the notation in \Ref{tint}. We insert \Ref{PiPhi from J} and obtain \Ref{Hcorr2_bosonized}. 

\begin{remark} 
The bosonization of higher derivative terms has been discussed in the literature before; see e.g.\ \cite{Kh} and references therein. 
\end{remark} 

\newsection{Index sets}
\label{Index sets}
We collect for easy reference the definitions of various index sets used throughout the text:
\begin{align}
\tag*{\Ref{Nodal position space}}
&\Lambda_s = \left\{\vx\, : \, x_s\in\mathbb{R},\;  x_{-s}\in \ta\mathbb{Z},\; -\frac{L}{2}\leq x_\pm < \frac{L}{2} \right\} \\ \tag*{\Ref{Continuum nodal momentum region}} &\Lambda_{s}^* = \left\{\vk\in \tPiL\Bigl(\mathbb{Z} +\frac{1}{2}\Bigr)^2 \; : \; -\frac{\pi}{\ta} \leq k_{-s}< \frac{\pi}{\ta} \right\} \\ \tag*{\Ref{Boson momentum set}} &\tilde\Lambda_{s}^* = \left\{\vp\in \tPiL\mathbb{Z}^2 \; : \; -\frac{\pi}{\ta} \leq p_{-s}< \frac{\pi}{\ta} \right\}\\ \tag*{\Ref{Momentum set with zero-component}} &\hat\Lambda_s^*=\left\{\vp\in \tilde\Lambda_s^*\; :\; p_s\neq 0\right\}\\
\tag*{\Ref{1D lattice}} &\Lambda_{1\rm D} = \left\{ x \in \ta\mathbb{Z}\; : \;  -\frac{L}{2}\leq x <\frac{L}{2} \right\} \\ \tag*{\Ref{1D Boson Fourier space}} &\tilde\Lambda_{1\rm D}^* = \left\{ p \in \tPiL\mathbb{Z}\, : \,  -\frac{\pi}{\ta} \leq p< \frac{\pi}{\ta} \right\} \\ \tag*{\Ref{1D Boson Fourier space with no zero}} &\hat\Lambda_{1\rm D}^* = \left\{ p \in \tilde\Lambda_{1\rm D}^*\, : \,  p \neq 0 \right\}
\end{align}

\end{document}